\documentclass[prd,superscriptaddress,amsfonts,amssymb,amsmath,twocolumn,floatfix]{revtex4-2}
\usepackage{bm}
\usepackage{amsfonts}
\usepackage{latexsym}
\usepackage{graphicx}
\usepackage{amsmath}
\usepackage{palatino}
\usepackage{bigints}
\usepackage{mathpazo}
\usepackage{textcomp}
\usepackage{float}
\usepackage{booktabs}
\usepackage{dcolumn}
\usepackage{booktabs}
\usepackage{multirow}
\usepackage{hyperref}
\hypersetup{colorlinks,citecolor=blue}
\usepackage{amsmath}
\usepackage{xcolor}
\usepackage{orcidlink}
\usepackage{commath}
\usepackage{subcaption}

\def\jnl@style{\it}
\def\aaref@jnl#1{{\jnl@style#1}}

\def\aaref@jnl#1{{\jnl@style#1}}

\def\aj{\aaref@jnl{AJ}}                   
\def\apj{\aaref@jnl{ApJ}}                 
\def\apjl{\aaref@jnl{ApJ}}                
\def\apjs{\aaref@jnl{ApJS}}               
\def\apss{\aaref@jnl{Ap\&SS}}             
\def\aap{\aaref@jnl{A\&A}}                
\def\aapr{\aaref@jnl{A\&A~Rev.}}          
\def\aaps{\aaref@jnl{A\&AS}}              
\def\mnras{\aaref@jnl{Mon.~Not.~Roy.~Astron.~Soc.}}             
\def\prd{\aaref@jnl{Phys.~Rev.~D}}        
\def\prc{\aaref@jnl{Phys.~Rev.~C}}  
\def\prl{\aaref@jnl{Phys.~Rev.~Lett.}}    
\def\qjras{\aaref@jnl{QJRAS}}             
\def\skytel{\aaref@jnl{S\&T}}             
\def\ssr{\aaref@jnl{Space~Sci.~Rev.}}     
\def\zap{\aaref@jnl{ZAp}}                 
\def\nat{\aaref@jnl{Nature}}              
\def\aplett{\aaref@jnl{Astrophys.~Lett.}} 
\def\apspr{\aaref@jnl{Astrophys.~Space~Phys.~Res.}} 
\def\physrep{\aaref@jnl{Phys.~Rep.}}      
\def\physscr{\aaref@jnl{Phys.~Scr}}       
\def\commat{\aaref@jnl{Comm.~Math.~Phys.}}              
\def\science{\aaref@jnl{Science}}               
\def\cqg{\aaref@jnl{Classical Quant.~Grav.}}            
\def\jpcs{\aaref@jnl{JPCS}}                                     
\def\ijmpd{\aaref@jnl{Int.~J.~Mod.~Phys.~D}}                    
\def\grg{\aaref@jnl{Gen.~Relat.~Gravit.}}               
\def\rpp{\aaref@jnl{Rep.~Prog.~Phys.}}          
\def\npa{\aaref@jnl{Nucl.~Phys.~A}}        
\def\lrr{\aaref@jnl{Living Rev.~Rel.}}                   
\def\jcap{\aaref@jnl{J.~Cosmology Astropart.~Phys.}}    
\def\rmp{\aaref@jnl{Rev.~Mod.~Phys.}}   
\def\epjc{\aaref@jnl{Eur.~Phys.~J.~C}}

\allowdisplaybreaks[1]

\begin{document}

\color{black}       

\title{Phase structure of charged AdS black holes surrounded by exotic fluid with modified Chaplygin equation of state}

\author{Y. Sekhmani\orcidlink{0000-0001-7448-4579}}
\email[Email: ]{sekhmaniyassine@gmail.com}
\affiliation{Ratbay Myrzakulov Eurasian International Centre for Theoretical Physics, Astana 010009, Kazakhstan.}
\author{ J. Rayimbaev\orcidlink{0000-0001-9293-1838}}
\email[Email: ]{javlon@astrin.uz}
\affiliation{
New Uzbekistan University, Mustaqillik Ave. 54, Tashkent 100007, Uzbekistan}
\affiliation{
Central Asian University, Tashkent 111221, Uzbekistan}
\affiliation{
Tashkent University of Applied Sciences, Gavhar Str. 1, Tashkent 100149, Uzbekistan }
\affiliation{Institute of Fundamental and Applied Research, National Research University TIIAME, Kori Niyoziy 39, Tashkent 100000, Uzbekistan}
\author{G. G. Luciano\orcidlink{0000-0002-5129-848X}}

\email[Email: ]{giuseppegaetano.luciano@udl.cat}

\affiliation{Department of Chemistry, Physics, Environmental and Soil Sciences, Escola Polit\`ecnica Superior, Universitat de Lleida, Av. Jaume
II, 69, 25001 Lleida, Spain.}

\author{R. Myrzakulov \orcidlink{0000-0002-5274-0815}}
\email[Email: ]{rmyrzakulov@gmail.com }
\affiliation{L. N. Gumilyov Eurasian National University, Astana 010008,
Kazakhstan.}
\affiliation{Ratbay Myrzakulov Eurasian International Centre for Theoretical
Physics, Astana 010009, Kazakhstan.}

\author{D. J. Gogoi\orcidlink{0000-0002-4776-8506}}
\email[Email: ]{moloydhruba@yahoo.in}

\affiliation{Department of Physics, Moran College, Moranhat, Charaideo 785670, Assam, India.}
\affiliation{Theoretical Physics Division, Centre for Atmospheric Studies, Dibrugarh University, Dibrugarh
786004, Assam, India.}

\begin{abstract}
By considering the concept of the modified Chaplygin gas (MCG) 
as a single fluid model unifying dark energy and dark matter, we construct a static, spherically charged black hole (BH) solution in the framework of General Relativity. The $P-V$ criticality of the charged anti-de Sitter (AdS) BH with a surrounding MCG is explored in the context of the extended phase space, where the negative cosmological constant operates as a thermodynamical pressure. This critical behavior shows that the small/large BH phase transition is analogous to the van der Waals liquid/gas phase transition. Accordingly, along the $P-V$ phase spaces, we derive the BH equations of state and then numerically evaluate the corresponding critical quantities. Similarly, critical exponents are identified, along with outcomes demonstrating the scaling behavior of thermodynamic quantities near criticality into a universal class. The use of  \emph{geometrothermodynamic} (GT) tools finally offers a new perspective on discovering the critical phase transition point. At this stage, we apply a class of GT tools, such as Weinhold, Ruppeiner, HPEM, and Quevedo classes I and II. The findings are therefore non-trivial, as each GT class metric captures at least either the physical limitation point or the phase transition critical point. Overall, this paper provides a detailed study of the critical behavior of the charged AdS BH with surrounding MCG.

\end{abstract}

\maketitle

\section{Introduction}

General relativity (GR) provides the best available description of gravity so far~\cite{Einstein:1915ca}. Among its most enthralling predictions, gravitational waves (GWs) and black holes (BHs) deserve special mention, as they represent the ultimate confirmation of Einstein's theory. On the one hand, the first direct detection of GWs dates back about a decade ago with the emission of signals from a binary BH merger and the subsequent ringdown of the single resulting BH~\cite{LIGOScientific:2016aoc}. On the other hand, X-rays coming from superheated material swirling around a dark object were recognized as distinctive evidence of a central BH (Cygnus X-1, located within the Milky Way) in the early sixties, although we had to wait another half-century to capture the first ever picture of a similar spacetime oddity - M87*~\cite{EventHorizonTelescope:2019dse}.
Since then, GWs and BHs have been extensively addressed as a new way to probe the cosmos at a fundamental level. In particular, it is commonly accepted that BH physics could offer valuable insight into the unification of GR, quantum theory, and statistical mechanics~\cite{Hawking:1975vcx,Bekenstein:1973ur,Bardeen:1973gs,Hawking:1982dh}, opening a novel route to quantum gravity. 

A class of solutions of Einstein's equations that have been attracting growing interest in recent years are anti-de Sitter (AdS) BHs. The observation that asymptotically AdS BHs can be modeled in the language of dual thermal field theory has motivated a fluid-like description of the underlying microphysics. Interestingly enough, such studies have revealed that the equilibrium thermodynamics of BHs can be investigated by considering the geometric properties of their event horizons and other relevant spacetime features~\cite{Davies:1977bgr,Cai:1998ep,Quevedo:2007mj,Quevedo:2008ry,Sahay:2010tx,Wei:2015iwa,Dehyadegari:2016nkd,KordZangeneh:2017lgs,Wei:2019yvs,Wei:2019uqg,Xu:2020gud,Ghosh:2019pwy}. In the ensuing  \emph{geometrothermodynamic} (GT) picture, the scalar curvature of the BH metric represents the thermodynamic (microstructure) interaction, with positive curvature indicating prevailing repulsion and vice versa. Moreover, singularities represent the breakdown of the classical theory because they correspond to an infinite gravitational interaction which cannot be treated properly in GR. The microscopic behavior of BHs has been analyzed in the geometrothermodynamic framework for a wide class of systems~\cite{Cai:1998ep,Wei:2015iwa,Wei:2019uqg,Xu:2020gud,Ghosh:2020kba} and in various entropic scenarios~\cite{Rani:2022xza,Jawad:2022lww,Luciano:2023fyr,Luciano:2023bai,Luciano:2023wtx,Luciano:2022hhy}.

Further suggestive features of AdS BHs are 
\emph{phase transitions} and critical phenomena, which were first demonstrated in the phase space of Schwarzschild-AdS BH non-rotating and uncharged~\cite{Hawking:1982dh}. This seminal discovery has opened up a new line of research in the field of BH thermodynamics. In this context, non-trivial results
have been achieved for spinning branes~\cite{Cvetic:1999ne,Cvetic:1999rb}
and charged Reissner-Nordstr\"om (RN) BHs~\cite{Chamblin:1999hg,Chamblin:1999tk}, whose 
first-order phase transitions display a critical behavior analogous to a van der Waals (vdW)-like (i.e. 
liquid-gas) change of phase. Recently, the correspondence between BHs and condensed matter systems has been explored further by incorporating the variation of the cosmological constant $\Lambda$ in the first law of BH thermodynamics~\cite{Caldarelli:1999xj,Kastor:2009wy,Dolan:2010ha,Dolan:2011xt,Dolan:2011jm,Cvetic:2010jb,Lu:2012xu} (see also~\cite{Kubiznak:2012wp} for more discussion), which allows maintaining consistency with the Smarr relation~\cite{Kastor:2009wy}. In this scenario, the BH mass is identified with enthalpy rather than with internal energy. Furthermore, since $\Lambda$ corresponds to pressure, it is natural to consider the thermodynamic volume of BH as its conjugate variable. 

Taking the above arguments seriously, various thermodynamic quantities (such as adiabatic compressibility, specific
heat at constant pressure, etc.) have been computed by using standard thermodynamic machinery~\cite{Dolan:2010ha,Dolan:2011xt,Dolan:2011jm}.
Remarkably, the critical behaviour of AdS BHs has been reconsidered in an \emph{extended phase space}, including pressure and volume as thermodynamic variables~\cite{Dolan:2011xt}. As a result of this approach, the $P = P(V, T)$ equation of state has been analyzed for a rotating charged AdS BH, emphasizing analogies with the vdW $P-V$ diagram. The recent study in~\cite{Kubiznak:2012wp} has gained further progress in the identification of charged BH first-order transitions with the standard liquid-gas phase transitions by examining the behavior of the Gibbs free energy of a RN-AdS BH in the canonical ensemble (i.e. fixed charge).  

Predicting critical points in an BH system is crucial for phase transition analysis. In this way, GT provides an alternative description using a thermodynamic metric, governed by a thermodynamic potential and its derivative with respect to extensive parameters. This metric is implemented in the thermodynamic equilibrium phase space. In this regard, the Weinhold geometry is considered the first attempt to analyze the critical phase transition of a thermodynamic system ~\cite{Weinhold1,Weinhold2}. An alternative approach, applied to the exploration of critical thermodynamic characteristics, is the Ruppeiner geometry  ~\cite{Rupp1,Rupp2}. As such, these two approaches are known to be closely linked via the inverse of temperature ~\cite{WR}. In fact, the Legendre transformation is vital in the context of GT due to its invariance on the thermodynamic potential ~\cite{Leg}. This invariance enables the use of the Legendre-invariant metric in thermodynamic phase space~\cite{Quevedo:2011np}. By the way, the Quevedo metric is a better example of the Legendre-invariant metric ~\cite{Que1, Que2}. Furthermore, Hendi et al.~\cite{Hendi:2015rja, Hendi:2015fya, Hendi:2015xya, Hendi:2015hoa, Hendi:2015pda} invented another representation base of a thermodynamic system, a metric that provides a perfect correlation between the curvature scalar and the phase transition point of the heat capacity. Similarly, Mansoori et al. ~\cite{Mansoori:2013pna, Mansoori:2014oia, Sekhmani:2023qqe} currently developing another thermodynamical metric. However, there were a few limitations to these thermodynamic metrics. In general, the main motivation behind the use of GT is that it offers an independent view of thermodynamic systems, aiding in the inspection of the bound points, patterns, and stability of phase transitions. It provides microscale behavior and the Ricci scalar sign, indicating repulsive or attractive interactions, i.e., negative or positive, respectively, along the transition curve, while $R = 0$ indicates the absence of interaction ~\cite{Rupp1}.

In the framework of modern cosmology, one of the most conspicuous problems is the explanation of the matter/energy content of the Universe. It is a fact that the amount of invisible dark sectors is about 95\% of the total density~\cite{Planck:2018vyg}. This dominance has stimulated great efforts to uncover the origin of these mysterious entities. A challenging possibility is that dark components may surround (or even be created inside) BHs. Along this direction, special focus has been devoted to analyzing static spherically symmetric BHs surrounded by quintessence matter (see~\cite{Kiselev:2002dx,Fernando:2012ue,Malakolkalami:2015cza,Hussain:2014fca,Ghosh:2016ddh,Ghosh:2017cuq} and references therein). Other solutions are in~\cite{Hassaine:2007py,Dey:2004yt,Cai:2004eh,Bizon:1990sr,Li:2018hhy,Toledo:2019amt,Gadbail:2022qnw}. Concretely, attempting to describe the acceleration of the universe attributable to a kind of exotic negative-pressure fluid, widely known as dark energy (DE), is a challenge in the fields of astrophysics and theoretical physics. Based on this challenge, various dark energy model candidates provide phenomenological and theoretical predictions to explain the acceleration of the universe  ~\cite{Bean:2003ae, Chiba:2002mw, Gasperini:2001pc, Horava:2000tb, Kamenshchik:2001cp, Chaubey:2014gja}. In addition to the familiar types of dark-energy model, there are other types of cosmological dark-energy model deemed interesting, such as the cosmological constant and the universe full of exotic fluids. Keep in mind that the acceleration phase in the late universe is broadly attributed to the cosmological constant $\Lambda$, or a constant energy density $\rho_{\Lambda} = 7.02 10^{-24} g/m^3$, dubbed dark energy. Indeed, the problem puts a frame on looking for a unification of the dark sector of the universe and considers it as a single component that behaves as both dark energy and dark matter. 

Recently, new models that mix dark matter and dark energy have been proposed as
candidates for the dark components. Among the suggested unified dark fluid models, the Chaplygin gas~\cite{Kamenshchik:2001cp} and related generalizations~\cite{Bilic:2001cg,Bento:2002ps} have been largely adopted to explain the accelerated expansion of the Universe~\cite{Carturan:2002si,Amendola:2003bz,Bean:2003ae,vomMarttens:2017cuz}. Further applications of the Chaplygin dark fluid (CDF) appear in relation to the Hubble tension~\cite{Sengupta:2023yxh} and the growth of cosmological perturbations~\cite{Abdullah:2021tee}, respectively. In~\cite{Li:2019lhr} the analytical solution and the related thermodynamic quantities have been addressed for a charged static spherically-symmetric
BH is surrounded by CDF in the Lovelock
gravity theory. Such a model has been later extended to the modified Chaplygin
gas (MCG) to study the stability of the MCG-surrounded BHs in Einstein-Gauss-Bonnet~\cite{Li:2019ndh} and Lovelock~\cite{Li:2022csn} gravity. On more thermodynamic shores, the phase
transitions and critical behavior of the static spherically-symmetric AdS BHs surrounded by CDF
in GR have been the subject of investigation in~\cite{Li:2023zfl}, while a preliminary geometro-thermodynamic analysis has been conducted in~\cite{Aviles:2012et}. All these studies indicate that the (generalized) CDF could not only be a theoretical model, but a naturally existing fluid deserving of further consideration.

Starting from the above premises, in this work, we aim to study the phase structures and transitions of a charged AdS BH surrounded by MCG. Toward this end, the analysis is structured as follows: the next Sec.~\ref{sec1} is devoted to discussing the charged BH solution surrounded by MCG. In Sec.~\ref{BH chemistry} we examine the general features of BH chemistry for the obtained solution, while Sec.~\ref{BHstab} contains the analysis of thermodynamic stability. The study of critical behavior and phase transitions on the basis of the geometrothermodynamic tools
are the main objectives of Sec.~\ref{PVcr} and~\ref{Geom}, respectively. 
Conclusions and perspectives are finally summarized in Sec.~\ref{Conc}.

\section{Charged black hole solution surrounded by Modified Chaplygin gas }
\label{sec1}
This section presents an examination of the MCG structure within the framework of GR theory. Thus, it is carefully noted that pure Chaplygin gas is said to have an exotic equation of a negative pressure state 
\begin{equation}
p= -\frac{B}{\rho},\,\,\,\,\,\, B>0\,.
\end{equation}
In addition, the extended version of this gas is the generalized Chaplygin gas (GCG), which is typified by the generalized equation of state 
\begin{equation}
p= -B/\rho^\beta\,,
\end{equation} 
with $0\leq\beta\leq1$ being a universal constant. In practical terms, MCG is a major extension of Chaplygin gas, which is an exotic fluid considered perfect, obeying the calorific EoS $p=A\rho -B/\rho^\beta$ where $A\geq0$, $\beta\geq0$ and the pair $A$ and $B$ are considered constant through an adiabatic process ~\cite{Debnath:2004cd, Benaoum:2002zs}. Parametrically speaking, conditions $\beta=1$ and $A=0$ lead to a characteristic of the lifting forces on a plane wing in the aerodynamic process. On the other hand, constraining $\beta>0$ and $A=0$ define the GCG. On top of that, there is plenty of evidence that the MCG can be rebuilt using $k$-quintessence (kinetic quintessence)~\cite{Chimento:2003ta, Sharif:2012cu}  and $f$-quintessence (fermionic quintessence)~\cite{Myrzakulov:2011cm}. 
Inspired by the exciting results of the MCG-specific BH chemistry method by assuming the negative cosmological constant as a thermodynamical pressure, revealing the critical behavior from the MCG background is needed ~\cite{Debnath:2019mzs, Debnath:2019yit}. In this regard, Ubbanah used a specific BH known as Chaplygin's BH and assumed the negative cosmological constant as the thermodynamic pressure. As a consequence, the work deserves further contributions, and the system is considered to be a heat engine ~\cite{Debnath:2019mzs}. Similarly, modified cosmic Chaplygin gas is examined in the context of the AdS feature by considering BH's chemistry formula. This work highlights several thermodynamic aspects, including the heat engine and Joule-Thomson expansion ~\cite{Debnath:2019yit}.

In what follows, we consider the system of a charged source with a MCG structure in the context of GR described by the following action:
\begin{equation}
    \mathcal{I}=\frac{1}{16\pi}\int \mathrm{d}^{4}x\sqrt{-g}\left[\mathcal{R}+6\ell^{-2}-\frac{1}{4}F_{\mu\nu}F^{\mu\nu} \right]+\mathcal{I}_M
    \label{1}
\end{equation}
where $\mathcal{R}$ is the Ricci scalar, $g=\det (g_{\mu\nu})$ is the determinant of the metric tensor $g_{\mu\nu}$, $\ell$ is the AdS length, $\mathcal{I}_M$ is the matter contribution arising from the MCG background, and $F_{\mu\nu}=\partial_\mu A_\nu-\partial_\nu A_\mu$ is the field strength of the electromagnetic field with $A_\mu$ is the gauge potential.Moreover, $\kappa=8\pi G$, where G is the Newtonian gravitational constant. Henceforth, we consider $G=c=1$. Varying the action $(\ref{1})$ leads to the following field equations:
\begin{align}
    G_{\mu\nu}-\frac{3}{\ell^2}g_{\mu\nu}&=T_{\mu\nu}^{\text{EM}}+T_{\mu\nu}^{\text{CMG}}\label{G},\\
    \partial_\mu(\sqrt{-g}F^{\mu\nu}) &=0
    \label{2}
\end{align}
where $G_{\mu\nu}$ is the Einstein tensor, $T_{\mu\nu}^{\text{CMG}}$ is the energy-momentum tensor for MCG, and $T_{\mu\nu}^{\text{EM}}$ is the energy-momentum tensor for the electromagnetic field which is explicitly given by
\begin{equation}
T_{\mu\nu}^{\text{EM}}=2\left(F_{\mu\lambda}F_\nu^\lambda-\frac{1}{4}g_{\mu\nu}F^{\lambda\delta}F_{\lambda\delta}\right).
    \label{3}
\end{equation}
We consider a static, spherically symmetric, four-dimensional space-time given by
\begin{equation} 
ds^2 = -f(r) \,\mathrm{d} t^2 + \dfrac{\mathrm{d} r^2}{f(r)} +r^2 \mathrm{d}\Omega^2,
\label{4}
\end{equation}
in this metric, $f(r)$ is the metric function that depends on $r$ and $d\Omega^2 = d\theta^2 + \sin^2\theta d\phi^2$.

The way to reveal the structure of the electromagnetic field, in particular an electrically charged BH solution, is to assume a radial electric field. Thus, in this choice, the gauge potential is expressed as
\begin{equation}
    A_\mu=h(r)\delta_\mu^0.
    \label{5}
\end{equation}
Exploiting Eqs. ($\ref{3}$) and ($\ref{4}$) provides a second-order differential equation given by
\begin{equation}
    r\,h''(r)+2 h'(r)=0
    \label{6}
\end{equation}
where the prime and double prime are the first and second derivatives with respect to $r$. To proceed to solve the differential equation yields a compact solution in the form of
\begin{equation}
    h(r)=\frac{q}{r}
    \label{7}
\end{equation}
where $q$ is an integration constant acting like an electric charge. Taking into account the structure of the gauge field $(\ref{7})$ in conjunction with the space-time metric $(\ref{4})$, the field strength component appears as follows:
\begin{equation}
    F_{tr}=\partial_tA_r-\partial_rA_t=\frac{q}{r^2}.
\end{equation}

We assign the MCG the equation of state $p= A\rho-B\rho^{-\beta}$ ~\cite{Debnath:2004cd, Benaoum:2002zs}, where $A$, $B$ are positive parameters and the parameter $\beta$ runs in the range $0\leq\beta\leq1$. In a $4D$ spherically symmetric spacetime scenario, the related energy-momentum tensor components of the MCG are expressed in the following way:
\begin{equation}
    T_t^t=\psi(r),\ \  T_t^i=0,\ \ T_i^j=\phi(r)\frac{r_ir^j}{r_nr^n}+\xi(r)\delta_i^j.
\end{equation}
It should be noted that the expression for the energy-momentum tensor was first given by Kiselev when studying the static spherically symmetric quintessence surrounding a BH. Due to the fact that spacetime is considered static and spherically symmetric, the $r-r$ component of the energy-momentum tensor should be equal to the $t-t$ component, so
\begin{equation}
    T_t^t=T_r^r=-\rho(r).
    \label{11}
\end{equation}
On the basis of the isotropic average over the angles, the following
\begin{equation}
    \langle r_ir^j\rangle=\frac{r_n r^n}{3}\delta_i^j
\end{equation}
where one can obtain
     \begin{eqnarray}\nonumber
         \langle T_i^j\rangle&=&\left(\frac{\phi(r)}{3}+\xi(r)\right)\delta_i^j \\ &=&p(r)\delta_i^j=\left(A\,\rho(r)-\frac{B}{[\rho(r)]^\beta}\right)\delta_i^j.
    \label{13}
     \end{eqnarray}
It is noteworthy that the given energy density and parameter set require an explicit and constraining statement. To this purpose, taking into account Eqs. ($\ref{11}$) and ($\ref{13}$) simplify the process and yield the following expressions:
\begin{align}
    \phi(r)&=\beta_1\,\rho(r)+\alpha_1 \,\rho(r) ^{-\beta},\\
    \xi(r)&=\beta_2\,\rho(r)+\alpha_2\,\rho(r) ^{-\beta}.
\end{align}
Furthermore, by exploiting equations ($\ref{11}$) and ($\ref{13}$), the parameters $\alpha_i$ and $\beta_i$ are constrained to be given by
    \begin{align}
    \beta_1&=-\frac{3(1+A)}{2}, \ \ \beta_2=\frac{1+3A}{2},\\
    \alpha_1&=\frac{3}{2}B,\ \ \alpha_2=-\frac{3}{2}B,
\end{align}
Thus, it is now simple to define the angular components of the energy-momentum tensor devoted to the MCG structure by
\begin{equation}
    T_\theta^\theta=T_\phi^\phi=\frac{1+3A}{2}\,\rho(r)-\frac{3}{2}B\,\rho(r)^{-\beta}.
    \label{17}
\end{equation}
By considering the spacetime metric  ($\ref{4}$) together with the field equations  ($\ref{G}$), the Einstein tensor components can be given by
\begin{align}
G_t^t&=G_r^r=\frac{1}{r^2}\left(f+r\, f'-1\right)+\frac{3}{\ell^2}\label{18}\\
G_\theta^\theta&=G_\phi^\phi=\frac{1}{2r}\left(2f'+r\, f''\right)+\frac{3}{\ell^2}\label{19}
\end{align} 
whereby combining the previous set with that of Eqs. ($\ref{11}$)-($\ref{17}$) produces a pair of differential equations such as:
\begin{align}
\frac{1}{r^2}&\left(f+r\, f'-1\right)+\frac{3}{\ell^2}=-\rho-\frac{q^2}{r^4}\label{20}\\
\frac{1}{2r}&\left(2f'+r\, f''\right)+\frac{3}{\ell^2}= \left(\frac{1+3A}{2}\,\rho-\frac{3}{2}\frac{B}{\rho^\beta}\right)+\frac{q^2}{ r^4}
\end{align} 

To properly look for the expression of the energy density, consider the set previously described above by using equations ($\ref{11}$)-($\ref{17}$) and, in a way, by imposing the conservation condition on the energy-momentum tensor. This consideration step produces an interesting result, as
\begin{equation}
    \rho(r)=\Bigg\{\frac{1}{1+A}\Bigg(B+\bigg(\frac{\gamma}{r^3}\bigg)^{(1+A)(1+\beta)}\Bigg)\Bigg\}^{\frac{1}{1+\beta}},
    \label{22}
\end{equation}
where an integration constant, $\gamma>0$, is used. It is important to keep in mind that the energy density is restricted to $\sim\left(\frac{B}{1+A}\right)^{\frac{1}{1+\beta}}$ at the asymptotic limit for $r$. As a result, MCG behaves as a cosmological constant far away from the BH, and as it approaches the BH, it gravitationally grows more densely.

Substituting Eq. ($\ref{22}$) into  Eq. ($\ref{20}$), one can obtain the analytical solution for $f(r)$ in the following compact form
    \begin{eqnarray}\nonumber
        f(r) &=& 1-\frac{2
   M}{r}+\frac{Q^2}{r^2}+\frac{r^2}{\ell^2}\\ &-&\frac{r^2}{3} \bigg(\frac{B}{A+1}\bigg)^{\frac{1}{\beta+1}} \,
   _2F_1[\alpha, \nu; \lambda; \xi]
   \label{23}
    \end{eqnarray}
where $M$ is the physical parameter standing for the mass of the BH. Whereas, the integration constant $q$ is mainly related to the physical parameter charge $Q$ over a two-sphere with an infinite radius according to the conservation law as
\begin{align}
    Q&=\frac{1}{4\pi}\int_{S^2_{\infty}}\star \mathbf{F}=\frac{1}{8\pi}\int_{S^2_{\infty}}F^{\mu\nu}\,\epsilon_{\mu\nu}\nonumber\\
    &=\frac{1}{4\pi}\int_{S^2_{\infty}}h'(r)\sqrt{-g}\,\mathrm{d}\theta\,\mathrm{d}\phi\nonumber\\
    &=\frac{1}{4\pi}\int_{S^2_{\infty}}\mathrm{d}\theta\,\mathrm{d}\phi\, r^2\sin\theta\,\frac{q}{r^2}= q
\end{align}
where $\epsilon_{tr}=n_t\sigma_r=1$ for a static spherically spacetime with $n_\mu$ is the unit normal and $\sigma_\nu$ is the unit normal to a unit two-sphere with an infinite radius. This proves that the physical electric charge $Q$ in our charged BH solution is the same as in Maxwell’s equation. Also, the electric potential evaluated at infinity by an observer with relation to the event horizon $r_+$ can be given as
\begin{equation}
\phi(r_+)=A_\mu\chi^\mu|_{r\rightarrow\infty}-A_\mu\chi^\mu|_{r=r_+}=\frac{Q}{r_+}
\end{equation}
where $\chi= C\,\partial_t$ is the null generator of the event horizon, while C acts as a fixed constant. Here, the hypergeometric function ${_2}F_1[\alpha, \nu; \lambda; \xi]$ representing the regular solution of the hypergeometric differential equation, is defined for $\lvert \xi\rvert<1$  by a power series of the form
\begin{equation}
{_2}F_1[\alpha, \nu; \lambda; \xi]=\sum_{k=0}^{\infty}\bigg[(\alpha)_k(\nu)_k/(\lambda)_k\bigg]\xi^k/k!
\end{equation}
with $(n)_k$ is the (rising) Pochhammer symbol~\cite{Abra}. Furthermore, the parameters set $(\alpha, \nu, \lambda, \xi)$ are given, respectively, by
    \begin{align}
      \alpha&=-\frac{1}{\beta+1},
      \,\,\,\nonumber
      \nu=-\frac{1}{1+A +\beta( A+1)}, \\
     \lambda&=1+\nu, \,\,\hspace{0.5cm}
     \xi=-\frac{1}{B}\left(\frac{\gamma}{r^3}\right)^{(A+1) (\beta+1)}.\nonumber
    \end{align}
    
To explore the asymptotic behavior of the metric function $f(r)$, the limit $r\rightarrow\infty$ is taken into account, giving
\begin{equation}
  f(r)  =1+r^2\Bigg(\frac{1}{\ell^2}-\bigg(\frac{B}{1+A}\bigg)^{\frac{1}{1 +\beta}}\Bigg),
\end{equation}
which implies that in the asymptotic limit, the behavior of the solution is controlled by means of the AdS length $\ell$ (the cosmological constant $\Lambda$) and the MCG parameters background. As a result, one can constrain $\frac{3}{\ell^2}<\left(\frac{B}{1+A}\right)^{\frac{1}{1 +\beta}}$. It may be further pointed out that the metric function can be close to RN-AdS BHs once conditions $\beta=0$, $\gamma= 0$, and $A>0$ are set after taking the limit $B\rightarrow0$.

\begin{figure*}[tbh!]
      	\centering{
       \includegraphics[scale=0.41]{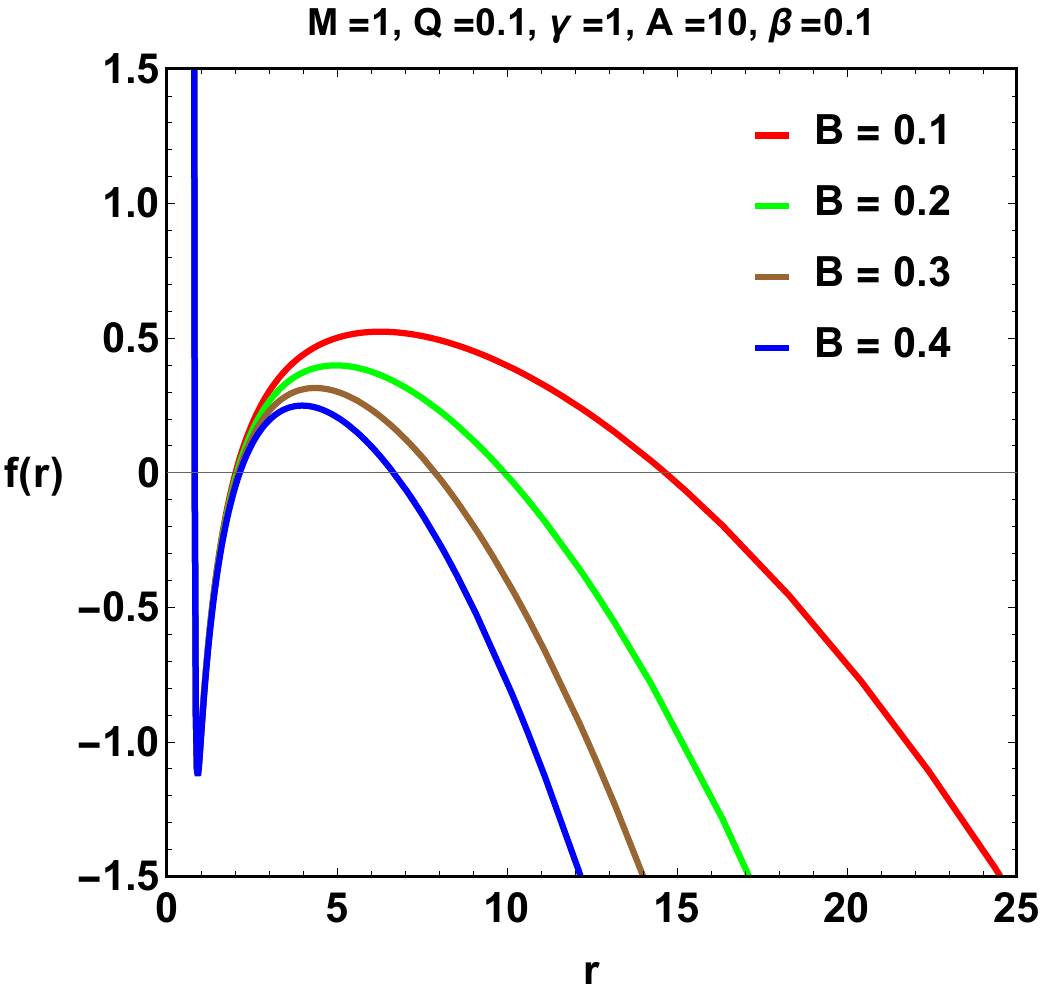} \hspace{2mm}
      	\includegraphics[scale=0.41]{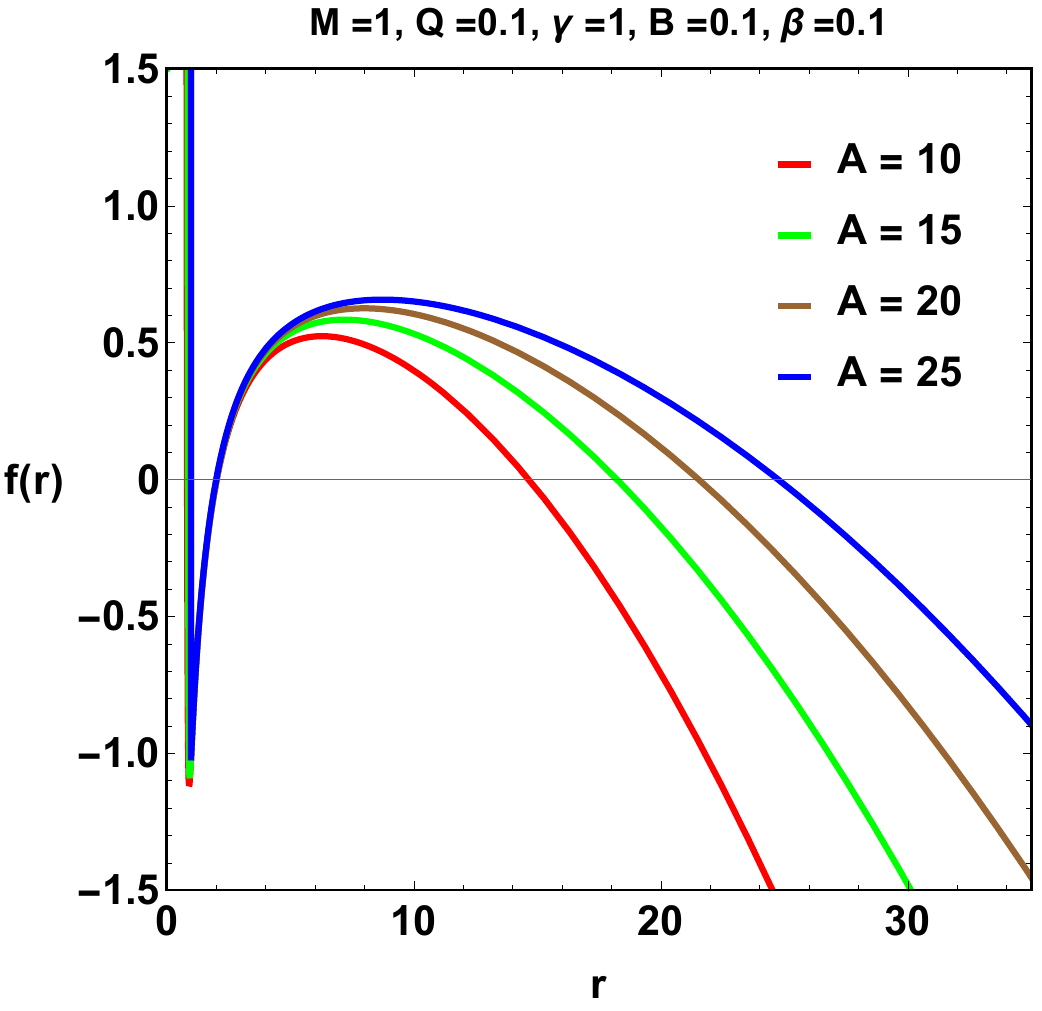} \hspace{2mm}
      }
       \centering{ 
       \includegraphics[scale=0.41]{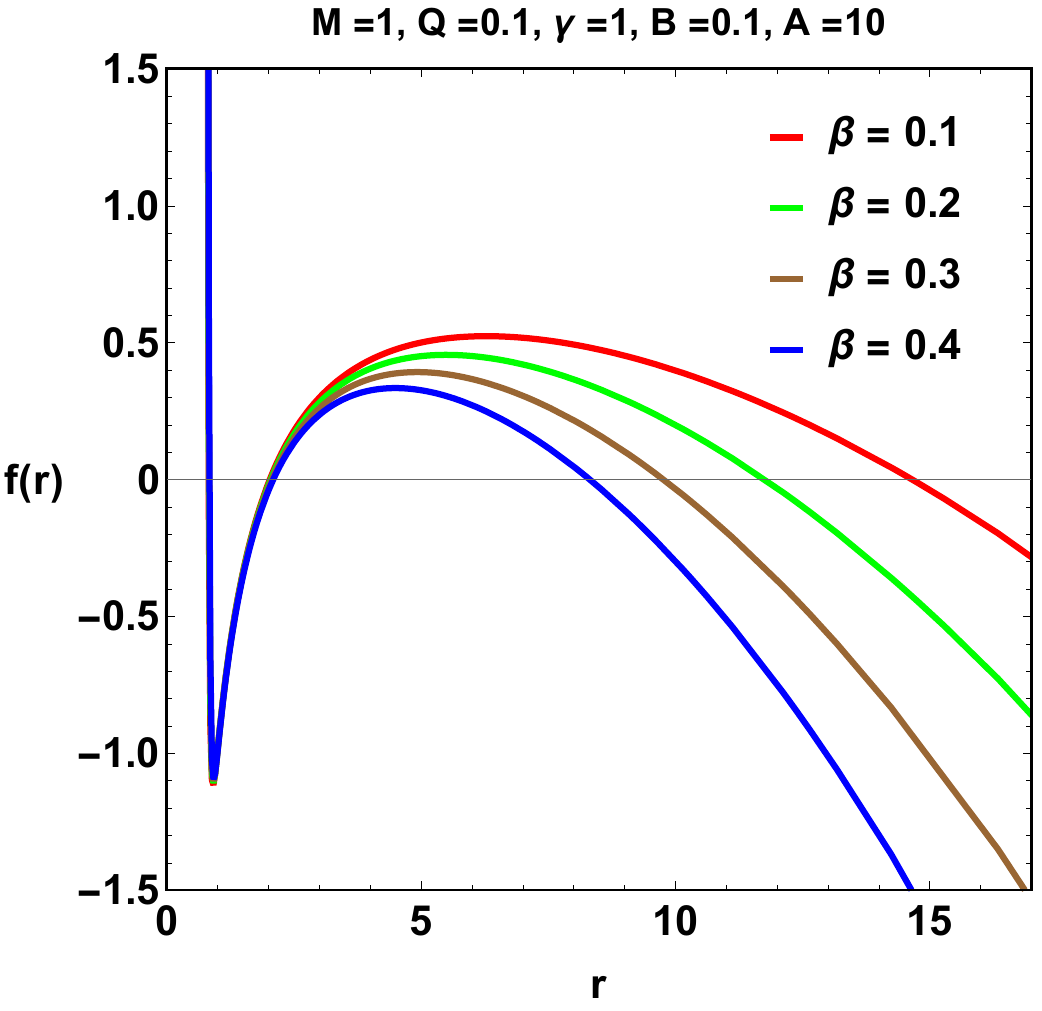}\hspace{2mm}
      \includegraphics[scale=0.41]{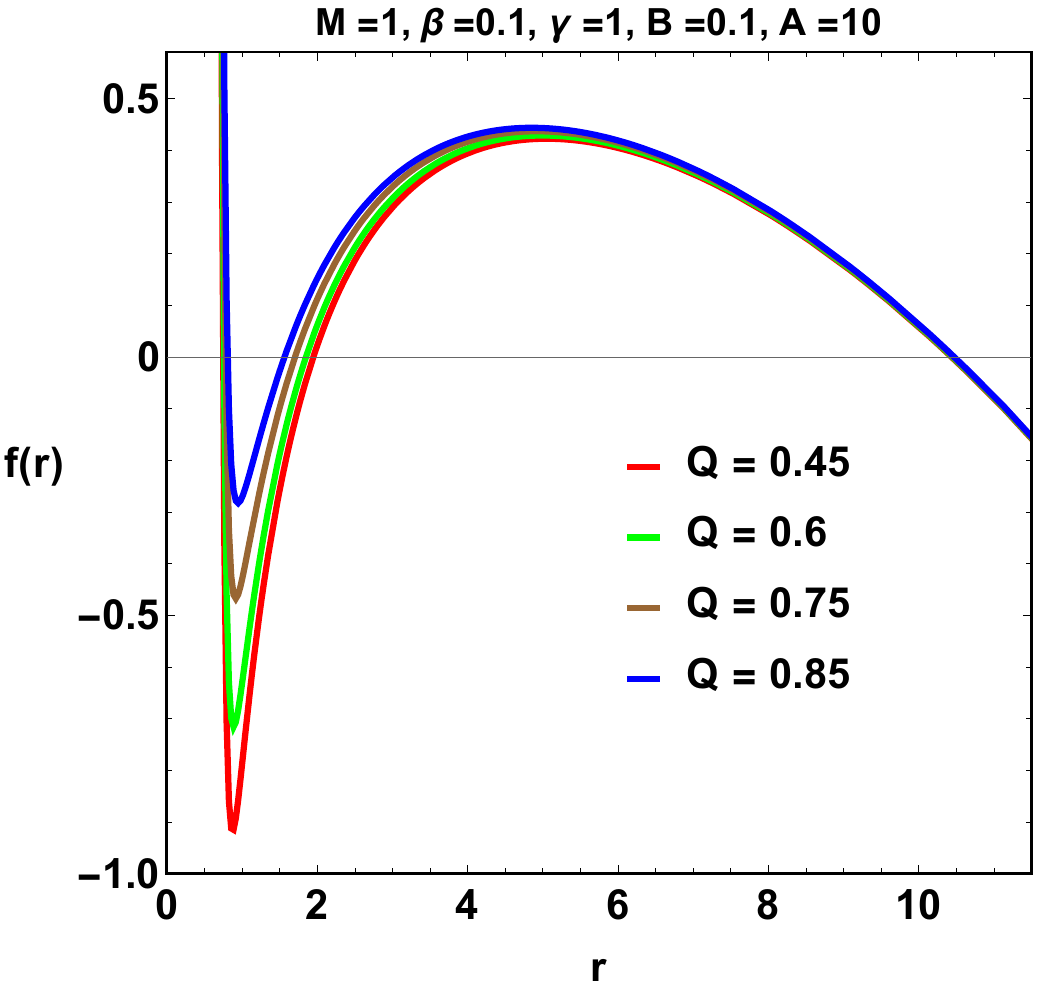} \hspace{2mm}
       }
      	\caption{Variation of the BH metric function (\ref{23}) with respect to $r$ for various values of the parameter space.}
      	\label{fig1}
      \end{figure*}
      To provide an examination regarding the behavior of the metric function for our BH solution, Fig. $\ref{fig1}$ depicts this behavior. For a certain choice in the parameter space, the metric function is being schematically considered. In particular, for all sets of the parameter space, the root of $f(r) $ is divided into two classes: the smallest root corresponds to two BH horizons, and the largest root is associated with a cosmological horizon. It appears that the two BH horizons, whether event or inner, are closely located at the same horizon radius for the variation of the parameter set $(A, B, \beta)$. On the other hand, varying the charge parameter at the level of the metric function generates the apparition of multiple event horizons. A closer observation shows that the behavior of the metric function is disproportional with respect to the pairs $(A, Q)$ and $(B, \beta)$. In what follows, our BH solution will be examined in part by studying its singular behavior and, secondly, by analyzing its violation or satisfaction with respect to the constraints of the Energy Conditions (EC).
      
To properly verify the singularity and uniqueness of our BH solution, we need to provide an analysis based on scalar invariants. These are the Ricci scalar, the Ricci square, and the Kretschmann scalar.

The Ricci scalar for the corresponding metric is given by
\begin{widetext}
\begin{align}\label{R}
    \mathcal{R}&=\frac{1}{B}\Biggl\{\left(\frac{B}{A+1}\right)^{\frac{1}{\beta+1}} \mathcal{T}_1^{-\frac{\beta}{\beta+1}} \bigg((1-3 A) \left(\frac{\gamma}{r^3}\right)^{(A+1)
   (\beta+1)}+4 B\bigg)\Biggl\}
   -\frac{12}{\ell^2}
\end{align}
\end{widetext}
The Ricci squared is given by
\begin{widetext}
\begin{align}\label{RR}
 \mathcal{R}_{\mu\nu}\mathcal{R}^{\mu\nu} &= \bigg(\frac{A+1}{B}\bigg)^{-\frac{2}{\beta +1}} \Biggl\{8 B^2 \Bigg(\bigg(\frac{A+1}{B}\bigg)^{\frac{2}{\beta +1}}
   \bigg(Q^4+\frac{9}{\ell^4} r^8\bigg)+r^8 \bigg(\mathcal{T}_1^{\frac{2}{\beta +1}}- \frac{6}{\ell^2}\,\mathcal{T}_2^{\frac{1}{\beta
   +1}}\bigg)\Bigg)+4 B \,\mathcal{T}_3 \bigg(\frac{\gamma}{r^3}\bigg)^{(A+1) (\beta +1)}\nonumber\\
   &+\mathcal{T}_4 \bigg(\frac{\gamma}{r^3}\bigg)^{2 (A+1)
   (\beta +1)}\Biggl\}\bigg(\frac{1}{2 B^2 \mathcal{T}_1^2\, r^8}\bigg)
\end{align}
\end{widetext}
Finally, the Kretschmann scalar is found to be
\begin{widetext}
    \begin{align}\label{RRR}
    \mathcal{R}_{\alpha\beta\mu\nu}\mathcal{R}^{\alpha\beta\mu\nu}& =
   4 \left(\frac{A+1}{B}\right)^{-\frac{2}{\beta +1}} \Bigg(r^4 \left(3\,
   \mathcal{T}_1^{\frac{1}{\beta +1}}-\, _2F_1(\alpha,\nu;\lambda;\xi)\right)+2 \left(\frac{A+1}{B}\right)^{\frac{1}{\beta +1}} \bigg(-3 M r+3
   Q^2-\frac{-3}{\ell^2}  r^4\bigg)\Bigg)^2\nonumber\\
   &
 +4 \Bigg(r^4 \left(\frac{A+1}{B}\right)^{-\frac{1}{\beta +1}} \, _2F_1(\alpha,\nu;\lambda;\xi)+6 M r-3
   Q^2-\frac{3}{\ell^2}  r^4\Bigg)^2 
   \nonumber\\
    &
 +\frac{(A+1)^{-\frac{2}{\beta +1}}}{B^2\, \mathcal{T}_1^2} \Bigg(-2 B \,\mathcal{T}_1 \bigg(r^4 B^{\frac{1}{\beta +1}} \, _2F_1(\alpha,\nu;\lambda;\xi)+(A+1)^{\frac{1}{\beta +1}} \bigg(6 M r-9 Q^2-\frac{3}{\ell^2} 
   r^4\bigg)\bigg)\nonumber\\
   &+9 (A+1)\, r\, \gamma (B\, \mathcal{T}_1)^{\frac{1}{\beta +1}} \,\left(\frac{\gamma}{r^3}\right)^{A(\beta +1)+\beta
   }\Bigg)^2
\end{align}
\end{widetext}
where
    \begin{align}
        \mathcal{T}_1&=\frac{1}{B}\bigg(\left(\frac{\gamma}{r^3}\right)^{(A+1) (\beta +1)}+B\bigg)\\
        \mathcal{T}_2& =\frac{(A+1) }{B^2}\left(\left(\frac{\gamma}{r^3}\right)^{(A+1) (\beta +1)}+B\right)\\
        \mathcal{T}_3&=r^4\,
   \mathcal{T}_2^{\frac{1}{\beta+1}} \left(3 (A+1) Q^2-(5-3 A)\frac{3}{\ell^2}  r^4\right)+r^8\,
   \mathcal{T}_1^{\frac{2}{\beta +1}}\nonumber\\
    &+4 Q^4 \left(\frac{A+1}{B}\right)^{\frac{2}{\beta +1}}+\frac{36}{\ell^4} r^8 \left(\frac{A+1}{B}\right)^{\frac{2}{\beta +1}}\nonumber\\
    &-3 A r^8 \mathcal{T}_1^{\frac{2}{\beta +1}}\\
   \mathcal{T}_4 &=\left(9 A^2+6 A+5\right) r^8 \,\mathcal{T}_1^{\frac{2}{\beta +1}}+8 \left(\frac{A+1}{B}\right)^{\frac{2}{\beta +1}} \nonumber\\
   &\times\left(Q^4+\frac{9}{\ell^4} 
   r^8\right)\nonumber\\
   &+4 r^4 \,\mathcal{T}_2^{\frac{1}{\beta +1}} \left(3 (A+1) Q^2-(1-3 A) \frac{3}{\ell^2}   r^4\right).
    \end{align}
    A closer look at the expressions \eqref{R}, \eqref{RR} and\eqref{RRR} shows that the BH solution represented by this metric is singular for any permissible value of the parameters $A$ and $\beta$. Practically speaking, the existence of the singularity results from the mass and charge terms in the BH metric. By imposing the constraints $\beta <0$ and $A>0$, the singularity will consequently disappear due to the product constant $\gamma$. Nevertheless, to rule out the singularity made from the mass and the charge terms, it may be useful to describe a procedure with a non-linear charge distribution function similar to Ref.  ~\cite{Balart:2014cga}. Throughout this work, we will not be thinking about such a situation and will stick to the metric function \eqref{23} for the rest of the analysis. A few remarks concerning the mentioned scalars show that the Ricci scalar is not a function of the BH charge $Q$. In contrast, Ricci squared and Kretschmann scalars are functions of the BH charge $Q$, so any variation in the BH charge may induce meaningful variations in these same scalars. Therefore, the considered scalars demonstrate that our BH solution is unique, and the AdS background, together with the MCG structure, changes the BH spacetime significantly.

\begin{figure*}[t!]
    \centering
    \begin{subfigure}[b]{0.5\textwidth}
        \centering
        \includegraphics[scale=0.55]{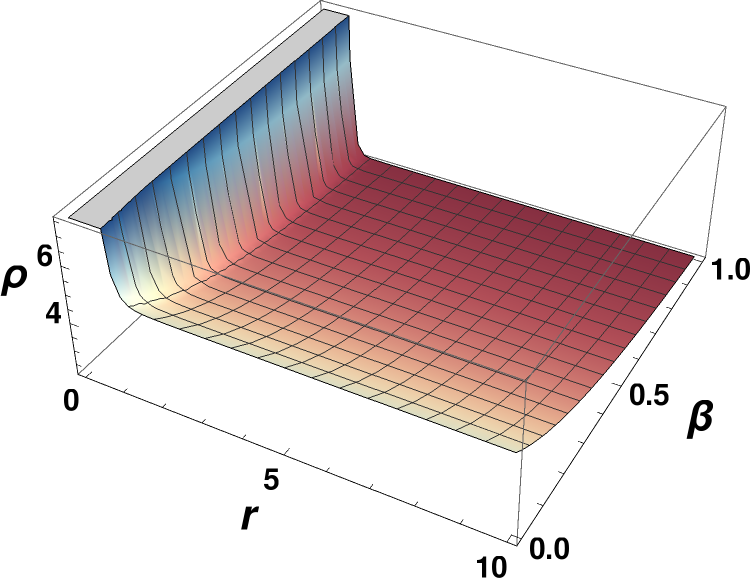}
        \caption{WEC}
        \label{subfig:wec}
    \end{subfigure}%
    \hfill
    \begin{subfigure}[b]{0.5\textwidth}
        \centering
        \includegraphics[scale=0.55]{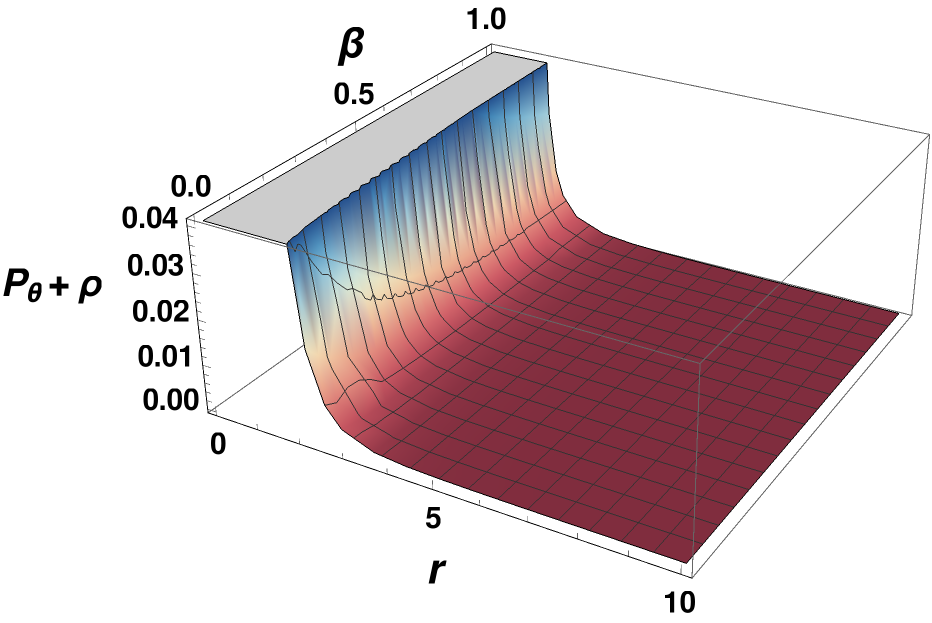}
        \caption{NEC}
        \label{subfig:nec}
    \end{subfigure}%
    \\
    \begin{subfigure}[b]{0.5\textwidth}
        \centering
        \includegraphics[scale=0.55]{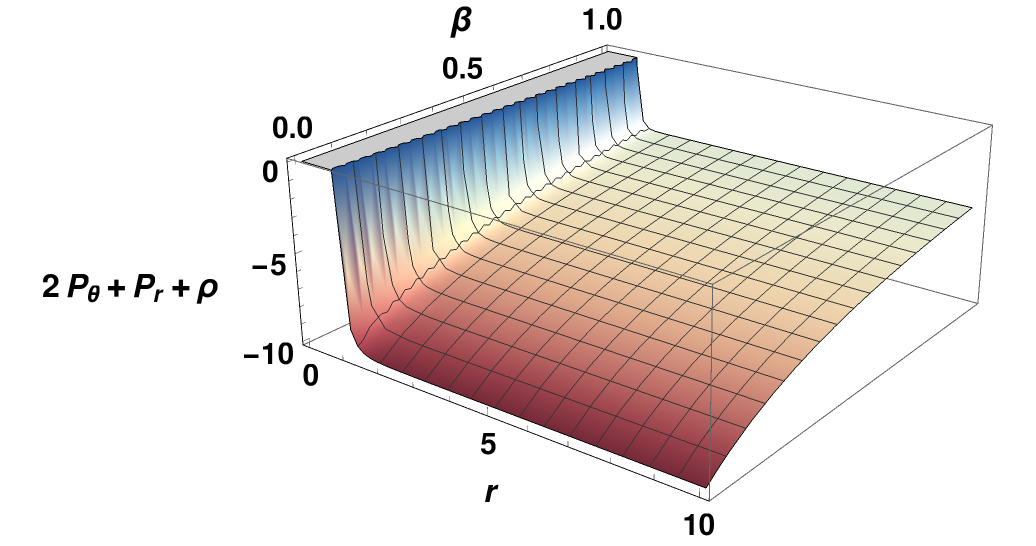}
        \caption{SEC}
        \label{subfig:sec}
    \end{subfigure}
    \caption{Energy conditions using $Q=0.2, B=10, \gamma=1, A=1.$}
    \label{fig2}
\end{figure*}

    In what follows, we take care to study the energy conditions for our BH solution ~\cite{Hawking:1973uf, Ghosh:2008zza,Kothawala:2004fy, Ghosh:2020syx}. The elements of the stress-energy tensor $T_{\mu\nu}$ governed by Einstein equations \eqref{G} for charged AdS BHs surrounding the CMG are as follows:
        \begin{align}
        &\rho= \mathcal{T}_1^{\frac{1}{\beta +1}} \left(\frac{B}{A+1}\right)^{\frac{1}{\beta +1}}+\frac{Q^2}{r^4}=-P_r\label{rho}\\
      &P_\theta=P_\phi=\left(\frac{A+1}{B}\right)^{-\frac{1}{\beta +1}} \Biggl\{2 B \bigg(Q^2 \left(\frac{A+1}{B}\right)^{\frac{1}{\beta
   +1}}-r^4\, \mathcal{T}_1^{\frac{1}{\beta +1}}\bigg)\nonumber\\
   &+\left(\frac{\gamma}{r^3}\right)^{(A+1) (\beta +1)} \bigg(2 Q^2
   \left(\frac{A+1}{B}\right)^{\frac{1}{\beta+1}}+(3 A+1) r^4 \,\mathcal{T}_1^{\frac{1}{\beta+1}}\bigg)\Biggl\}.\label{P}
    \end{align}
    \begin{itemize}
        \item The weak energy condition (WEC) requires that $T_{\mu\nu}\, t^\mu t^\nu\geqslant0$ everywhere, for any time vector $t^\mu$, which is equivalent to
    \begin{equation}
        \rho>0,\quad \rho+ P_i>0\quad (i=r, \theta, \phi)
    \end{equation}
    and so
    \begin{align}
        \rho&+P_\theta=\frac{1}{2 \mathcal{T}_1\, r^4}\Bigg(\frac{4 Q^2 }{B}\bigg(\left(\frac{\gamma}{r^3}\right)^{(A+1) (\beta +1)}+B\bigg)   \nonumber\\
   &+3 r \gamma \,\mathcal{T}_1^{\frac{1}{\beta +1}} \left(\frac{A+1}{B}\right)^{\frac{\beta }{\beta +1}} \left(\frac{\gamma}{r^3}\right)^{A (\beta
   +1)+\beta }\Bigg).
   \label{3-}
    \end{align}
The WEC is satisfied since $A$, $B$ and $\gamma$ are positive parameters and $0\leqslant \beta \leqslant 1$. Fig. $\ref{fig2}$  clearly shows that the energy density $\rho$ along with $\rho+P_\theta$ is positive for a span of horizon radii.
\item The zero energy condition (NEC) stipulates that $T_{\mu\nu}\, t^\mu t^\nu\geqslant0$ in the overall spacetime for any null vector $t^\mu$. The NEC predicts $\rho+P_r \geqslant 0$ which is identically zero, and $\rho+P_\theta \geqslant 0$ which is satisfied for equation \eqref{3-} whenever $\lvert A \rvert>1$.

\item The strong energy condition (SEC) asserts that  $T_{\mu\nu}\, t^\mu t^\nu\geqslant 1/2 \,T_{\mu\nu} t^\nu t_\nu$ globally, for any time vector $t^\mu$ which assumes that
\begin{equation}
        \rho+P_r+2\,P_\theta>0.
        \label{37}
    \end{equation}
    \end{itemize}
    Consequently, it is obvious that the MCG does not satisfy or violate the SEC, as shown in Fig. $\ref{fig2}$. This is similar to the situation with the quintessence of dark energy. Indeed, a violation of the SEC is interpreted as a violation of the attractive behavior of gravity, as evidenced by the dark energy that accelerates the expansion of the universe in cosmological studies, along with the matter content of the background of a regular BH, whose singularity has been superseded by a Sitter core.

\section{Black hole chemistry with Modified Chaplygin Gas}
\label{BH chemistry}
Investigating the thermodynamic quantities of our BH solution is the main task of this section. Therefore, at the radius of the $r_+$ horizon of the metric solution \eqref{23}, the mass of the BH is expressed in such a way as
    \begin{eqnarray}\label{Mass}\nonumber
   M=\frac{1}{6 r_+}\Bigg\{
  &-&r_+^4\,\bigg(\frac{A+1}{B}\bigg)^{-\frac{1}{\beta +1}}   \, _2F_1[\alpha, \nu; \lambda; \xi] \\
  &+&\frac{3}{\ell^2}r_+^4+ 3( Q^2 +r_+^2) \Bigg\}.
  \end{eqnarray}
Subsequently, in order to determine the Hawking temperature, it is first necessary to consider the surface gravity ~\cite{Kubiznak:2016qmn}, which is provided by 
\begin{equation}\label{r30}
    \kappa=\left(-\frac{1}{2}\nabla_\mu\xi_\nu\nabla^\mu\xi^\nu\right)^{1/2}=\frac{1}{2}f'\left(r_+\right),
\end{equation}
with $\xi^\mu=\partial/\partial t$ is a killing vector. So, the formula $T_+=\kappa/2\pi$ is the Hawking temperature, expressed in terms of the BH system parameters as
\begin{widetext}
    \begin{equation}\label{43}
        T_+=\frac{1}{4 \pi  \,r_+^3} \Bigg\{r_+^2\bigg(1+\frac{3}{\ell^2}\,r_+^2\bigg)-Q^2 -r_+^4\bigg(\frac{A+1}{B}\bigg)^{-\frac{1}{\beta +1}}\, \bigg(\frac{1}{B}\bigg(\frac{\gamma}{r_+^3}\bigg)^{(A+1) (\beta
   +1)}\,+1\bigg)^{\frac{1}{\beta +1}}\Bigg\}.
    \end{equation}
    \end{widetext}
It is worth mentioning that all the parameters set in the BH solution effectively contribute to the behavior of the Hawking temperature.
\begin{figure*}[tbh!]
      	\centering{
       \includegraphics[scale=0.43]{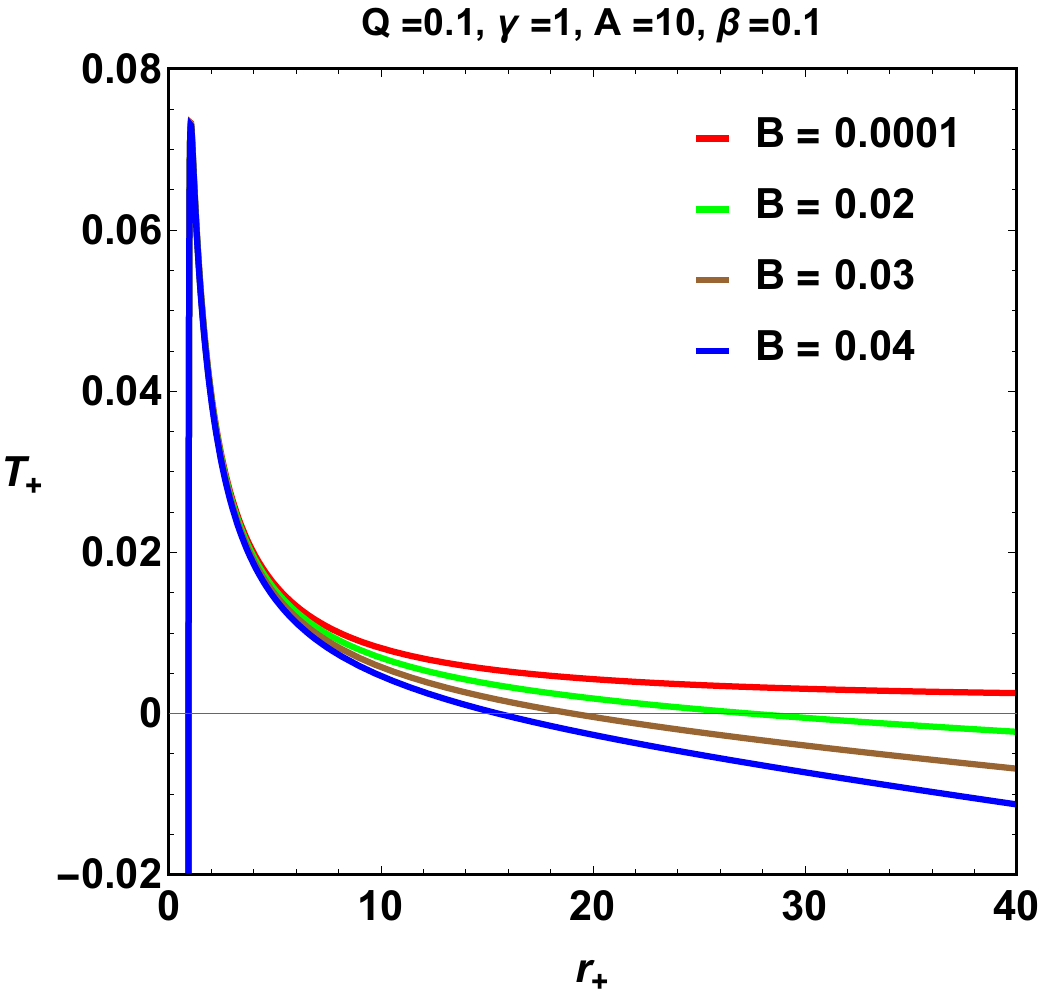} \hspace{2mm}
      	\includegraphics[scale=0.43]{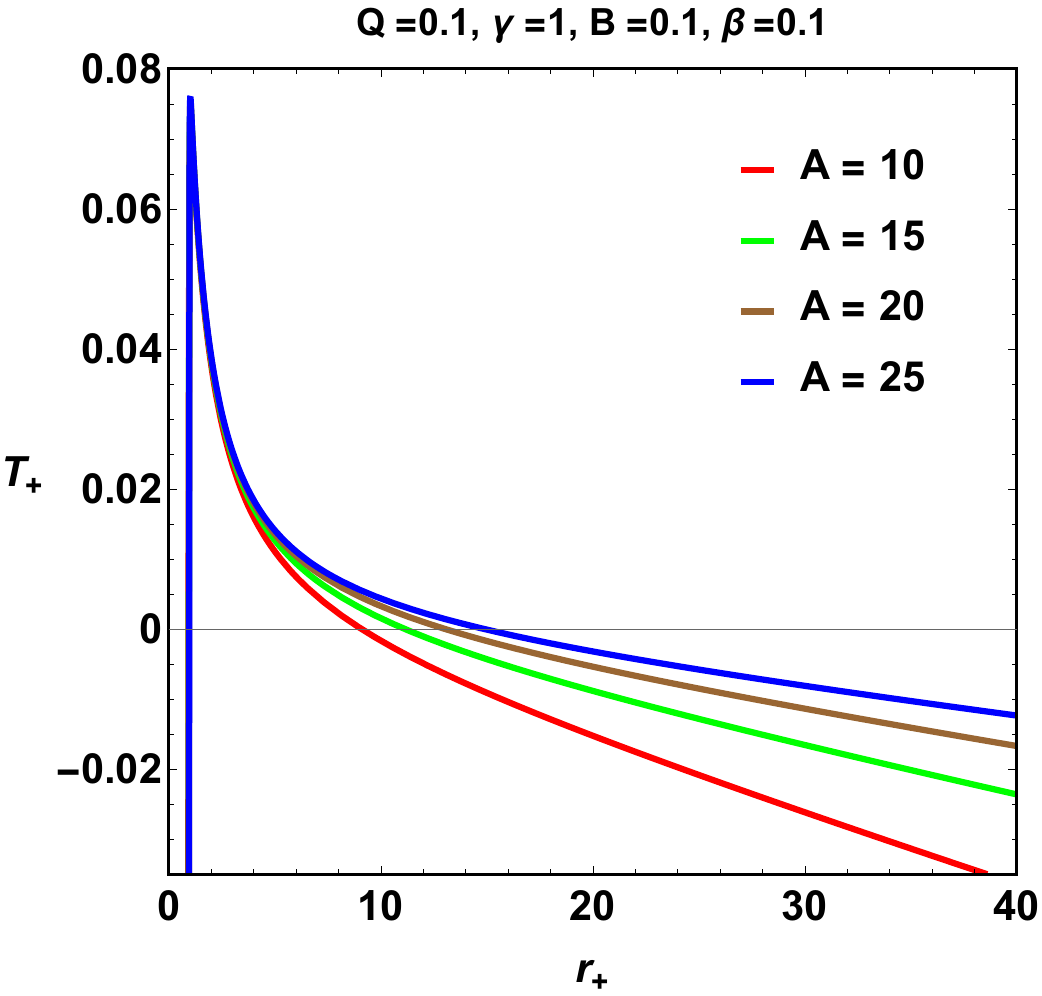} \hspace{2mm}
      }
       \centering{ 
       \includegraphics[scale=0.45]{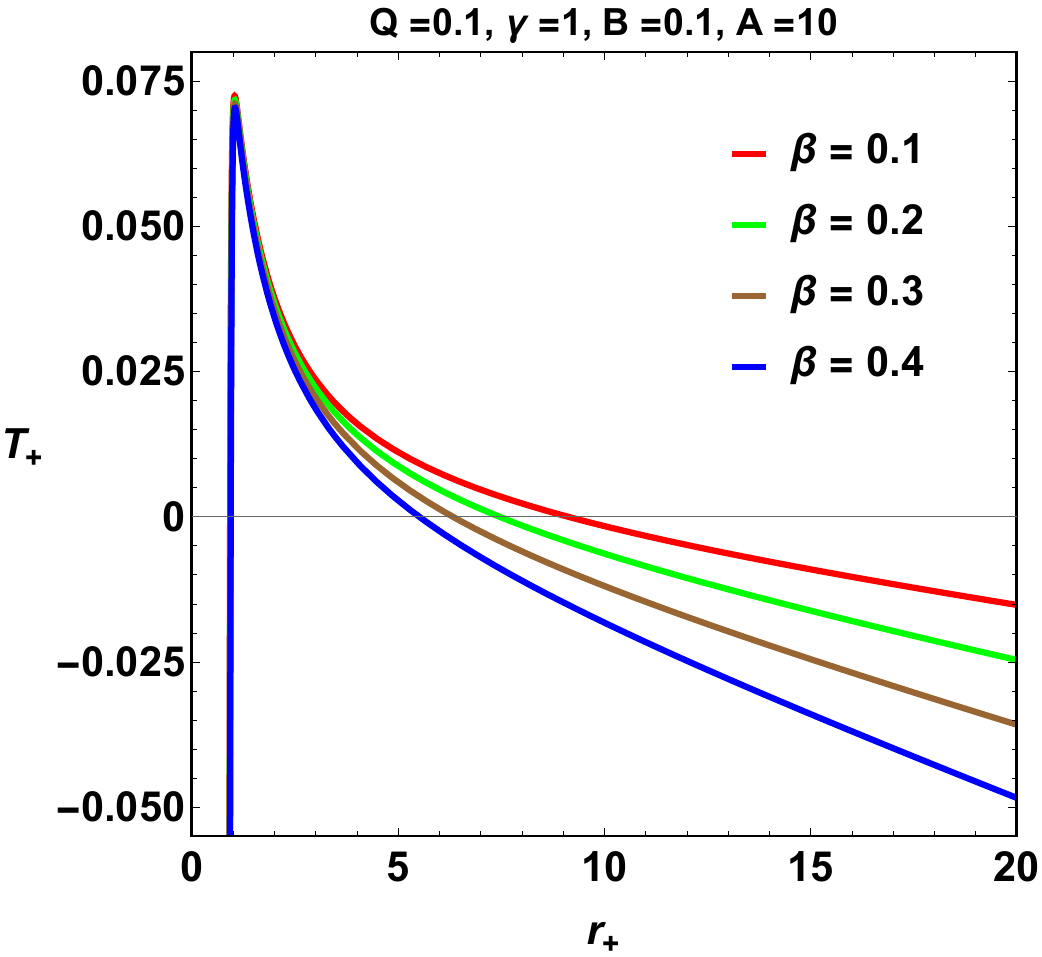}\hspace{2mm}
      \includegraphics[scale=0.43]{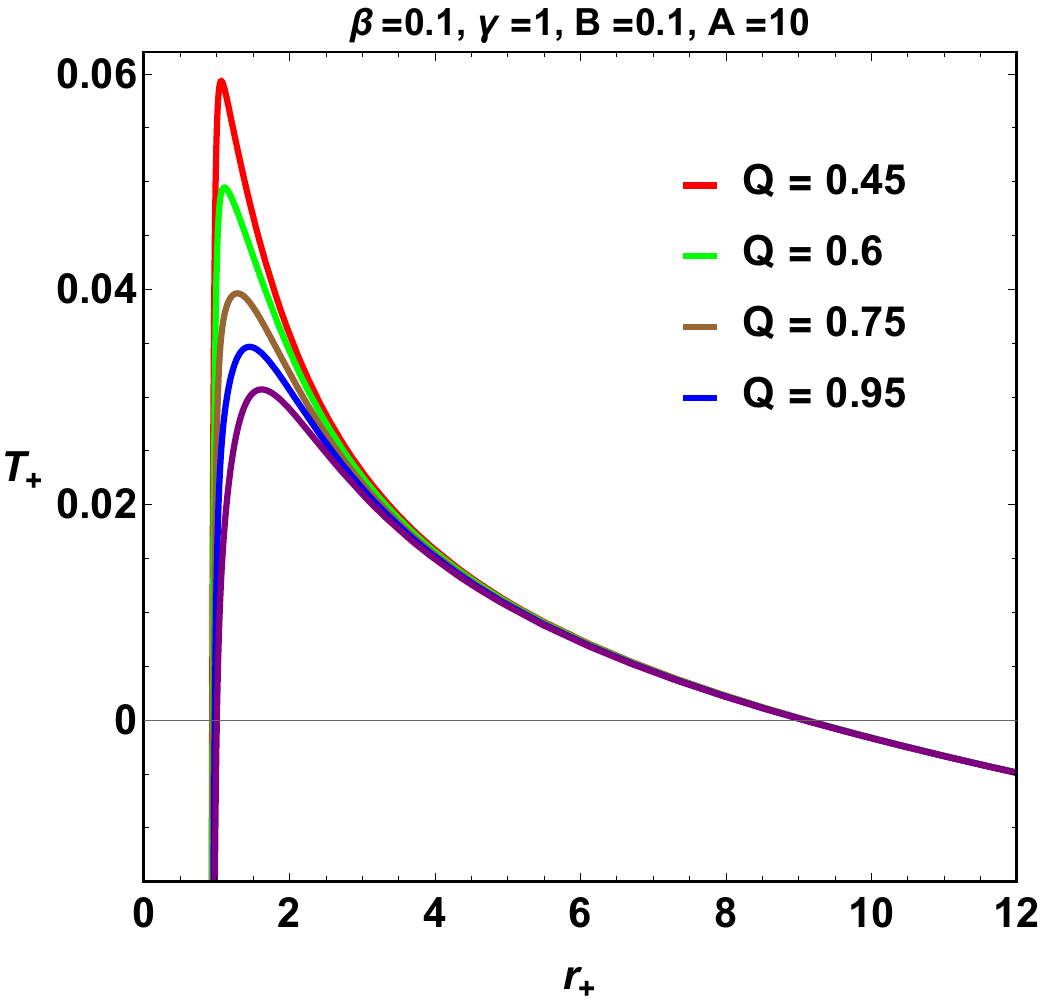} \hspace{2mm}
       }
      	\caption{Variation of the Hawking temperature $T_+$ (\ref{43}) with respect to the horizon radius $r_+$.}
      	\label{fig3}
      \end{figure*}
      
To describe the behavior of the Hawking temperature, Fig. $\ref{fig3}$ depicts graphically the Hawking temperature against the horizon radius. It is worth noting that all the parameter variations in the BH system contribute to the effect on the Hawking temperature. Remarkably, the Hawking temperature rises to a maximum at $T^{max}_+$ for a specific horizon radius $r_+$. In particular, the maximum related to the Hawking temperature spanned the interval $1.04269<T^{max}_+<1.06535$ for all the considered parameter spaces, and that maximum grows significantly with the charge parameter variation, which is the largest maximum among the other parameter variations. After reaching the maximum, all the Hawking temperature configurations start to decrease monotonically, up to being negative in the end. A distinct remark schematically shows that the variation of the charge parameter at the level of the Hawking temperature could generate multiple peaks in contrast to the other parameter variations. In addition, a closer examination shows that the behaviors of the Hawking temperature for all parameters set finished being unphysically due to the negativity referred to as $T_+<0$.

To carry out a nice generation of the corresponding thermodynamic quantities, it is useful to apply the first law of BH thermodynamics. In particular, the first law can be given as ~\cite{Cai:1998vy}
\begin{equation}\label{r32}
    \mathrm{d} M=T_+ \mathrm{d}S+\sum_i \mu_i \,\mathrm{d}\mathcal{N}_i
\end{equation}
where $\mu_i$ are the chemical potentials corresponding to the conserved charges $\mathcal{N}_i$.

Holding parameters constant, with the exception of entropy, one can find
\begin{equation}\label{r33}
    S=\frac{1}{T_+}\int\frac{\partial M_+}{\partial r_+}\,\mathrm{d}r_+=\pi\, r_+^2,
\end{equation}
which is similar compared to multiple background studies~\cite{Bekenstein:1973ur, Bekenstein:1972tm, Bekenstein:1974ax, Bardeen:1973gs}.

On the other hand, by considering the extended phase space, certain critical processes as a thermodynamic aspect are achieved. At this point, the $P-V$ criticality, which will be a main feature of this work, is treated once the negative cosmological constant behaves like pressure, also known as "black hole chemistry". So, one has ~\cite{Kastor:2009wy, Gunasekaran:2012dq}
\begin{equation}\label{r49}
P=-\frac{\Lambda}{8\pi} =\frac{3}{8\pi\,\ell^2}.
\end{equation} 
In light of this objective, the first modified law of thermodynamics is formulated as follows ~\cite{Dolan:2010ha, Brown:1987dd}:
\begin{equation}\label{r40}
\mathrm{d}H=\mathrm{d}M=T\mathrm{d}S+V\mathrm{d}P+\Phi \mathrm{d}Q
\end{equation}
where mass resembles enthalpy. Whereas, to construct the thermodynamic phase space framework, in particular, the parameter function $f(r_+, M, P, Q)$ must always vanish under any transformation of the parameter. Further remarks on this subject make similar arguments to consider the constraints $f(r_+, M, P, Q)= 0$ and $\delta f(r_+, M, P, Q)= 0$ on the evolution along the space of parameters. Nevertheless, an alternative way is considered: taking the mass parameter, $M$, as a function of the parameters $M(r_+,\ell, Q)$ too.

The thermodynamic parameters are $S$, $P$, and $Q$. It is then convenient to redefine $M=M(S, P, Q)$ for the possibility of explicitly obtaining
\begin{widetext}
\begin{equation}\label{r41}
\mathrm{d}M=\bigg(\frac{\partial M}{\partial S}\bigg)_{P,Q}\mathrm{d}S+\bigg(\frac{\partial M}{\partial P}\bigg)_{S,Q}\mathrm{d}P+\bigg(\frac{\partial M}{\partial Q}\bigg)_{S,P}\mathrm{d}Q.
\end{equation}
\end{widetext}
This is similar to such a differential 1-form in the space of parameters. Accordingly, all components are nothing more than thermodynamic quantities expressed in the context of extended phase space. So, one has
\begin{align}\label{r42}
T&=\bigg(\frac{\partial M}{\partial S}\bigg)_{P,Q} \\
V&=\bigg(\frac{\partial M}{\partial P}\bigg)_{S,Q}\\
\phi&=\bigg(\frac{\partial M}{\partial Q}\bigg)_{S,P}.
\end{align}

In an alternative way, the same results can be expressed in accordance with the variation along the space of the parameters of the condition described by $f(r_+, M, P, Q)$,
\begin{widetext}
\begin{equation}\label{r43}
\mathrm{d}f(r_+, M, P, Q)=0=\frac{\partial f}{\partial r_+}\mathrm{d}r_++\frac{\partial f}{\partial M}\mathrm{d}M+\frac{\partial f}{\partial P}\mathrm{d}P+\frac{\partial f}{\partial Q}\mathrm{d}Q
\end{equation}
\end{widetext}
reshape another term for $\mathrm{d}M$ giving as follows:
\begin{widetext}
\begin{align}\label{r44}
\mathrm{d}M&=\bigg(\frac{1}{4\pi}\frac{\partial f}{\partial r_+}\bigg)\bigg(-\frac{1}{4\pi}\frac{\partial f}{\partial M}\bigg)^{-1}\mathrm{d}r_++\bigg(-\frac{\partial f}{\partial M}\bigg)^{-1}\bigg(\frac{\partial f}{\partial P}\bigg)\mathrm{d}P+\bigg(-\frac{\partial f}{\partial M}\bigg)^{-1}\bigg(\frac{\partial f}{\partial Q}\bigg)\mathrm{d}Q
\end{align}
\end{widetext}
which must be in conformity with equation (\ref{r40}). Eq. (\ref{r44}) embraces the presence of temperature, which is geometrically defined as
\begin{equation}\label{r45}
T=\frac{1}{4\pi}\frac{\partial f}{\partial r_+},
\end{equation}
and which is a well-known finding, providing
\begin{eqnarray}\label{r46}
\mathrm{d}S=\bigg(-\frac{1}{4\pi}\frac{\partial f}{\partial M}\bigg)^{-1}\,\mathrm{d}r_+.
\end{eqnarray}
It should be pointed out that this expression can also be derived using Wald's formalism; basically, $\delta S = \delta \int\frac{\partial L}{\partial R}$ as long as $\mathrm{d}f = 0$ is satisfied.

Furthermore, the thermodynamic volume and the conjugate potential are defined by the following formula:
\begin{align}\label{r47}
V&=\bigg(\frac{\partial M}{\partial P}\bigg)_{S,Q}=\bigg(-\frac{\partial f}{\partial M}\bigg)^{-1}\bigg(\frac{\partial f}{\partial P}\bigg),\\
\phi&=\bigg(\frac{\partial M}{\partial Q}\bigg)_{S,P}=\bigg(-\frac{\partial f}{\partial M}\bigg)^{-1}\bigg(\frac{\partial f}{\partial Q}\bigg).
\end{align}
In the case of the BH system, the enthalpy is defined by the total mass of the system. Consequently, in terms of the parameters BH system and in the context of the extended phase space, the thermodynamic volume and the electric potential can thus be formulated as
\begin{align}\label{r52}
V&=\left(\frac{\partial M}{\partial P}\right)_{S,Q}=\frac{4\pi r_+^3}{3}.
\\
\phi&=\left(\frac{\partial M}{\partial Q}\right)_{S,P}=\frac{Q}{r_+}.
\end{align}
while the Hawking temperature is given by
\begin{widetext}
    \begin{equation}
        T=\bigg(\frac{\partial M}{\partial S}\bigg)_{P,Q}=\frac{1}{4
   \pi  r_+^3}\Bigg(r_+^4 \bigg(8 \pi  P-\left(\frac{A+1}{B}\right)^{-\frac{1}{\beta +1}} \bigg(\frac{1}{B}\left(\frac{\gamma }{r_+^3}\right)^{(A+1) (\beta +1)}+1\bigg)^{\frac{1}{\beta +1}}\bigg)-Q^2+r_+^2\Bigg).
    \end{equation}
\end{widetext} 
According to Euler's theorem~\cite{Kastor:2009wy,Altamirano:2014tva}, with $M(S, P, Q)$, the Smarr formula can be constructed for the charged source in the framework of GR as 
\begin{equation}\label{r48}
M=2TS-2PV+\phi\,Q
\end{equation}
in which another combination arises between thermodynamic quantities. These amounts in the classical limit $(Q \rightarrow 0)$ are consistent with the corresponding amounts for the Schwarzchild AdS BH ~\cite{Kumar:2020cve}.
\begin{figure*}[tbh!]
      	\centering{
       \includegraphics[scale=0.415]{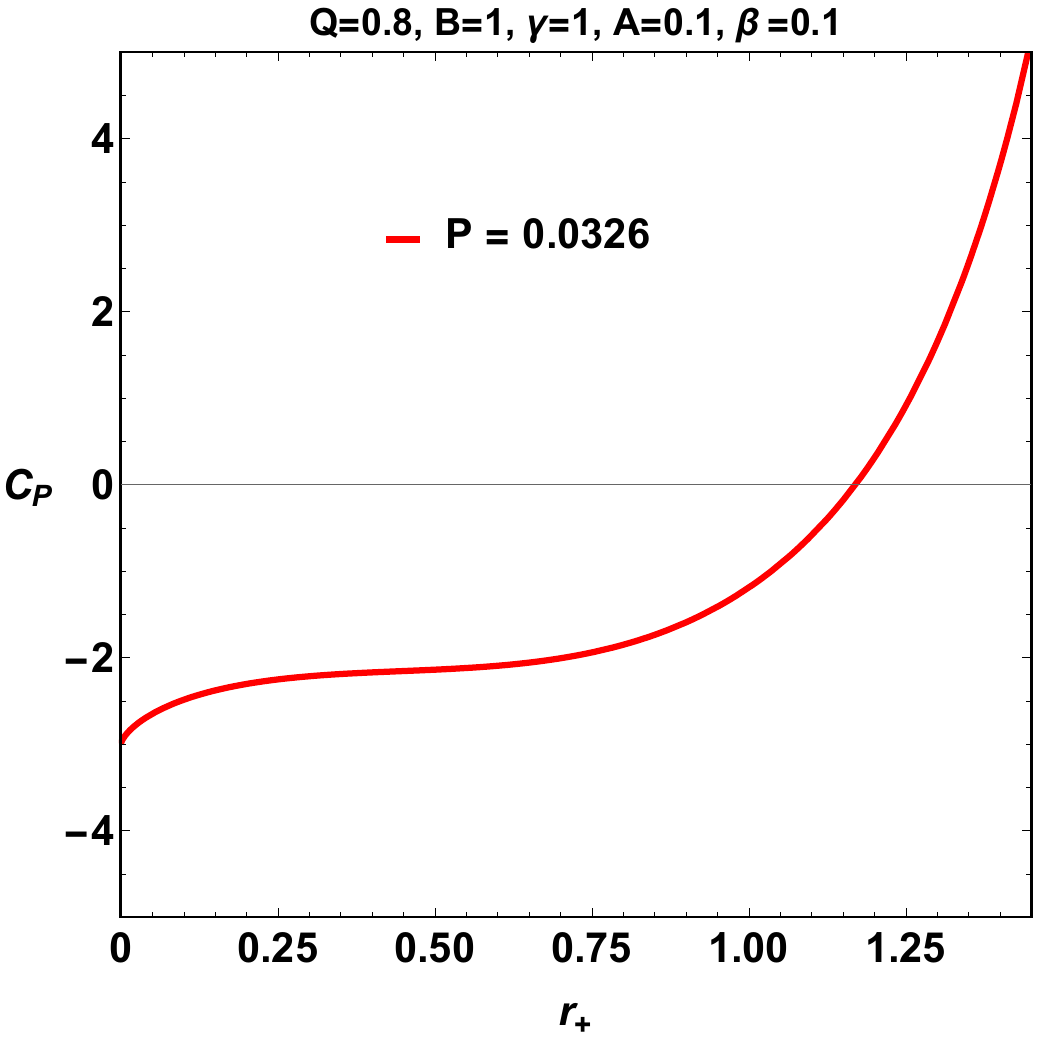} \hspace{2mm}
      	\includegraphics[scale=0.8]{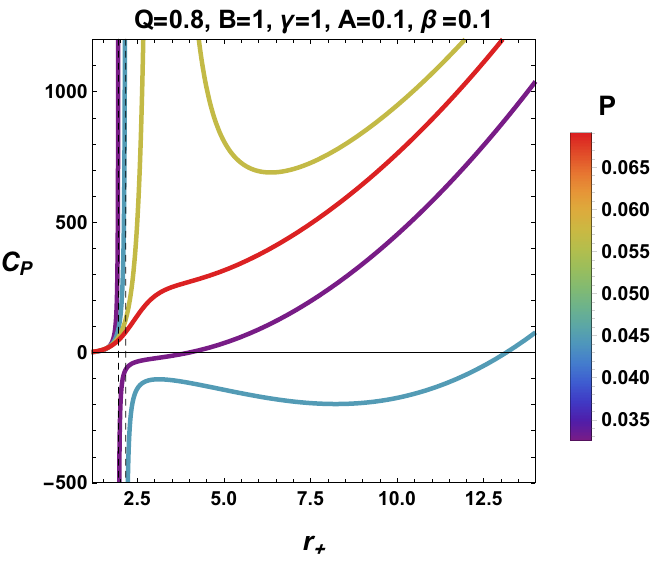} \hspace{2mm}
      }
       \centering{ 
       \includegraphics[scale=0.44]{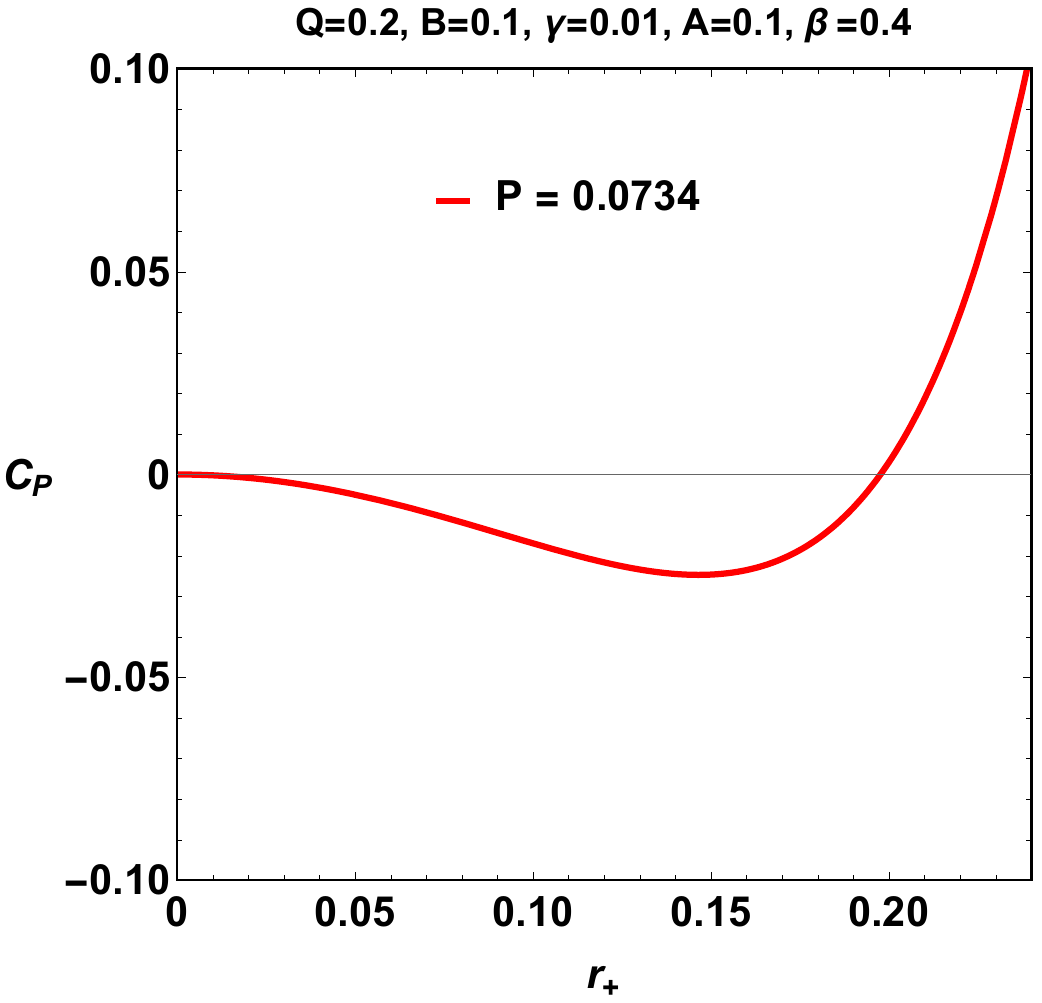}\hspace{2mm}
      \includegraphics[scale=0.8]{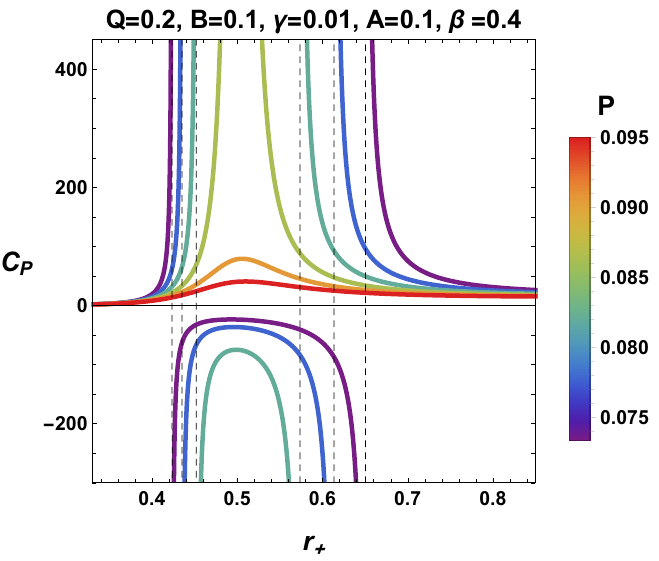} \hspace{2mm}
       }
      	\caption{Variation of heat capacity $C_P$ (\ref{ra3}) as a function of $r_+$ for different values of pressure and for fixed values in parameter space.}
      	\label{fig4}
      \end{figure*}
      
\section{Black hole stability}
\label{BHstab}
Within the canonical ensemble, heat capacity is a further thermodynamical quantity providing information on the thermal state of the BHs. indeed, the heat capacity involves three specific and fascinating pieces of information. Firstly, the discontinuous behavior of this quantity means the existence of possible thermal phase transitions that the system may undergo. Second, another feature deals with the sign of heat capacity. In essence, the sign shows whether the system is thermally stable or not. In other words, positivity generates thermal stability, whereas the opposite indicates instability  ~\cite{Sahabandu:2005ma, Cai:2003kt}. Third, the corresponding possible set of roots is useful given leads to indicate the sign change, which may show stable/unstable states or bound points. As a consequence of these points, this section and the next are devoted to computing the relevant heat capacity in the extended phase space and inspecting the relation between this quantity and the GT tools. One can compute the heat capacity according to the following form 
 ~\cite{Kubiznak:2016qmn} :
\begin{equation}\label{r34}
C_P=\left(\frac{\partial H}{\partial T_+}\right)_P=T_+\left(\frac{\partial S}{\partial T_+}\right)_P.
\end{equation}
Likewise, the function is expressed in terms of the parameter space as
 \begin{equation}\label{ra3}
            C_P=2 \pi \, r_+^2 \left(\left(\frac{\gamma}{r_+^3}\right)^{(A+1) (\beta +1)}+B\right)\textit{frac}
        \end{equation}
        where
\begin{widetext}
    \begin{eqnarray}
     \textit{frac}= \frac{\left(\frac{A+1}{B}\right)^{\frac{1}{\beta +1}} \left(-8 \pi  P r^4+Q^2-r^2\right)+r^4 \mathcal{T}_1^{\frac{1}{\beta+1}}}{\left(\frac{A+1}{B}\right)^{\frac{1}{\beta +1}} \left(-8 \pi  P r^4-3 Q^2+r^2\right) B\,\mathcal{T}_1+r^4
  \mathcal{T}_1^{\frac{1}{\beta +1}} \left(B-(3 A+2) \left(\frac{\gamma}{r^3}\right)^{(A+1) (\beta +1)}\right)}\nonumber
    \end{eqnarray}
\end{widetext}
Within the realm of BH physics, it is claimed that the associated roots of the heat capacity $(C_{P} = T = 0)$ indicate a one-dimensional line between physical $(T > 0)$ and non-physical $(T < 0)$ BHs, known as the physical limit. In this region, the system exhibits a sign change in heat capacity. On the other hand, the set of divergence points of the heat capacity represents the phase transition critical points of BHs~\cite{Tzikas:2018cvs}. Thus, both the phase transition critical and limitation points of the BHs are explicitly computed with consideration of the following constraints:
  \begin{itemize}
    \item $\left(\frac{\partial M}{\partial S}\right)_{P,Q}=0$ \hspace{0.4cm} \text{physical limitation points} \\
    \vspace*{1mm}
   \item    $\left(\frac{\partial^2 M}{\partial S^2}\right)_{P,Q}=0$\hspace{0.3cm}\text{phase transition critical points}
    \end{itemize}
So as to find the physical limitation and phase transition critical points, we consider Eq. ($\ref{Mass}$) and solve the following equations for the entropy or in terms of the horizon radius:
\begin{widetext}
    \begin{align}
   &\left(\frac{\partial M}{\partial S}\right)_{P,Q}= r_+^4 \Biggl\{8 \pi  P-\left(\frac{A+1}{B}\right)^{-\frac{1}{\beta +1}} \left(\frac{1}{B}\Bigg(\left(\frac{\gamma
   }{r_+^3}\right)^{(A+1) (\beta +1)}+B\Bigg)\right)^{\frac{1}{\beta +1}}\Biggl\}-Q^2+r_+^2=0,\\
   &\left(\frac{\partial^2 M}{\partial S^2}\right)_{P,Q}  = B \left(\frac{A+1}{B}\right)^{\frac{1}{\beta +1}} \left(8 \pi  P r_+^4+3 Q^2-r_+^2\right)+8 \pi 
   \gamma  P r_+ \left(\frac{A+1}{B}\right)^{\frac{1}{\beta +1}} \left(\frac{\gamma
   }{r_+^3}\right)^{A \beta +A+\beta }\nonumber\\
   &+\left(\frac{\gamma }{r_+^3}\right){}^{(A+1) (\beta +1)}
   \left(\left(3 Q^2-r_+^2\right) \left(\frac{A+1}{B}\right)^{\frac{1}{\beta +1}}+(3 A+2) r_+^4
   \left(\frac{1}{B}\left(\frac{\gamma }{r_+^3}\right)^{(A+1) (\beta +1)}+1\right)^{\frac{1}{\beta
   +1}}\right)\nonumber\\
   &-B r_+^4 \left(\frac{1}{B}\left(\frac{\gamma }{r_+^3}\right)^{(A+1) (\beta
   +1)}+1\right)^{\frac{1}{\beta +1}}=0
\end{align}
\end{widetext}
Attempting to solve these equations appears to be analytically difficult. This is why we apply the numerical approach to find physical limit points and phase transition points for a given parameter space. The presentation of the heat capacity root point and singularity point is illustrated for two given parameter spaces in Tabs. $\ref{Tab1}$- $\ref{Tab2}$.
\begin{widetext}
\begin{center}
      \begin{table}[!h]      
    \begin{tabular}{lccccccc} 
    \hline\hline
       $Q$  \,&\, $\beta$ \,&\, $B$ \,&\, $\gamma$ \,&\,  $P$ \,&\, $r_1^\star$ \,&\,  $r_2^\star$ \,&\,  \text{Number of points}
       \\
       \hline
       0.8 \,&\, 0.1 \,&\, 1 \,&\, 1 \,&\,   0.0306$<P_c$ \,&\, 1.63453 \,&\,1.94108 \,&\, 2\\
       0.8 \,&\, 0.1 \,&\, 1 \,&\, 1 \,&\,   0.0326$<P_c$ \,&\, 1.39086 \,&\, 2.79912 \,&\, 2\\
        0.8 \,&\, 0.1 \,&\, 1 \,&\, 1 \,&\,   0.0383 $=P_c$\,&\,1.19141\,&\, $\emptyset$\,&\, 1
        \\
        0.8 \,&\, 0.1 \,&\, 1 \,&\, 1 \,&\,   0.0387$>P_c$\,&\,1.18307\,&\, $\emptyset$\,&\, 1
        \\
       0.2 \,&\, 0.4 \,&\, 0.1 \,&\, 0.01 \,&\,  0.0734$<P_c$\,&\,0.197083\,&\, $\emptyset$ \,&\, 1\\
     0.2 \,&\, 0.4 \,&\, 0.1 \,&\, 0.01 \,&\,   0.0863$=P_c$\,&\,0.196779\,&\, $\emptyset$ \,&\, 1\\
      0.2 \,&\, 0.4 \,&\, 0.1 \,&\, 0.01 \,&\,   0.0950$>P_c$\,&\,0.196066\,&\, $\emptyset$ \,&\, 1\\
     \\
           \hline\hline
    \end{tabular}
    \caption{The physical limitation points with $A=0.1$.}
    \label{Tab1}
\end{table}
\end{center}
\end{widetext}

To get an appropriate description of the heat capacity behavior at constant pressure, Fig. $\ref{fig4}$  presents the variation of $C_P$ against the horizon radius$r_+$. The appearance of divergent points and intersecting points both correspond to a change of sign, which indicates a phase transition, either a second-order phase transition or a first-order one, respectively. Graphically, the analysis of the behavior of the corresponding heat capacity is present in Fig. \ref{fig4}. Thus, the upper panel concerns a fixed-valued parameter space that generates the behavior of heat capacity as a function of horizon radius. It should be noted that for pressure less than the critical pressure $(P<P_c)$, especially $P=0.0306$, there are two roots for the heat capacity (Tab. \ref{Tab1}). In this case, these two roots are in fact two critical horizon radii, i.e., $r_1^\star$ and $r_2^\star$ (the second root exists only for the upper panel with a certain parameter space), at which, for $r_+ < r_1^\star$, the heat capacity of BHs is negative. So, the BHs with a small horizon radius are unstable. In addition, for $r_2^\star> r_+$, the heat capacity is positive which claims that the BHs are stable. It can be clearly observed that there is an important region due to $r_1^\star < r_+ < r_1^\text{div}$, where $r_1^\text{div}$ is a divergence point. The heat capacity of BHs in this region is positive. In other words, BHs have thermal stability when their horizon radius lies within the interval $r_1^\star < r_+ < r_1^\text{div}$. Similarly, for the region $r_1^\text{div} < r_+ < r_2^\star$, the heat capacity is negative, which means that our BHs are unstable. On the other hand, for $(P<P_c)$ in the bottom panel with fixed values in parameter space, there are two roots and two critical phase transition points generating four intervals along the horizon radius. Clearly, BHs are thermally stable for regions $r_1^\star < r_+ < r_1^\text{div}$ and $r_+ > r_2^\text{div}$, while they are thermally unstable for regions $r_+ < r_1^\star$ and $r_1^\text{div} < r_+ < r_2^\text{div}$. Furthermore, as the pressure closes to critical pressure $(P=P_c)$, the number of physical limitation points and divergent points reduces, and the heat capacity is positive, which means that the BHs are thermally stable and have a one-phase transition. This behavior is similar for the two sets of parameter spaces regarding the heat capacity behavior. For $(P>P_c)$, the critical behavior is no longer present due to the absence of such a discontinuity provided by the divergent point. This situation remains the BHs thermally stable. By contrast, physical limitations are always present. Roughly speaking, our BH solution clearly remains in a thermally stable state.
      \begin{widetext}
\begin{center}
      \begin{table}[ht]     
    \begin{tabular}{lccccccc} 
    \hline\hline
       $Q$  \,&\, $\beta$ \,&\, $B$ \,&\, $\gamma$ \,&\,  $P$ \,&\,$r_1^\text{div}$ \,&\,  $r_2^\text{div}$ \,&\,  \text{Number of points}
       \\
       \hline
        0.8 \,&\, 0.1 \,&\, 1 \,&\, 1 \,&\,   0.0306$<P_c$ \,&\,1.94108 \,&\, $\emptyset$ \,&\, 1\\
       0.8 \,&\, 0.1 \,&\, 1 \,&\, 1 \,&\,   0.0326$<P_c$ \,&\, 1.86425 \,&\, $\emptyset$ \,&\, 1\\
       0.8 \,&\, 0.1 \,&\, 1 \,&\, 1 \,&\,   0.0383 $=P_c$\,&\,3.1466312\,&\,$\emptyset$\,&\, 1\\
       0.8 \,&\, 0.1 \,&\, 1 \,&\, 1 \,&\,   0.0387$>P_c$\,&\,$\emptyset$\,&\,$\emptyset$\,&\, 0
       \\
       0.2 \,&\, 0.4 \,&\, 0.1 \,&\, 0.01 \,&\,   0.0734$<P_c$\,&\, 0.4237 \,&\,0.650683\,&\, 2\\
       0.2 \,&\, 0.4 \,&\, 0.1 \,&\, 0.01 \,&\,   0.0863$=P_c$\,&\, 0.5022368 \,&\,$\emptyset$\,&\, 1\\
       0.2 \,&\, 0.4 \,&\, 0.1 \,&\, 0.01 \,&\,   0.0950$>P_c$\,&\, $\emptyset$ \,&\,$\emptyset$\,&\, 0 \\
      \\
           \hline\hline
    \end{tabular}
    \caption{The phase transition critical points with $A=0.1$.}
     \label{Tab2}
\end{table}
\end{center}
\end{widetext}

\section{$P-V$ criticality}
\label{PVcr}
In this section, special care is taken to study the critical $P-V$ for charged AdS BHs surrounded by an MCG structure. As such, by exploiting the expression for the Hawking temperature with (\ref{r49}), we can obtain the corresponding equation of state as follows:
\begin{widetext}
\begin{align}\label{r51}
P&= -\frac{1}{8 \pi\,  r_+^2}+\frac{T}{2r_+} +\frac{Q^2}{8 \pi  \,r_+^4}+\frac{1}{8 \pi }\bigg(\frac{A+1}{B}\bigg)^{-\frac{1}{\beta +1}} \Bigg(\frac{1}{B}\bigg(\frac{\gamma}{r_+^3}\bigg)^{(A+1) (\beta
   +1)}+1\Bigg)^{\frac{1}{\beta +1}}
\end{align}
\end{widetext}
\begin{figure*}[tbh!]
      	\centering{
       \includegraphics[scale=0.79]{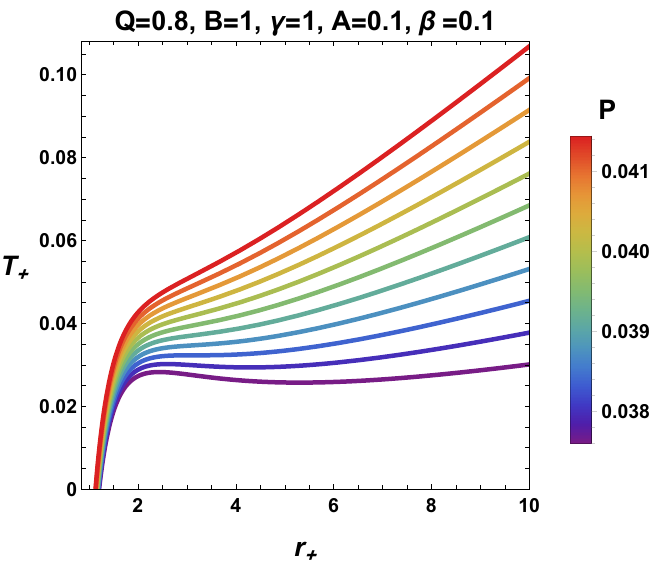} \hspace{2mm}
      	\includegraphics[scale=0.79]{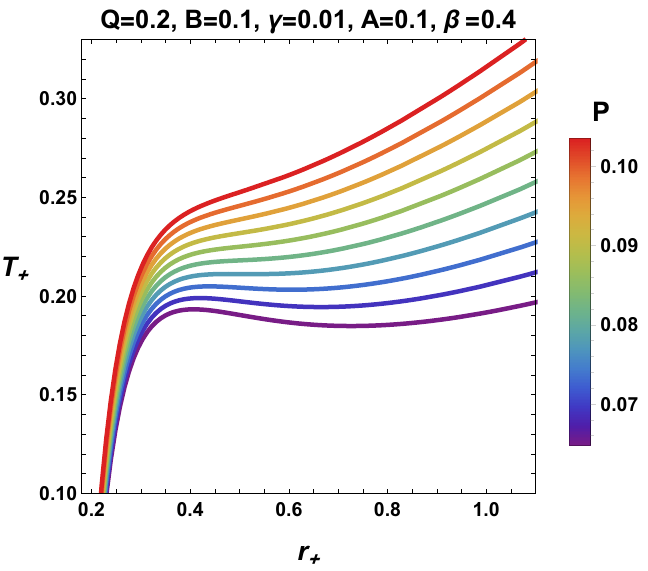} \hspace{2mm}
      }
       
      	\caption{Isobaric curve $T$-$r_+$ diagrams of the BH system for certain values of the parameter space.}
      	\label{fig5}
      \end{figure*}
Of course, the parameters $\beta$, $A$, $B$, $\gamma$ and $Q$ potentially affect this equation. As intended, we can define the specific volume $v = 2r_+$, wherewith the pressure is cast in the standard form $P = \frac{1}{v}\, T + \mathcal{O}(v)$.Moreover, since the thermodynamic volume $V \propto r^3_+$ the critical point can be inspected considering the following constraints:
\begin{equation}\label{r53}
\bigg(\frac{\partial P}{\partial r_+}\bigg)_T=0,\quad \bigg(\frac{\partial^2 P}{\partial r_+^2}\bigg)_T=0
\end{equation}
or alternatively,
\begin{equation}\label{r54}
\bigg(\frac{\partial T}{\partial r_+}\bigg)_P=0,\quad \bigg(\frac{\partial^2 T}{\partial r_+^2}\bigg)_P=0.
\end{equation}
The previous constraints set the triply critical point $(T_c, P_c, r_c)$ with the unknown critical horizon radius $r_c$ which needs numerical solving. one has
\begin{figure*}[tbh!]
      	\centering{
       \includegraphics[scale=0.8]{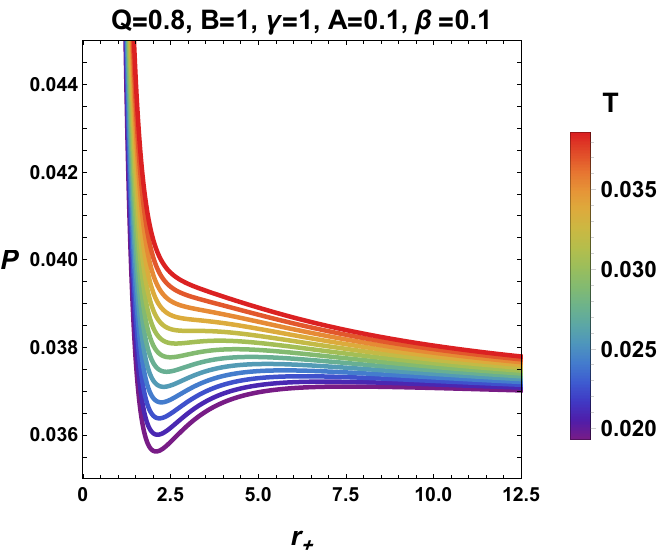} \hspace{2mm}
      	\includegraphics[scale=0.77]{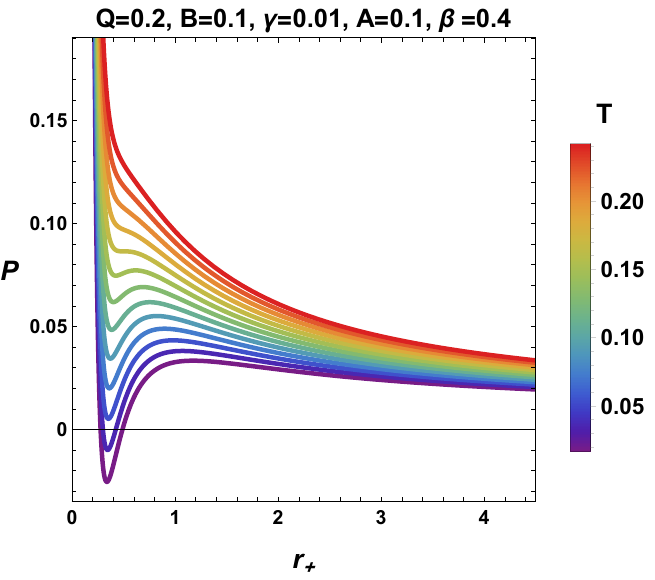} \hspace{2mm}
      }
       
      	\caption{Isotherme curve $P$-$r_+$ diagrams of the BH system for certain values of the parameter space. }
      	\label{fig6}
      \end{figure*}
      \begin{widetext}
          \begin{align}
        T_c&=  \frac{1}{4 \pi  r_c^3}\Biggl\{r_c \bigg(2 r_c-3 \gamma  \left(\frac{A+1}{B}\right)^{\frac{\beta }{\beta +1}} \left(\frac{\gamma }{r_c^3}\right){}^{A \beta +A+\beta } \bigg(\frac{1}{B}\left(\frac{\gamma }{r_c^3}\right){}^{(A+1)
   (\beta +1)}+1\bigg){}^{\frac{1}{\beta +1}-1}\bigg)-4 Q^2\Biggl\},
    \\
        P_c&=  \frac{1}{8 \pi  r_c^4}\Biggl\{r_c^2 \bigg(r_c^2 \left(\frac{A+1}{B}\right)^{-\frac{1}{\beta +1}} \left(\frac{1}{B}\left(\frac{\gamma
   }{r_c^3}\right){}^{(A+1) (\beta +1)}+1\right){}^{\frac{1}{\beta +1}}-1\bigg)+Q^2\nonumber\\
  &+ \frac{1}{r_c^2}\Bigg(r_c \bigg(2 r_c-3 \gamma  \left(\frac{A+1}{B}\right)^{\frac{\beta }{\beta +1}} \left(\frac{\gamma }{r_c^3}\right){}^{A \beta +A+\beta } \bigg(\frac{1}{B}\left(\frac{\gamma
   }{r_c^3}\right){}^{(A+1) (\beta +1)}+1\bigg){}^{\frac{1}{\beta +1}-1}\bigg)-4 Q^2\Bigg)\Biggl\}
      \end{align}
      \end{widetext}
      and
     \begin{widetext}
         \begin{align}
        &2 B^2 \left(\frac{A+1}{B}\right)^{\frac{1}{\beta +1}} \left(6 Q^2-r_c^2\right)+\frac{1}{r_c^6}\Bigg(\gamma ^2 \left(\frac{\gamma }{r_c^3}\right)^{2 (A
   \beta +A+\beta )} \bigg(3 (A+1) (3 A+2) r_c^4 \bigg(\frac{1}{B}\left(\frac{\gamma }{r_c^3}\right)^{(A+1) (\beta +1)}+1\bigg){}^{\frac{1}{\beta +1}}\nonumber\\
   &-2 \left(\frac{A+1}{B}\right)^{\frac{1}{\beta +1}}
   \left(r_c^2-6 Q^2\right)\bigg)\Bigg)+B \left(\frac{\gamma }{r_c^3}\right)^{(A+1) (\beta +1)} \bigg(3 (A+1) (3 A (\beta +1)+3 \beta +2) r_c^4 \bigg(\frac{1}{B}\left(\frac{\gamma
   }{r_c^3}\right)^{(A+1) (\beta +1)}+1\bigg)^{\frac{1}{\beta +1}}\nonumber\\
   &-4 \left(\frac{A+1}{B}\right)^{\frac{1}{\beta +1}} \left(r_c^2-6 Q^2\right)\bigg)=0
         \end{align}
     \end{widetext} 
In Tab. $\ref{Tab3}$, the numerical sets of critical points are shown as a function of various BH system parameter values. Thus, the universal constant ratio $\frac{P_c\,r_c}{T_c}$ clearly varies with the variation of the BH system parameter, being proportional to the set of parameters ($Q$, $B$, and $\gamma$) and disproportionate to the $\beta$ parameter.
     
     To highlight the key properties of $P$-$V$ criticality, the $T$-$r_+$ and $P$-$r_+$ diagrams are useful at this stage, Fig. $\ref{fig5}$  shows the isobaric curve on the $T-r_+$ diagram, which is classified according to multiple pressure values. Indeed, the appropriate case $P<P_c$ implies two extreme points along the sides of three branches, namely the small BH branch, the intermediate BH branch, and the large BH branch. The small and large BH branches are characterized by a positive slope, meaning that the heat capacity is positive, and the system is thermally stable. While the intermediate BH branch stands on a negative slope, the BH is thermally unstable due to the negativity of the heat capacity. On the other hand, the case $P>P_c$ indicates the absence of any extremal point which proves that the system displays one stable BH branch in this situation. Alternatively, Fig. $\ref{fig6}$  displays, throughout a spectrum of temperature $T$, the $P$-$r_+$ criticality behavior via the isotherm curve $P$-$r_+$. It is obvious all too well that the diagram shown is very similar to the VdW liquid-gas system (\ref{fig6}). In this way, the situation $T<T_c$ predicts the existence of a small-large BH phase transition underlying the liquid-gas phase transition of the VdW. As the temperature reaches the critical temperature $(T=T_c)$, the number of phase transitions is limited to one, for which the first-order phase transition merges with the second-order phase transition like in a real gas system. On the other hand, once $(T>T_c)$, the system displays one-phase behavior, and it is in fact an ideal with an empty set of phase transitions.
\begin{widetext}
    \begin{center}    
      \begin{table}[ht]
    \begin{tabular}{lccccccccccc} 
    \hline\hline
       $Q$  \,& $\beta$ \,& $B$ \,& $\gamma$  \,& $r_c$ \,& $T_c$ \,& $P_c$ \,& $\frac{P_c\, r_c}{T_c}$ \,& $\mathcal{C}$ \,&  $\mathcal{C}_1$ \,& $\mathcal{C}_3$ \,& $\mathcal{C}_5$
       \\
       \hline
       0.2 \,&\, 0.4 \,&\, 0.1 \,&\, 0.01  \,&\, 0.50223\,&\, 0.21186\,&\, 0.08637\,&\,  0.20474\,&\, 1\,&\, 2.44204\,&\, -2.44204\,&\,-1.22199  \\
       0.4 \,&\, 0.3 \,&\, 0.3 \,&\, 0.03  \,&\, 1.00185\,&\, 0.10609\,&\, 0.03452\,&\,  0.32596\,&\, 1\,&\, 1.5339\,&\, -1.5339\,&\, -0.76960 \\
       0.6 \,&\, 0.2 \,&\, 0.6 \,&\, 0.06  \,&\, 1.51043\,&\, 0.07021\,&\, 0.0327248\,&\,  0.70392\,&\, 1\,&\, 0.7103\,&\, -0.7103\,&\, -0.3548\\
       0.8 \,&\, 0.1 \,&\, 1 \,&\, 1  \,&\, 3.14663\,&\, 0.03224\,&\, 0.03836\,&\,3.7441 \,&\, 1\,&\,0.1335\,&\, -0.1335\,&\, -0.06166 \\
       1 \,&\, 0.05 \,&\, 1.8 \,&\, 1.5  \,&\, 4.31299\,&\, 0.02278\,&\,0.06455\,&\, 12.2182 \,&\, 1\,&\, 0.04092\,&\, -0.04092 \,&\, -0.01806\\
           \hline\hline
    \end{tabular}
     \caption{Numerical sets for critical physical quantities and coefficients $\mathcal{C}$ and $\mathcal{C}_i$ in $P - V$ critical behavior with $A =0.1$}
      \label{Tab3}
    \end{table}
    \end{center}
\end{widetext}
   
The present step applies a different thermodynamic potential to achieve perfect disclosure of the phase transition of the BH system. In particular, the Gibbs free energy is a thermodynamic quantity calculated from the Euclidean action with an appropriate limit term. A fruitful feature resulting from the sign of the Gibbs free energy enables a global stability analysis. In the extended phase space, the thermodynamic potential is now the Gibbs free energy $G=M-T\, S=H-T\, S$. In practice, it should be noted that any discontinuous behavior in the first- or second-order derivatives of the Gibbs energy leads to a first- or second-order phase transition in the system. The Gibbs free energy is therefore given by
\begin{widetext}
\begin{align}\label{r56}
G&=G(P, T)=H-TS=M-TS
   \nonumber
   \\
&=\frac{12}{r_+}\Bigg\{-r_+^2 \bigg( -3 +8 P\,\pi\,r_+^2\,  \bigg)+9 Q^2+r_+^4 \bigg(\frac{B}{A+1}\bigg)^{\frac{1}{\beta +1}} \Bigg(3
   \bigg(\frac{1}{B}\bigg(\frac{\gamma}{r_+^3}\bigg)^{(A+1) (\beta +1)}+1\bigg)^{\frac{1}{\beta +1}}-2\, \,
   _2F_1[\alpha,\nu;\lambda;\xi]
\Bigg) \Bigg\}
\end{align}
\end{widetext}
\begin{figure*}[tbh!]
      	\centering{
       \includegraphics[scale=0.78]{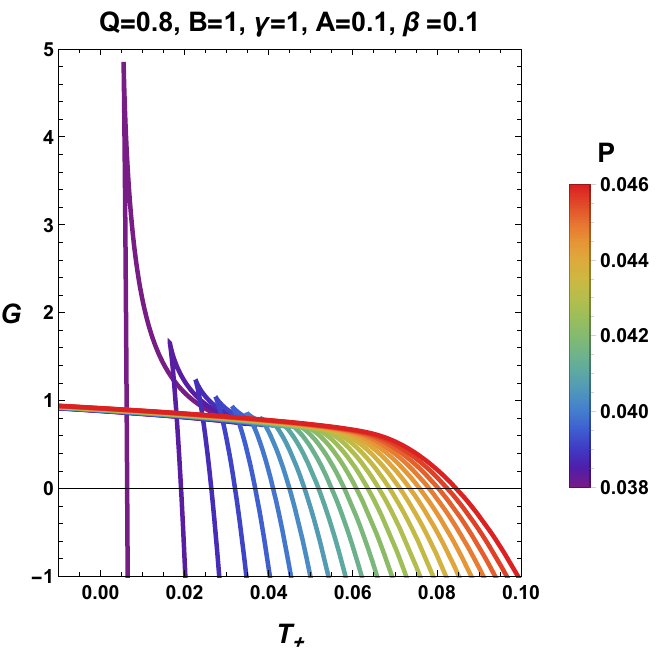} \hspace{2mm}
      	\includegraphics[scale=0.77]{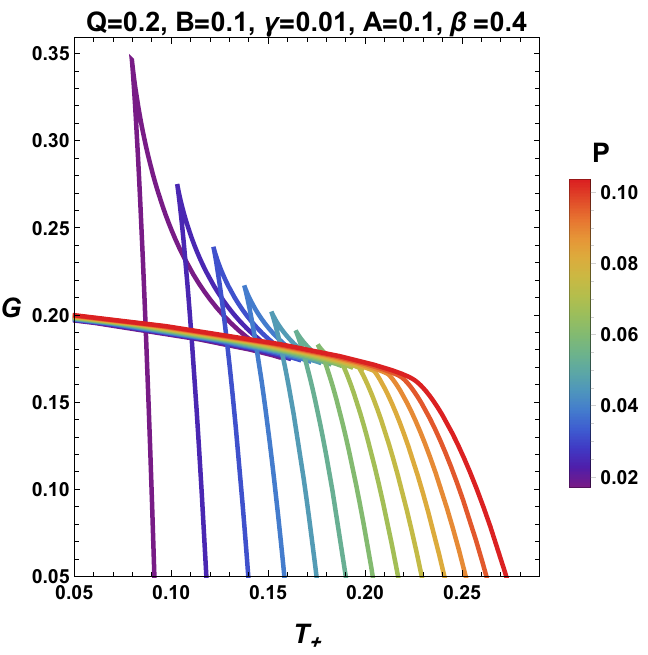} \hspace{2mm}
      }
       
      	\caption{The $G-T$ diagram of the black hole system for certain values of the parameter space.}
      	\label{fig7}
      \end{figure*}
      \begin{figure*}[tbh!]
      	\centering{
       \includegraphics[scale=0.78]{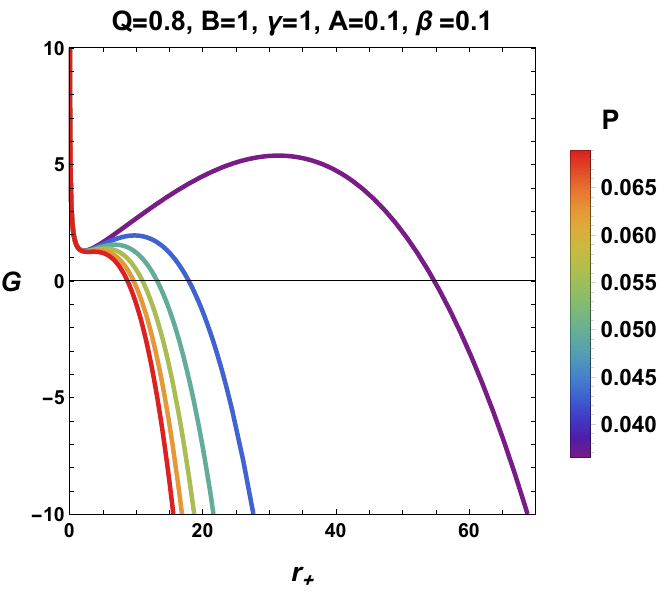} \hspace{2mm}
      	\includegraphics[scale=0.785]{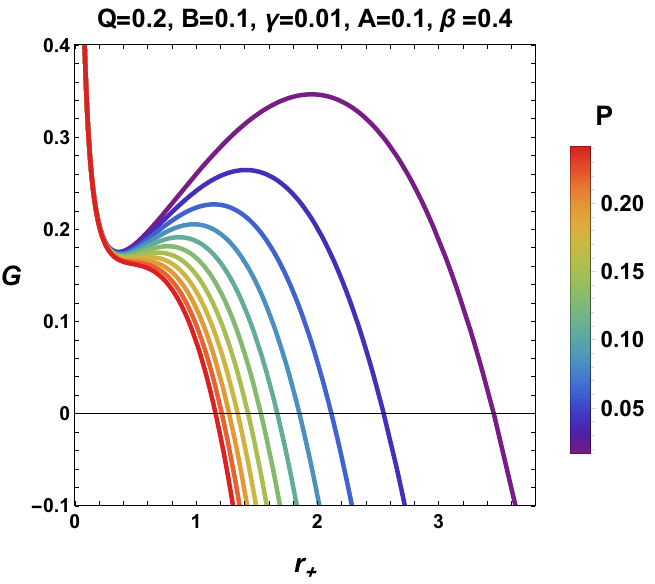} \hspace{2mm}
      }
\caption{The $G-r_+$ diagram of the black hole system for certain values of the parameter space.}
      	\label{fig8}
      \end{figure*}
     To provide an appropriate analysis of the behavior of the Gibbs free energy, either as a function of temperature or in terms of the horizon radius, Figs. $\ref{fig7}$ and $\ref{fig8}$  easily show this behavior. To start with, it is well-known to mention that Figs. $\ref{fig5}$ and $\ref{fig7}$ represent the same BH branches, where the same spectrum of values corresponds to the same pressures. Hence, the case $(P<P_c)$ generates a classic swallow-tail phenomenon on the $G$-$T$ diagram (Fig. $\ref{fig7})$, implying a first-order small/large BH phase transition. In other words, the non-smooth points on the isobaric curves related to the $G-T$ diagram are those of the extreme points on the isobaric curve in the $T-r$ diagram. In the monitoring phase of the considered scenario, the swallowtail gradually decreases in size, eventually disappearing as the pressure increases. Due to the increasing pressure, the extreme points of the $T$ on the isobaric curve move closer to each other to converge at the critical pressure $(P=P_c)$, and hence, any kind of first-order phase transition in the system disappears completely. For $P>P_c$, $G$ becomes a monotonic function of $T$, indicating that no phase transition occurs in the system. Some meaningful information from Gibbs Free Energy is predicting the local stable BH system. By the way, the sign of $G$ claims the system is locally stable when $(G<0)$ and unstable when $(G>0)$. For that reason,  Fig. $\ref{fig8}$ shows graphically the behavior of Gibbs free energy as a function of the horizon radius. Thus, it is observed for all the parameter space values that the BH system, in the overall view of the horizon radius, is locally stable and, in contrast, unstable for small values of the horizon radius. Alternatively, one can similarly inspect the same critical study from  Fig. $\ref{fig7}$ as previously mentioned.
\subsection{Critical exponents}
Critical exponents perfectly describe the behavior of physical quantities in the vicinity of the critical point. Indeed, critical exponents are independent of such physical systems and can be considered quasi-universal parameters. It is handy to introduce the following notations:
\begin{equation}
    t=\frac{T}{T_c}-1,\quad\omega=\frac{V}{V_c}-1,\quad p=\frac{P}{P_c},
\end{equation}
in which the critical thermodynamic volume $V_c$ is associated with the critical event horizon radius $r_c$ by $V_c=\frac{r_c^3}{6}$. The critical exponents are explicitly defined in the following way:
\begin{align}
    C_V&\propto\lvert t\rvert^{-\alpha}\\ \eta&\propto\lvert t\rvert^{\lambda}\\
    \kappa_T&\propto\lvert t\rvert^{-\gamma}\\ \lvert P-P_c\rvert&\propto\lvert V-V_c\rvert^\delta.\label{71}
    \end{align}

    The exponent $\alpha$ describes the behavior of specific heat at a constant volume. It is easy to conclude that the entropy $S$ is independent of the Hawking temperature $T$, thus
    \begin{equation}
        C_V= T\left(\frac{\partial S}{\partial T}\right)_V=0,
    \end{equation}
    hence it can be deduced that $\alpha = 0$.

     The exponent $\beta$ describes the behavior of the order parameter near the critical point. It is then possible to expand the equation of state near the critical point in 
     \begin{equation}
         p=\mathcal{C}+\mathcal{C}_1\, t+\mathcal{C}_2\, \omega+\mathcal{C}_3\,t\omega+\mathcal{C}_4\,\omega^2+\mathcal{C}_5\,\omega^3+O(t\omega^2,\omega^4),
     \end{equation}
     where
     \begin{widetext}
           \begin{align}
         \mathcal{C}& =\frac{1}{8 \pi  P_c \,r_c^4}\,\,\Bigg\{r_c^2 \bigg(r_c^2 \bigg(\frac{A+1}{B}\bigg)^{-\frac{1}{\beta +1}} \bigg(\frac{1}{B}\bigg(\frac{\gamma}{r_c^3}\bigg){}^{(A+1)
   (\beta +1)}+1\bigg)^{\frac{1}{\beta +1}}+4 \pi  r_c T_c-1\bigg)+Q^2\Bigg\},\\
   \mathcal{C}_1& =-\mathcal{C}_3=\frac{T_c}{2 P_c \,r_c},\\
   \mathcal{C}_2& =\mathcal{C}_4 =0,\\
   \mathcal{C}_5& = -\frac{1}{16 \pi  P_c \,r_c^4 \bigg(\left(\frac{\gamma}{r_c^3}\right)^{(A+1) (\beta +1)}+B\bigg)^3}\left(\frac{A+1}{B}\right)^{-\frac{1}{\beta +1}} \Bigg\{8 B^3 \left(\frac{A+1}{B}\right)^{\frac{1}{\beta +1}}
   \left(r_c^2 \left(\pi  r_c T_c-1\right)+5 Q^2\right)\nonumber\\
   &+B^2 \left(\frac{\gamma}{r_c^3}\right)^{(A+1) (\beta +1)} \Bigg(24
   \left(\frac{A+1}{B}\right)^{\frac{1}{\beta +1}} \left(r_c^2 \left(\pi  r_c T_c-1\right)+5 Q^2\right)+\mathcal{F} \bigg(\frac{1}{B}\left(\frac{\gamma}{r_c^3}\right)^{(A+1) (\beta
   +1)}+1\bigg)^{\frac{1}{\beta +1}}\Bigg)\nonumber\\
   &+B \left(\frac{\gamma}{r_c^3}\right)^{2 (A+1) (\beta +1)} \Bigg(24
   \left(\frac{A+1}{B}\right)^{\frac{1}{\beta +1}} \left(r_c^2 \left(\pi  r_c T_c-1\right)+5 Q^2\right)+\mathcal{G} \bigg(\frac{1}{B}\left(\frac{\gamma}{\text{rc}^3}\right)^{(A+1) (\beta
   +1)}+1\bigg)^{\frac{1}{\beta +1}}\Bigg)\nonumber\\
   &+\left(\frac{\gamma}{r_c^3}\right)^{3 (A+1) (\beta +1)} \Bigg(8
   \left(\frac{A+1}{B}\right)^{\frac{1}{\beta +1}} \left(r_c^2 \left(\pi  r_c T_c-1\right)+5 Q^2\right)+\mathcal{H}\, \bigg(\frac{1}{B}\left(\frac{\gamma}{r_c^3}\right)^{(A+1) (\beta +1)}+1\bigg)^{\frac{1}{\beta
   +1}}\Bigg)\Bigg\},
     \end{align}
     \end{widetext}
     where
         \begin{eqnarray}
         \mathcal{F}&=&(A+1)(3 A (\beta
   +1)\nonumber\\
   &+&3 \beta +4)(3 A (\beta +1)+3 \beta +5)\, r_c^4\nonumber\\
   \mathcal{G}&=&-(A+1) (3 A (\beta
   -2)+3 \beta -10) (3 A (\beta +1)\nonumber\\
   &+&3 \beta +4)\, r_c^4\nonumber\\
   \mathcal{H} &=& (A+1) (3 A+4) (3
   A+5) \,r_c^4\nonumber
     \end{eqnarray}
     Accordingly, in terms of numerical results, the dependencies of the $\mathcal{C}_i$ coefficients on the $Q$ and $\beta$ parameters are presented in Tab. $\ref{Tab3}$. 

     As the pressure remains constant during the phase transition, one can have 
     \begin{equation}
         \mathcal{C}+\mathcal{C}_1 \,t+\mathcal{C}_3 \,t\omega_l+\mathcal{C}_5\,\omega_l^3=\mathcal{C}+\mathcal{C}_1\, t+\mathcal{C}_3 \,t\omega_s+\mathcal{C}_5\,\omega_s^3.\label{78}
     \end{equation}
     where $\omega_s$ and $\omega_l$ are the reduced volumes of the small and large BHs, respectively.

     In addition, Maxwell's equal area law is simply given by the following formula
   \begin{equation}
       \int_{\omega_l}^{\omega_l}\omega\,\frac{\mathrm{d}p}{\mathrm{d}\omega}\,\mathrm{d}\omega=0,\label{79}
   \end{equation}
and considering the first derivative so that
\begin{equation}
    \frac{\mathrm{d}p}{\mathrm{d}\omega}=\mathcal{C}_3\, t+3\,\mathcal{C}_5\,\omega^2.\label{80}
\end{equation}
Exploiting Eqs. \eqref{79} and \eqref{80} yield the following finding:
\begin{equation}
    \mathcal{C}_3\, t(\omega_s^2-\omega_l^2)+ \frac{3}{2}\,\mathcal{C}_5(\omega_s^4-\omega_l^4)=0.
\end{equation}
from which, with Eq. \eqref{78}, it is possible to find an explicit link between $\omega_l$ and $\omega_s$ in the following form:
\begin{equation}
    \omega_l=-\omega_s=\sqrt{-\frac{\mathcal{C}_3}{\mathcal{C}_5}t}\label{82}
\end{equation}
where the argument under the square root function remains positive. A quick look at Eq. \eqref{82}  yields the desired results, namely
\begin{equation}
    \eta=V _l-V_s=V_c(\omega_l-\omega_s)=2 V_c\,\omega_l\propto\sqrt{-t}
\end{equation}
which provides $\lambda=1/2$.

The exponent $\gamma$ describes the critical behavior of the isothermal compressibility $\kappa_T$ given explicitly by
\begin{equation}
    \kappa_T=-\frac{1}{V}\frac{\partial V}{\partial P}\biggr\rvert_{V_c}=-\frac{1}{P_c}\frac{1}{\frac{\partial p}{\partial \omega}}\biggr\rvert_{\omega =0}\propto\, \frac{2r_c}{T_c}t^{-1}
\end{equation}
giving rise to $\gamma=1$. 

The exponent $\delta$ is in charge of describing the critical behavior of Eq. \eqref{71} on the critical isotherm $T=T_c$. So, the shape of the critical isotherm is defined at $t=0$ providing the following finding:
\begin{equation}
    \lvert P-P_c\rvert= P_c  \lvert p-1\rvert= P_c  \,\lvert \mathcal{C}_5\, \omega^3\rvert=\frac{P_c\,\lvert \mathcal{C}_5\rvert}{V_c^3} \lvert V-V_c\rvert^3
\end{equation}
which easily proves $\delta=3$.

From the aforementioned findings, it is clear that the four critical exponents are precisely the ones obtained previously for charged AdS BHs. This in fact demonstrates that the MCG does not alter the critical exponents, similar to the quintessential dark energy. Thus, the universality profile of VdW-like phase transitions and the values of critical exponents for AdS BHs have been ascertained.
\section{Thermal geometries}
\label{Geom}
Based on the arguments set out in the introduction, this section focuses on providing an appropriate analysis of the phase transition for our charged BH solution using GT tools. Attention is paid to examining the shape of the metric in regard to Ruppeiner, Weinhold, HPEM and Quevedo theoretic. So, the Weinhold shape can be given in mass representation as~\cite{Soroushfar:2016nbu}
\begin{equation}
    g_{jk}^W=\partial_j\partial_k\, M(S, Q, \ell).
\end{equation}
For a charged AdS BH, the line element appears as follows:
\begin{align}
\mathrm{d}s_W^2&=M_{SS}\,\mathrm{d}S^2+M_{\ell\ell}\,\mathrm{d}\ell^2+M_{QQ}\,\mathrm{d}Q^2
+2M_{S\ell}\,\mathrm{d}S\mathrm{d}\ell\nonumber\\
&+2M_{SQ}\, \mathrm{d}S\mathrm{d}Q+2M_{\ell Q}\,\mathrm{d}\ell\mathrm{d}Q
\end{align}
Or, in terms of mass matrix representation, the formulation is given explicitly as
\begin{equation}
    g^W= \begin{pmatrix}
M_{SS} & M_{S\ell} & M_{SQ} \\
M_{\ell S} & M_{\ell\ell} & 0 \\
M_{Q S} & 0 & M_{Q Q}
\end{pmatrix}
\end{equation}
Similarly, in Ruppeiner formalism one considers entropy as basic thermodynamic potential,
\begin{equation}
    g_{jk}^R=\partial_j\partial_k\, S.
\end{equation}
It is worth noting that the Ruppeiner metric is linked to the Weinhold metric through a conformal transformation, yielding the following defining expression ~\cite{Mrugala1984OnEO}:
\begin{equation}
    \mathrm{d}s_R^2=\frac{1}{T}\mathrm{d}s_W^2
\end{equation}
Or, in terms of mass matrix representation, one has 
\begin{equation}
    g^R=\frac{1}{T} \begin{pmatrix}
M_{SS} & M_{S\ell} & M_{SQ} \\
M_{\ell S} & M_{\ell\ell} & 0 \\
M_{Q S} & 0 & M_{Q Q}
\end{pmatrix}.
\end{equation}
It should be noted that the treatment will be carried out in terms of pressure, which is related to the derivative of the AdS length by the following expression :
\begin{eqnarray}
    \partial\ell=-\frac{3}{16\pi\ell}\frac{\partial P}{P^2}.
\end{eqnarray}
      
\begin{figure*}[ht!]
      	\centering{
       \includegraphics[scale=0.5]{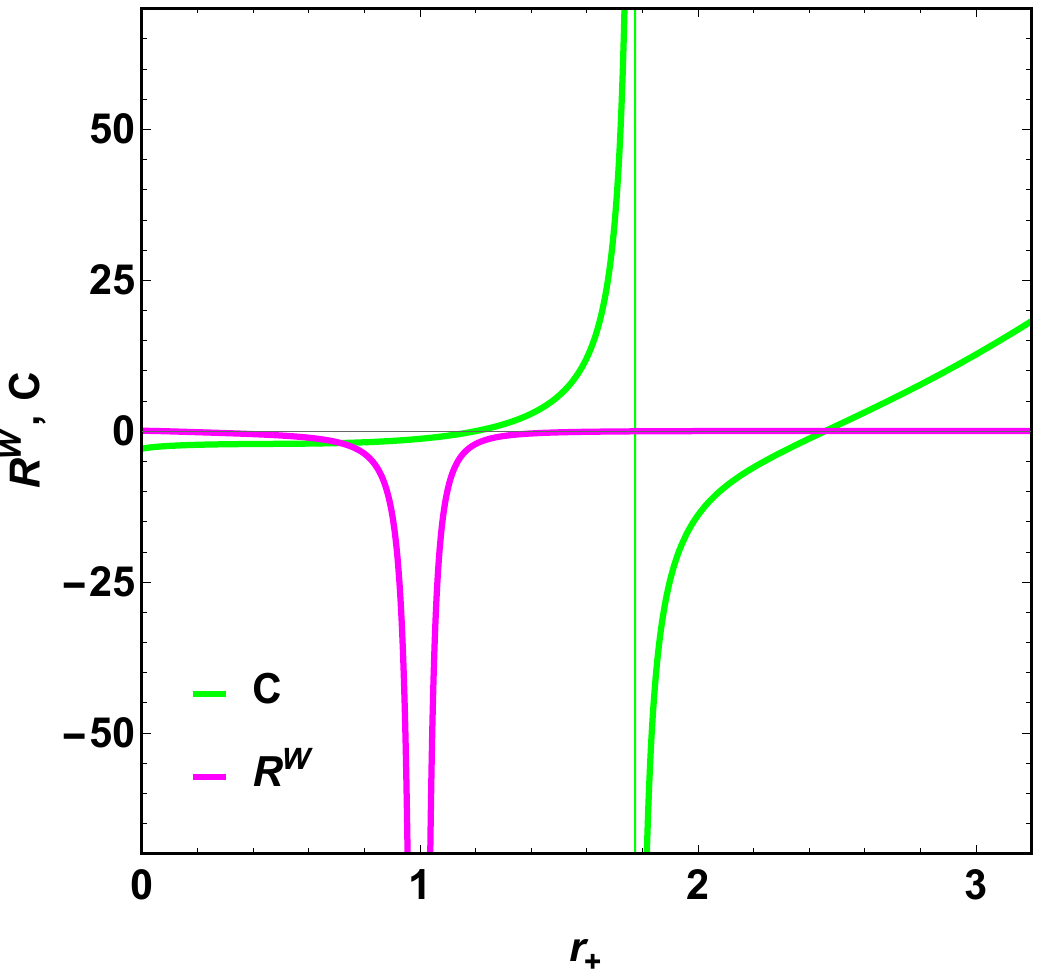}\hspace{2mm}
      	\includegraphics[scale=0.5]{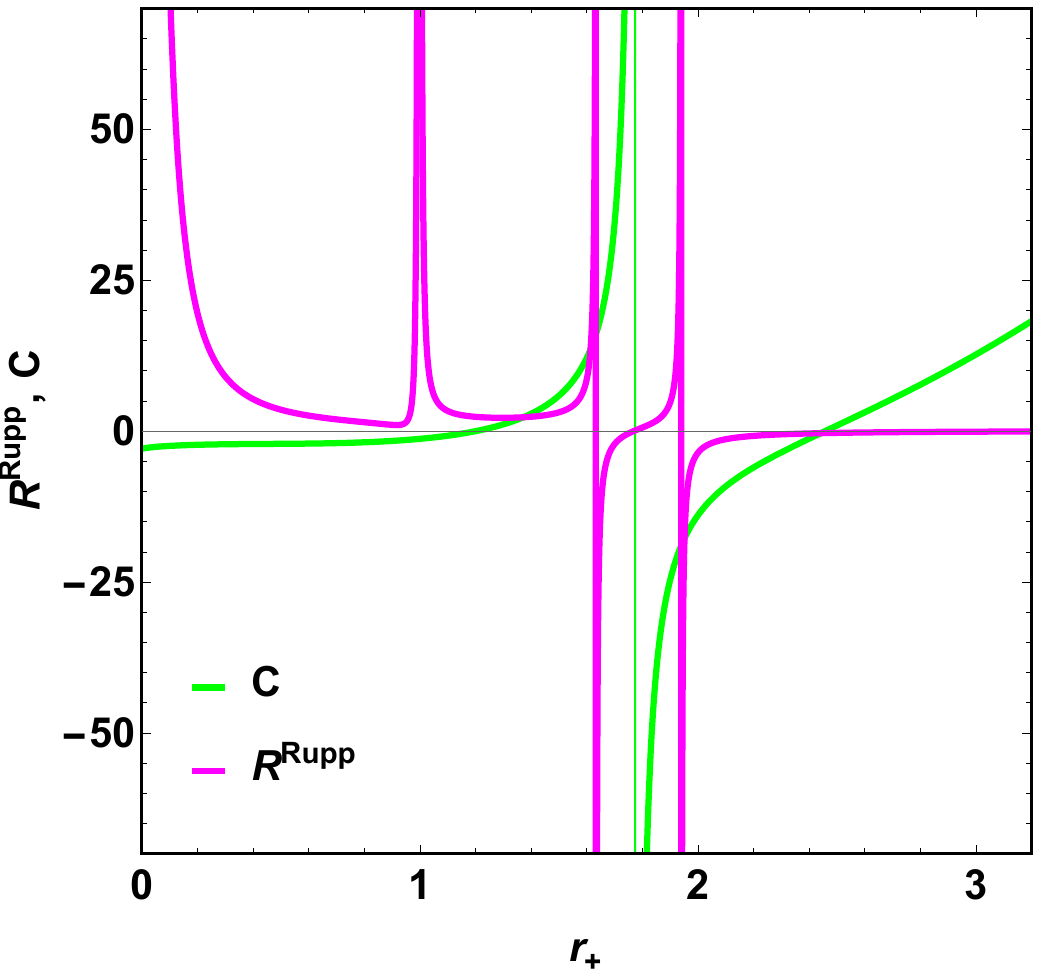} \hspace{2mm}
      	}
      	\caption{Curvature scalar variation of Weinhold and Ruppeiner metrics for $Q=0.8$, $B=1$, $\gamma=1$, $A=0.1$, $\beta=0.1$ and $P=0.0306\, (P<P_c)$}
      	\label{fig9}
      \end{figure*}

\begin{figure*}[ht!]
      	\centering{
       \includegraphics[scale=0.47]{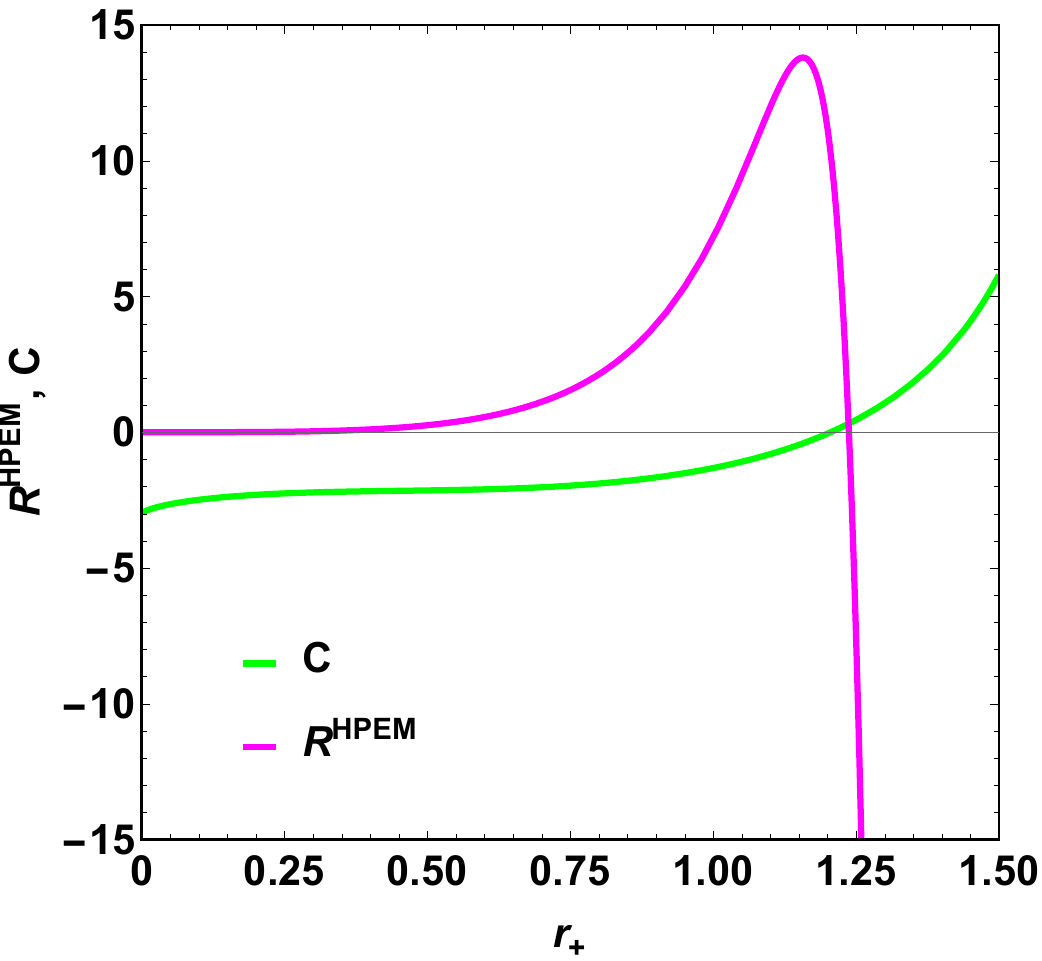} \hspace{2mm}
      	\includegraphics[scale=0.52]{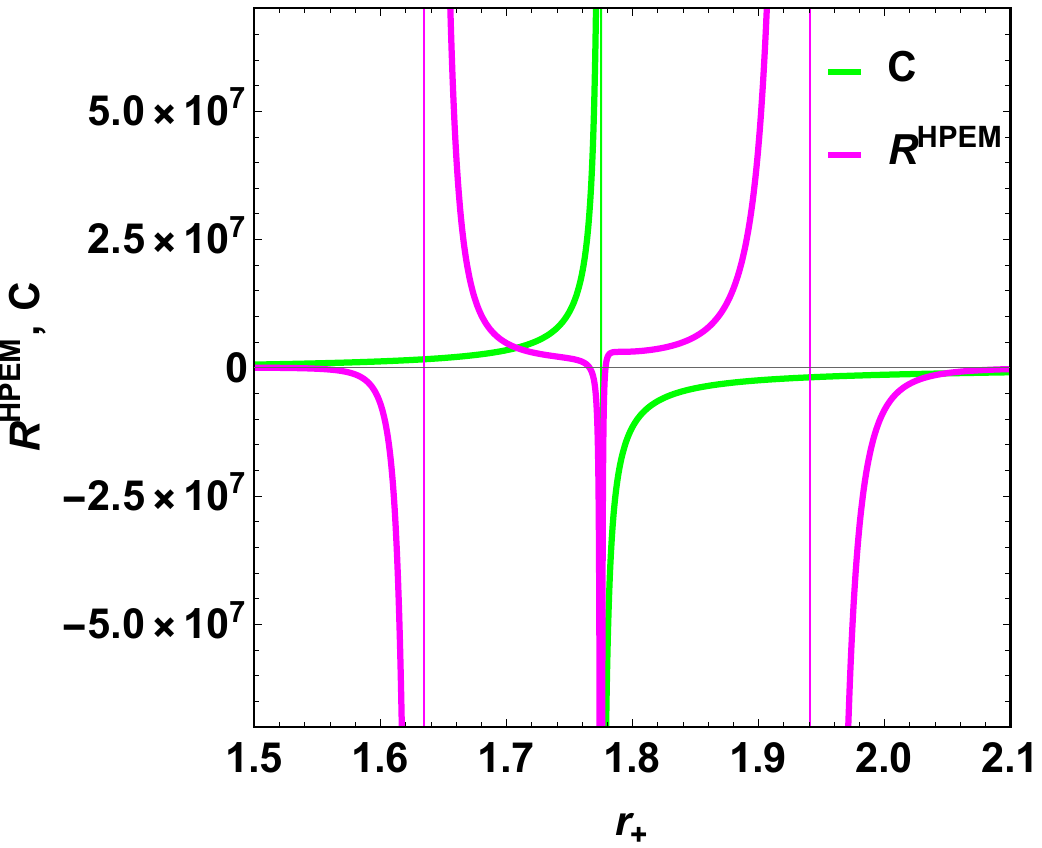} \hspace{2mm}
      	}
      	\centering{
       \includegraphics[scale=0.45]{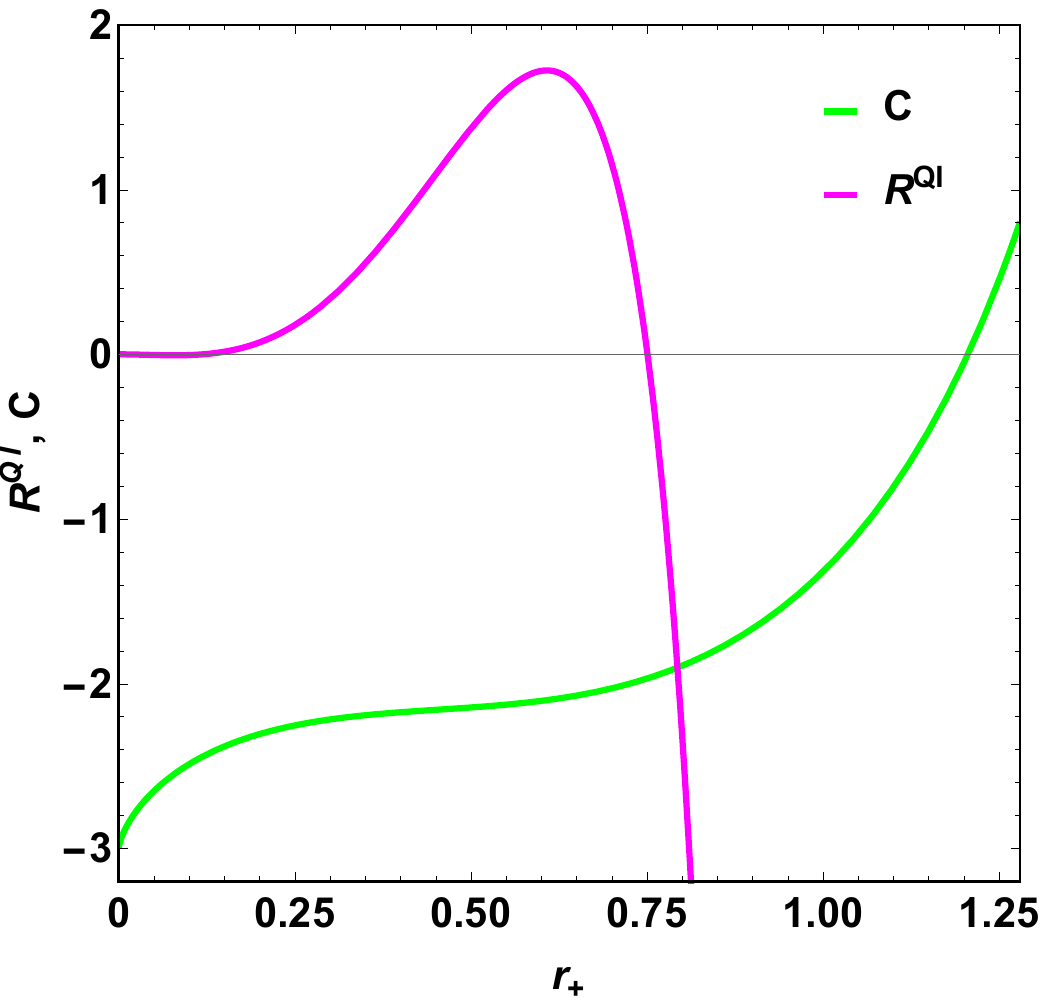} \hspace{2mm}
      	\includegraphics[scale=0.51]{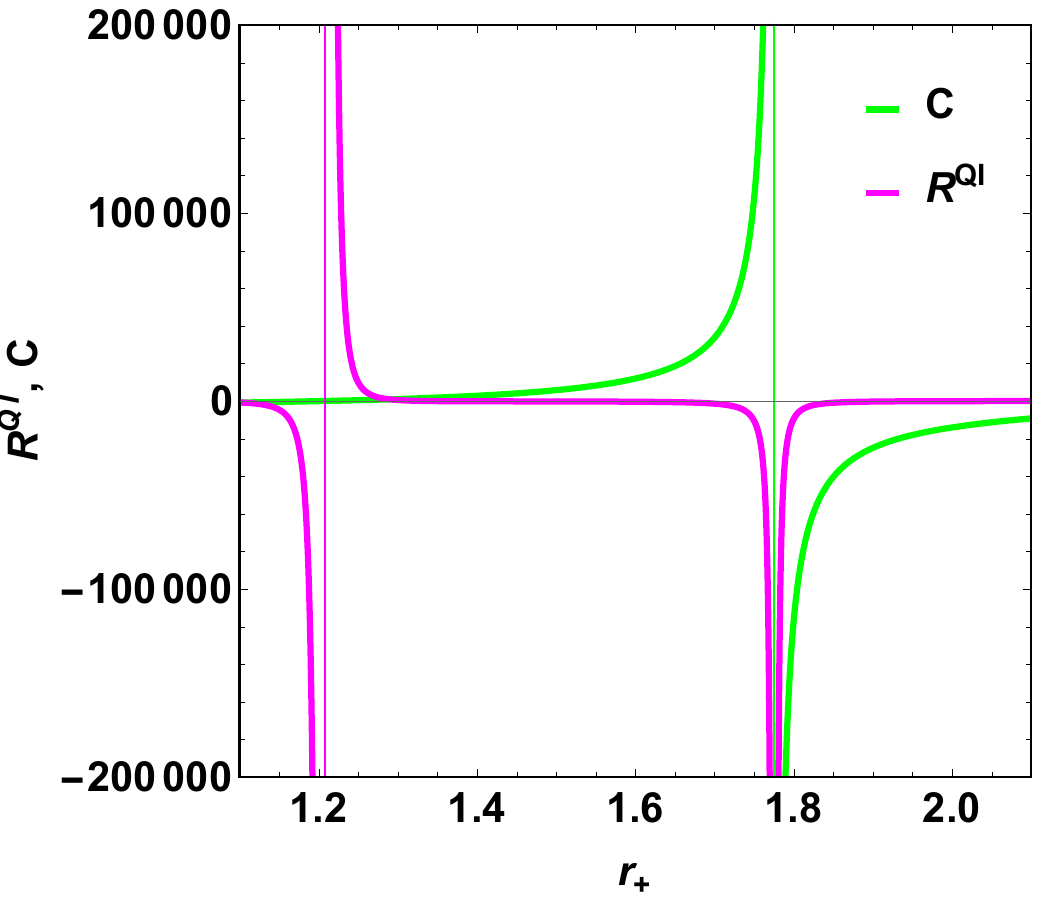} \hspace{2mm}
      }

      	\centering{
       \includegraphics[scale=0.52]{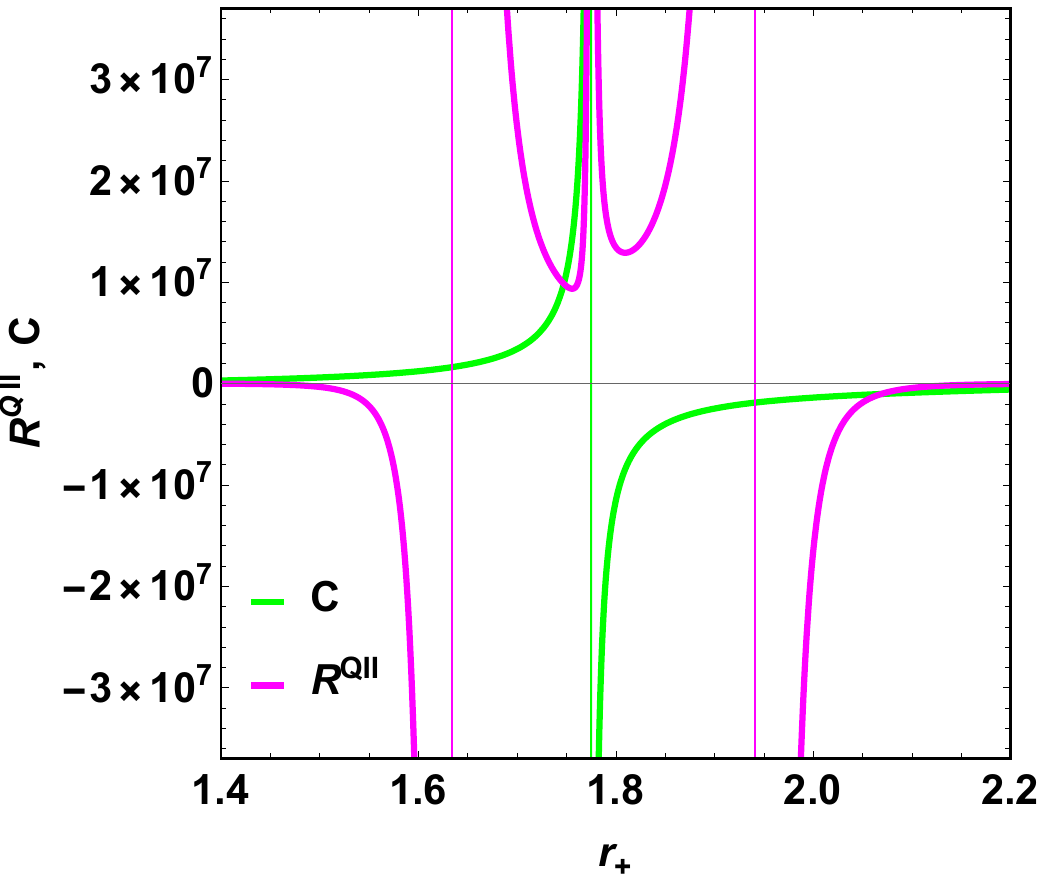} \hspace{2mm}
      	}
      	\caption{Curvature scalar variation of HPEM and Quevedo metrics for $Q=0.8$, $B=1$, $\gamma=1$, $A=0.1$, $\beta=0.1$ and $P=0.0306\, (P<P_c)$}
      	\label{fig10}
      \end{figure*}

      In order to properly examine an interesting discussion of the GT tools in the essence of our BH solution in extended phase space,  Figs. $\ref{fig9}$-$\ref{fig10}$ provide the necessary analysis in this regard. To begin with, the Weinhold and Ruppeiner structures provide an analysis concerning a certain gauge parameter space, which is illustrated in Fig. $\ref{fig9}$. It is observed that the zero of the heat capacity yields a divergent scalar curvature, as in the case where the Weinhold scalar curvature has a negative divergent point. This negative divergent point coincides with the physical limitation point of the heat capacity at $r_1^\star= 1.63453$. Moreover, the application of the Weinhold geometry is compatible with the realm interpretation of the GTs tools. However, the Ruppeiner scalar curvature has three divergent points, one of which is positive. With care, the positive divergence point is shown to coincide with a physical limitation point (the root of the heat capacity at 0) at $r_1^\text{div}= 1.94108$. Furthermore, another coincidence is demonstrated between the Ruppeiener scalar curvature and the phase transition critical point at  $r_1^\star= 1.63453$. Therefore, the Ruppeiener structure offers valuable information.
      
      To get a better approach to the BH phase transition analysis, it might be useful to apply geometrical thermodynamics, such as the HPEM and Quevedo tools. In particular, the Quevedo metric is expressed as follows
\begin{equation}
    g=\left( E^c\frac{\partial\Phi}{\partial E^c}\right)\left(\eta_{ab}\delta^{bc}\frac{\partial ^2\Phi}{\partial E^c\partial E^d}\mathrm{d}E^a\,\mathrm{d}E^d\right)
\end{equation}
where
\begin{equation}
    \frac{\partial \Phi}{\partial E^c}=\delta_{cb}\, I^b.
\end{equation}
    
Within this structure, $\Phi$, $I^b$, and $E^a$ represent the thermodynamic potential and intense and extended variables, respectively, On the other hand, the generalized HPEM metric labeled by n extended variables is given in such a way as ~\cite{Mrugala1984OnEO, Hendi:2015rja,Hendi:2015xya}
\begin{equation}
    \mathrm{d}S^2_{HPEM}=\frac{S\,M_S}{\left(\prod_{i=2}^n\frac{\partial^2M}{\partial \xi_i^2}\right)^3}\left(-M_{SS}\mathrm{d}S^2+\sum_{i=2}^n\left(\frac{\partial^2M}{\partial \xi_i^2}\right)\mathrm{d}\phi_i^2\right).
\end{equation}
Here, $\xi(\xi\neq S)$, $M_S=\frac{\partial M}{\partial S}$, and $M_{SS}=\frac{\partial^2 M}{\partial S^2}$ are extensive parameters, respectively. So, the background of metrics is easily expressed as follows~\cite{Mrugala1984OnEO, Hendi:2015rja,Hendi:2015xya}:
\begin{widetext}
    \begin{align}
    \mathrm{d}S^2_{HPEM}&=\frac{S\,M_S}{\left(\frac{\partial^2M}{\partial \ell^2}\frac{\partial^2M}{\partial Q^2}\right)^3}\left(-M_{SS}\mathrm{d}S^2+M_{\ell\ell}\mathrm{d}\ell^2+M_{QQ}\mathrm{d}Q^2\right)\\
    \mathrm{d}S^2_{QI}&=\left(S M_S+\ell M_{\ell}+Q M_{Q} \right)\left(-M_{SS}\mathrm{d}S^2+M_{\ell\ell}\mathrm{d}\ell^2+M_{QQ}\mathrm{d}Q^2\right)\\
    \mathrm{d}S^2_{QII}&=S M_S\left(-M_{SS}\mathrm{d}S^2+M_{\ell\ell}\mathrm{d}\ell^2+M_{QQ}\mathrm{d}Q^2\right)
\end{align}
\end{widetext}
By the way, Ricci scalars employed in making these measurements have the following denominator~\cite{Mrugala1984OnEO,Hendi:2015xya}
\begin{widetext}
    \begin{equation}
   \text{denom}(R) = \begin{cases}
     2M_{SS}^2 S^3 M_{S}^3\hspace{4.5cm} \text{HPEM} \\
      2M_{SS}^2M_{QQ}^2M_{\ell\ell}^2\left(S M_S+Q M_Q+ \ell M_{\ell}\right) \quad \text{Quevedo Classe I}\\
      2S^3 M_{SS}^2M_{QQ}^2M_{\ell\ell}^2 M_S^3 \hspace{3.05cm}\text{Quevedo Classe II}
    \end{cases}\,.
\end{equation}
\end{widetext}
      
To examine, at a graphical level, the behavior of the three applying GT tools, Fig. $\ref{fig10}$ presents the relevant results. With regard to the application of the HPEM geometry, the results seem very interesting. This fascinating observation is due to the presence of three divergent points of HPEM scalar curvature, one of which is negative. It is clear that the negative divergent point is perfectly aligned with the phase transition critical point. Furthermore, it is clearly demonstrated that the HEPM scalar curvature reaches zeros of the heat capacity at r. In turn, the application of the Quevedo class provides two divergent points, one of which is negative. Unlike the HPEM scalar curvature, the divergence of the scalar curvature is at the zero point of the heat capacity. Moreover, the negative divergent point coincides with the phase transition critical point. What remains to be discussed is the application of Quevedo Class II. It can be seen that the critical point of the phase transition coincides with the positive divergent point of the corresponding scalar curvature.
\section{Conclusion}
\label{Conc}
A number of cosmological challenges aim to reveal the unified dark fluid as an unknown energy component of the universal dark sector. What is more, the unified dark fluid is a hybrid of dark matter and dark energy, making it a candidate for the so-called Chaplygin gas. Spurred by this insight, we constructed an exact static, spherically symmetrically charged AdS BH solution with a surrounding MCG. Taking advantage of the MCG equation of state $p=A\rho -B/\rho^\beta$, we established its energy density with respect to the radial coordinate. To verify the singularity and uniqueness of the BH, we have performed an examination based on scalar invariants. This showed that the BH solution is singular for a given parameter space and absolutely unique. Furthermore, we have investigated the classical energy conditions for the MCG and concluded that it admits the null and weak energy conditions but violates the strong energy condition. Further, the BH solution is graphically represented according to the permitted values on the parameter space.  

As part of this work, we have inspected, within the extended phase space, different thermodynamic quantities, and the modified first law of thermodynamics, along with the Smarr relations, have been established. The frame of the extended phase space, together with the statement of BH chemistry, is helpful to consider the $P-V$ criticality. Therefore, we have managed the $P-V$ criticality for the charged AdS BH endowed with MCG structure and discovered a generic small/large BH phase transition, which is analogous to the liquid/gas phase transition of VdW. Furthermore, the critical exponents are shown to be similar to those of the VdW fluid. Furthermore, the heat capacity at constant pressure is said to show the thermal local stability of the BH solution. Indeed, we have shown, according to the sign of the heat capacity, that our BH solution is locally thermally stable.

Among ordinary predictions of the phase transition critical point, GT tools are alternatively used to predict certain correlations. In this way, we have applied in the extended phase space some tools from the GT background, such as Weinhlold, Ruppeiner, HPEM, and Quevedo classes I and II. The examination showed consistent results between the GT scalar curvature and either the physical limitation point (the root of the heat capacity at $0$) or the critical phase transition point of the heat capacity (divergent point). In essence, the microstructures are thereby defined precisely for all classes of GT tools and according to the curvature scalar sign.

This work raises certain questions and will be a subject for revealing thermodynamic topology or discovering aspects such as the Joule-Thomson expansion. On the other hand, quasi-normal modes and shadow behaviors, as well as the angle of deflection, are topics that could be addressed in the future.
\section*{Acknowledgments}
GGL acknowledges the Spanish “Ministerio de Universidades” for the awarded Maria Zambrano fellowship and funding received from the European Union NextGenerationEU. He also acknowledges participation in the COST Association Action CA18108 “Quantum Gravity Phenomenology in the Multimessenger Approach” and LISA Cosmology Working group. J.R. acknowledges the grant from the Ministry of Higher Education, Science, and Innovation of the Republic of Uzbekistan F-FA-2021-510.

\bibliographystyle{apsrev4-1}
\bibliography{Library}

\begin{thebibliography}{113}%
\makeatletter
\providecommand \@ifxundefined [1]{%
 \@ifx{#1\undefined}
}%
\providecommand \@ifnum [1]{%
 \ifnum #1\expandafter \@firstoftwo
 \else \expandafter \@secondoftwo
 \fi
}%
\providecommand \@ifx [1]{%
 \ifx #1\expandafter \@firstoftwo
 \else \expandafter \@secondoftwo
 \fi
}%
\providecommand \natexlab [1]{#1}%
\providecommand \enquote  [1]{``#1''}%
\providecommand \bibnamefont  [1]{#1}%
\providecommand \bibfnamefont [1]{#1}%
\providecommand \citenamefont [1]{#1}%
\providecommand \href@noop [0]{\@secondoftwo}%
\providecommand \href [0]{\begingroup \@sanitize@url \@href}%
\providecommand \@href[1]{\@@startlink{#1}\@@href}%
\providecommand \@@href[1]{\endgroup#1\@@endlink}%
\providecommand \@sanitize@url [0]{\catcode `\\12\catcode `\$12\catcode `\&12\catcode `\#12\catcode `\^12\catcode `\_12\catcode `\%12\relax}%
\providecommand \@@startlink[1]{}%
\providecommand \@@endlink[0]{}%
\providecommand \url  [0]{\begingroup\@sanitize@url \@url }%
\providecommand \@url [1]{\endgroup\@href {#1}{\urlprefix }}%
\providecommand \urlprefix  [0]{URL }%
\providecommand \Eprint [0]{\href }%
\providecommand \doibase [0]{http://dx.doi.org/}%
\providecommand \selectlanguage [0]{\@gobble}%
\providecommand \bibinfo  [0]{\@secondoftwo}%
\providecommand \bibfield  [0]{\@secondoftwo}%
\providecommand \translation [1]{[#1]}%
\providecommand \BibitemOpen [0]{}%
\providecommand \bibitemStop [0]{}%
\providecommand \bibitemNoStop [0]{.\EOS\space}%
\providecommand \EOS [0]{\spacefactor3000\relax}%
\providecommand \BibitemShut  [1]{\csname bibitem#1\endcsname}%
\let\auto@bib@innerbib\@empty
\bibitem [{\citenamefont {Einstein}(1915)}]{Einstein:1915ca}%
  \BibitemOpen
  \bibfield  {author} {\bibinfo {author} {\bibfnamefont {A.}~\bibnamefont {Einstein}},\ }\href@noop {} {\bibfield  {journal} {\bibinfo  {journal} {Sitzungsber. Preuss. Akad. Wiss. Berlin (Math. Phys. )}\ }\textbf {\bibinfo {volume} {1915}},\ \bibinfo {pages} {844} (\bibinfo {year} {1915})}\BibitemShut {NoStop}%
\bibitem [{\citenamefont {Abbott}\ \emph {et~al.}(2016)\citenamefont {Abbott} \emph {et~al.}}]{LIGOScientific:2016aoc}%
  \BibitemOpen
  \bibfield  {author} {\bibinfo {author} {\bibfnamefont {B.~P.}\ \bibnamefont {Abbott}} \emph {et~al.} (\bibinfo {collaboration} {LIGO Scientific, Virgo}),\ }\href {\doibase 10.1103/PhysRevLett.116.061102} {\bibfield  {journal} {\bibinfo  {journal} {Phys. Rev. Lett.}\ }\textbf {\bibinfo {volume} {116}},\ \bibinfo {pages} {061102} (\bibinfo {year} {2016})},\ \Eprint {http://arxiv.org/abs/1602.03837} {arXiv:1602.03837 [gr-qc]} \BibitemShut {NoStop}%
\bibitem [{\citenamefont {Akiyama}\ \emph {et~al.}(2019)\citenamefont {Akiyama} \emph {et~al.}}]{EventHorizonTelescope:2019dse}%
  \BibitemOpen
  \bibfield  {author} {\bibinfo {author} {\bibfnamefont {K.}~\bibnamefont {Akiyama}} \emph {et~al.} (\bibinfo {collaboration} {Event Horizon Telescope}),\ }\href {\doibase 10.3847/2041-8213/ab0ec7} {\bibfield  {journal} {\bibinfo  {journal} {Astrophys. J. Lett.}\ }\textbf {\bibinfo {volume} {875}},\ \bibinfo {pages} {L1} (\bibinfo {year} {2019})},\ \Eprint {http://arxiv.org/abs/1906.11238} {arXiv:1906.11238 [astro-ph.GA]} \BibitemShut {NoStop}%
\bibitem [{\citenamefont {Hawking}(1975)}]{Hawking:1975vcx}%
  \BibitemOpen
  \bibfield  {author} {\bibinfo {author} {\bibfnamefont {S.~W.}\ \bibnamefont {Hawking}},\ }\href {\doibase 10.1007/BF02345020} {\bibfield  {journal} {\bibinfo  {journal} {Commun. Math. Phys.}\ }\textbf {\bibinfo {volume} {43}},\ \bibinfo {pages} {199} (\bibinfo {year} {1975})},\ \bibinfo {note} {[Erratum: Commun.Math.Phys. 46, 206 (1976)]}\BibitemShut {NoStop}%
\bibitem [{\citenamefont {Bekenstein}(1973)}]{Bekenstein:1973ur}%
  \BibitemOpen
  \bibfield  {author} {\bibinfo {author} {\bibfnamefont {J.~D.}\ \bibnamefont {Bekenstein}},\ }\href {\doibase 10.1103/PhysRevD.7.2333} {\bibfield  {journal} {\bibinfo  {journal} {Phys. Rev. D}\ }\textbf {\bibinfo {volume} {7}},\ \bibinfo {pages} {2333} (\bibinfo {year} {1973})}\BibitemShut {NoStop}%
\bibitem [{\citenamefont {Bardeen}\ \emph {et~al.}(1973)\citenamefont {Bardeen}, \citenamefont {Carter},\ and\ \citenamefont {Hawking}}]{Bardeen:1973gs}%
  \BibitemOpen
  \bibfield  {author} {\bibinfo {author} {\bibfnamefont {J.~M.}\ \bibnamefont {Bardeen}}, \bibinfo {author} {\bibfnamefont {B.}~\bibnamefont {Carter}}, \ and\ \bibinfo {author} {\bibfnamefont {S.~W.}\ \bibnamefont {Hawking}},\ }\href {\doibase 10.1007/BF01645742} {\bibfield  {journal} {\bibinfo  {journal} {Commun. Math. Phys.}\ }\textbf {\bibinfo {volume} {31}},\ \bibinfo {pages} {161} (\bibinfo {year} {1973})}\BibitemShut {NoStop}%
\bibitem [{\citenamefont {Hawking}\ and\ \citenamefont {Page}(1983)}]{Hawking:1982dh}%
  \BibitemOpen
  \bibfield  {author} {\bibinfo {author} {\bibfnamefont {S.~W.}\ \bibnamefont {Hawking}}\ and\ \bibinfo {author} {\bibfnamefont {D.~N.}\ \bibnamefont {Page}},\ }\href {\doibase 10.1007/BF01208266} {\bibfield  {journal} {\bibinfo  {journal} {Commun. Math. Phys.}\ }\textbf {\bibinfo {volume} {87}},\ \bibinfo {pages} {577} (\bibinfo {year} {1983})}\BibitemShut {NoStop}%
\bibitem [{\citenamefont {Davies}(1977)}]{Davies:1977bgr}%
  \BibitemOpen
  \bibfield  {author} {\bibinfo {author} {\bibfnamefont {P.~C.~W.}\ \bibnamefont {Davies}},\ }\href {\doibase 10.1098/rspa.1977.0047} {\bibfield  {journal} {\bibinfo  {journal} {Proc. Roy. Soc. Lond. A}\ }\textbf {\bibinfo {volume} {353}},\ \bibinfo {pages} {499} (\bibinfo {year} {1977})}\BibitemShut {NoStop}%
\bibitem [{\citenamefont {Cai}\ and\ \citenamefont {Cho}(1999)}]{Cai:1998ep}%
  \BibitemOpen
  \bibfield  {author} {\bibinfo {author} {\bibfnamefont {R.-G.}\ \bibnamefont {Cai}}\ and\ \bibinfo {author} {\bibfnamefont {J.-H.}\ \bibnamefont {Cho}},\ }\href {\doibase 10.1103/PhysRevD.60.067502} {\bibfield  {journal} {\bibinfo  {journal} {Phys. Rev. D}\ }\textbf {\bibinfo {volume} {60}},\ \bibinfo {pages} {067502} (\bibinfo {year} {1999})},\ \Eprint {http://arxiv.org/abs/hep-th/9803261} {arXiv:hep-th/9803261} \BibitemShut {NoStop}%
\bibitem [{\citenamefont {Quevedo}(2008{\natexlab{a}})}]{Quevedo:2007mj}%
  \BibitemOpen
  \bibfield  {author} {\bibinfo {author} {\bibfnamefont {H.}~\bibnamefont {Quevedo}},\ }\href {\doibase 10.1007/s10714-007-0586-0} {\bibfield  {journal} {\bibinfo  {journal} {Gen. Rel. Grav.}\ }\textbf {\bibinfo {volume} {40}},\ \bibinfo {pages} {971} (\bibinfo {year} {2008}{\natexlab{a}})},\ \Eprint {http://arxiv.org/abs/0704.3102} {arXiv:0704.3102 [gr-qc]} \BibitemShut {NoStop}%
\bibitem [{\citenamefont {Quevedo}\ and\ \citenamefont {Sanchez}(2009)}]{Quevedo:2008ry}%
  \BibitemOpen
  \bibfield  {author} {\bibinfo {author} {\bibfnamefont {H.}~\bibnamefont {Quevedo}}\ and\ \bibinfo {author} {\bibfnamefont {A.}~\bibnamefont {Sanchez}},\ }\href {\doibase 10.1103/PhysRevD.79.024012} {\bibfield  {journal} {\bibinfo  {journal} {Phys. Rev. D}\ }\textbf {\bibinfo {volume} {79}},\ \bibinfo {pages} {024012} (\bibinfo {year} {2009})},\ \Eprint {http://arxiv.org/abs/0811.2524} {arXiv:0811.2524 [gr-qc]} \BibitemShut {NoStop}%
\bibitem [{\citenamefont {Sahay}\ \emph {et~al.}(2010)\citenamefont {Sahay}, \citenamefont {Sarkar},\ and\ \citenamefont {Sengupta}}]{Sahay:2010tx}%
  \BibitemOpen
  \bibfield  {author} {\bibinfo {author} {\bibfnamefont {A.}~\bibnamefont {Sahay}}, \bibinfo {author} {\bibfnamefont {T.}~\bibnamefont {Sarkar}}, \ and\ \bibinfo {author} {\bibfnamefont {G.}~\bibnamefont {Sengupta}},\ }\href {\doibase 10.1007/JHEP07(2010)082} {\bibfield  {journal} {\bibinfo  {journal} {JHEP}\ }\textbf {\bibinfo {volume} {07}},\ \bibinfo {pages} {082} (\bibinfo {year} {2010})},\ \Eprint {http://arxiv.org/abs/1004.1625} {arXiv:1004.1625 [hep-th]} \BibitemShut {NoStop}%
\bibitem [{\citenamefont {Wei}\ and\ \citenamefont {Liu}(2015)}]{Wei:2015iwa}%
  \BibitemOpen
  \bibfield  {author} {\bibinfo {author} {\bibfnamefont {S.-W.}\ \bibnamefont {Wei}}\ and\ \bibinfo {author} {\bibfnamefont {Y.-X.}\ \bibnamefont {Liu}},\ }\href {\doibase 10.1103/PhysRevLett.115.111302} {\bibfield  {journal} {\bibinfo  {journal} {Phys. Rev. Lett.}\ }\textbf {\bibinfo {volume} {115}},\ \bibinfo {pages} {111302} (\bibinfo {year} {2015})},\ \bibinfo {note} {[Erratum: Phys.Rev.Lett. 116, 169903 (2016)]},\ \Eprint {http://arxiv.org/abs/1502.00386} {arXiv:1502.00386 [gr-qc]} \BibitemShut {NoStop}%
\bibitem [{\citenamefont {Dehyadegari}\ \emph {et~al.}(2017)\citenamefont {Dehyadegari}, \citenamefont {Sheykhi},\ and\ \citenamefont {Montakhab}}]{Dehyadegari:2016nkd}%
  \BibitemOpen
  \bibfield  {author} {\bibinfo {author} {\bibfnamefont {A.}~\bibnamefont {Dehyadegari}}, \bibinfo {author} {\bibfnamefont {A.}~\bibnamefont {Sheykhi}}, \ and\ \bibinfo {author} {\bibfnamefont {A.}~\bibnamefont {Montakhab}},\ }\href {\doibase 10.1016/j.physletb.2017.02.064} {\bibfield  {journal} {\bibinfo  {journal} {Phys. Lett. B}\ }\textbf {\bibinfo {volume} {768}},\ \bibinfo {pages} {235} (\bibinfo {year} {2017})},\ \Eprint {http://arxiv.org/abs/1607.05333} {arXiv:1607.05333 [gr-qc]} \BibitemShut {NoStop}%
\bibitem [{\citenamefont {Kord~Zangeneh}\ \emph {et~al.}(2018)\citenamefont {Kord~Zangeneh}, \citenamefont {Dehyadegari}, \citenamefont {Sheykhi},\ and\ \citenamefont {Mann}}]{KordZangeneh:2017lgs}%
  \BibitemOpen
  \bibfield  {author} {\bibinfo {author} {\bibfnamefont {M.}~\bibnamefont {Kord~Zangeneh}}, \bibinfo {author} {\bibfnamefont {A.}~\bibnamefont {Dehyadegari}}, \bibinfo {author} {\bibfnamefont {A.}~\bibnamefont {Sheykhi}}, \ and\ \bibinfo {author} {\bibfnamefont {R.~B.}\ \bibnamefont {Mann}},\ }\href {\doibase 10.1103/PhysRevD.97.084054} {\bibfield  {journal} {\bibinfo  {journal} {Phys. Rev. D}\ }\textbf {\bibinfo {volume} {97}},\ \bibinfo {pages} {084054} (\bibinfo {year} {2018})},\ \Eprint {http://arxiv.org/abs/1709.04432} {arXiv:1709.04432 [hep-th]} \BibitemShut {NoStop}%
\bibitem [{\citenamefont {Wei}\ \emph {et~al.}(2019{\natexlab{a}})\citenamefont {Wei}, \citenamefont {Liu},\ and\ \citenamefont {Mann}}]{Wei:2019yvs}%
  \BibitemOpen
  \bibfield  {author} {\bibinfo {author} {\bibfnamefont {S.-W.}\ \bibnamefont {Wei}}, \bibinfo {author} {\bibfnamefont {Y.-X.}\ \bibnamefont {Liu}}, \ and\ \bibinfo {author} {\bibfnamefont {R.~B.}\ \bibnamefont {Mann}},\ }\href {\doibase 10.1103/PhysRevD.100.124033} {\bibfield  {journal} {\bibinfo  {journal} {Phys. Rev. D}\ }\textbf {\bibinfo {volume} {100}},\ \bibinfo {pages} {124033} (\bibinfo {year} {2019}{\natexlab{a}})},\ \Eprint {http://arxiv.org/abs/1909.03887} {arXiv:1909.03887 [gr-qc]} \BibitemShut {NoStop}%
\bibitem [{\citenamefont {Wei}\ \emph {et~al.}(2019{\natexlab{b}})\citenamefont {Wei}, \citenamefont {Liu},\ and\ \citenamefont {Mann}}]{Wei:2019uqg}%
  \BibitemOpen
  \bibfield  {author} {\bibinfo {author} {\bibfnamefont {S.-W.}\ \bibnamefont {Wei}}, \bibinfo {author} {\bibfnamefont {Y.-X.}\ \bibnamefont {Liu}}, \ and\ \bibinfo {author} {\bibfnamefont {R.~B.}\ \bibnamefont {Mann}},\ }\href {\doibase 10.1103/PhysRevLett.123.071103} {\bibfield  {journal} {\bibinfo  {journal} {Phys. Rev. Lett.}\ }\textbf {\bibinfo {volume} {123}},\ \bibinfo {pages} {071103} (\bibinfo {year} {2019}{\natexlab{b}})},\ \Eprint {http://arxiv.org/abs/1906.10840} {arXiv:1906.10840 [gr-qc]} \BibitemShut {NoStop}%
\bibitem [{\citenamefont {Xu}\ \emph {et~al.}(2020)\citenamefont {Xu}, \citenamefont {Wu},\ and\ \citenamefont {Yang}}]{Xu:2020gud}%
  \BibitemOpen
  \bibfield  {author} {\bibinfo {author} {\bibfnamefont {Z.-M.}\ \bibnamefont {Xu}}, \bibinfo {author} {\bibfnamefont {B.}~\bibnamefont {Wu}}, \ and\ \bibinfo {author} {\bibfnamefont {W.-L.}\ \bibnamefont {Yang}},\ }\href {\doibase 10.1103/PhysRevD.101.024018} {\bibfield  {journal} {\bibinfo  {journal} {Phys. Rev. D}\ }\textbf {\bibinfo {volume} {101}},\ \bibinfo {pages} {024018} (\bibinfo {year} {2020})},\ \Eprint {http://arxiv.org/abs/1910.12182} {arXiv:1910.12182 [gr-qc]} \BibitemShut {NoStop}%
\bibitem [{\citenamefont {Ghosh}\ and\ \citenamefont {Bhamidipati}(2020{\natexlab{a}})}]{Ghosh:2019pwy}%
  \BibitemOpen
  \bibfield  {author} {\bibinfo {author} {\bibfnamefont {A.}~\bibnamefont {Ghosh}}\ and\ \bibinfo {author} {\bibfnamefont {C.}~\bibnamefont {Bhamidipati}},\ }\href {\doibase 10.1103/PhysRevD.101.046005} {\bibfield  {journal} {\bibinfo  {journal} {Phys. Rev. D}\ }\textbf {\bibinfo {volume} {101}},\ \bibinfo {pages} {046005} (\bibinfo {year} {2020}{\natexlab{a}})},\ \Eprint {http://arxiv.org/abs/1911.06280} {arXiv:1911.06280 [gr-qc]} \BibitemShut {NoStop}%
\bibitem [{\citenamefont {Ghosh}\ and\ \citenamefont {Bhamidipati}(2020{\natexlab{b}})}]{Ghosh:2020kba}%
  \BibitemOpen
  \bibfield  {author} {\bibinfo {author} {\bibfnamefont {A.}~\bibnamefont {Ghosh}}\ and\ \bibinfo {author} {\bibfnamefont {C.}~\bibnamefont {Bhamidipati}},\ }\href {\doibase 10.1103/PhysRevD.101.106007} {\bibfield  {journal} {\bibinfo  {journal} {Phys. Rev. D}\ }\textbf {\bibinfo {volume} {101}},\ \bibinfo {pages} {106007} (\bibinfo {year} {2020}{\natexlab{b}})},\ \Eprint {http://arxiv.org/abs/2001.10510} {arXiv:2001.10510 [hep-th]} \BibitemShut {NoStop}%
\bibitem [{\citenamefont {Rani}\ \emph {et~al.}(2022)\citenamefont {Rani}, \citenamefont {Jawad}, \citenamefont {Moradpour},\ and\ \citenamefont {Tanveer}}]{Rani:2022xza}%
  \BibitemOpen
  \bibfield  {author} {\bibinfo {author} {\bibfnamefont {S.}~\bibnamefont {Rani}}, \bibinfo {author} {\bibfnamefont {A.}~\bibnamefont {Jawad}}, \bibinfo {author} {\bibfnamefont {H.}~\bibnamefont {Moradpour}}, \ and\ \bibinfo {author} {\bibfnamefont {A.}~\bibnamefont {Tanveer}},\ }\href {\doibase 10.1140/epjc/s10052-022-10655-9} {\bibfield  {journal} {\bibinfo  {journal} {Eur. Phys. J. C}\ }\textbf {\bibinfo {volume} {82}},\ \bibinfo {pages} {713} (\bibinfo {year} {2022})}\BibitemShut {NoStop}%
\bibitem [{\citenamefont {Jawad}\ and\ \citenamefont {Fatima}(2022)}]{Jawad:2022lww}%
  \BibitemOpen
  \bibfield  {author} {\bibinfo {author} {\bibfnamefont {A.}~\bibnamefont {Jawad}}\ and\ \bibinfo {author} {\bibfnamefont {S.~R.}\ \bibnamefont {Fatima}},\ }\href {\doibase 10.1016/j.nuclphysb.2022.115697} {\bibfield  {journal} {\bibinfo  {journal} {Nucl. Phys. B}\ }\textbf {\bibinfo {volume} {976}},\ \bibinfo {pages} {115697} (\bibinfo {year} {2022})}\BibitemShut {NoStop}%
\bibitem [{\citenamefont {Luciano}\ and\ \citenamefont {Sheykhi}(2023)}]{Luciano:2023fyr}%
  \BibitemOpen
  \bibfield  {author} {\bibinfo {author} {\bibfnamefont {G.~G.}\ \bibnamefont {Luciano}}\ and\ \bibinfo {author} {\bibfnamefont {A.}~\bibnamefont {Sheykhi}},\ }\href {\doibase 10.1016/j.dark.2023.101319} {\bibfield  {journal} {\bibinfo  {journal} {Phys. Dark Univ.}\ }\textbf {\bibinfo {volume} {42}},\ \bibinfo {pages} {101319} (\bibinfo {year} {2023})},\ \Eprint {http://arxiv.org/abs/2304.11006} {arXiv:2304.11006 [hep-th]} \BibitemShut {NoStop}%
\bibitem [{\citenamefont {Luciano}\ and\ \citenamefont {Saridakis}()}]{Luciano:2023bai}%
  \BibitemOpen
  \bibfield  {author} {\bibinfo {author} {\bibfnamefont {G.~G.}\ \bibnamefont {Luciano}}\ and\ \bibinfo {author} {\bibfnamefont {E.}~\bibnamefont {Saridakis}},\ }\href@noop {} {\bibinfo  {journal} {arXiv:2308.12669 [gr-qc]}\ }\BibitemShut {NoStop}%
\bibitem [{\citenamefont {Luciano}(2023)}]{Luciano:2023wtx}%
  \BibitemOpen
\bibfield  {journal} {  }\bibfield  {author} {\bibinfo {author} {\bibfnamefont {G.~G.}\ \bibnamefont {Luciano}},\ }\href {\doibase 10.1016/j.dark.2023.101237} {\bibfield  {journal} {\bibinfo  {journal} {Phys. Dark Univ.}\ }\textbf {\bibinfo {volume} {41}},\ \bibinfo {pages} {101237} (\bibinfo {year} {2023})},\ \Eprint {http://arxiv.org/abs/2301.12488} {arXiv:2301.12488 [gr-qc]} \BibitemShut {NoStop}%
\bibitem [{\citenamefont {Luciano}\ and\ \citenamefont {Gin\'e}(2023)}]{Luciano:2022hhy}%
  \BibitemOpen
  \bibfield  {author} {\bibinfo {author} {\bibfnamefont {G.~G.}\ \bibnamefont {Luciano}}\ and\ \bibinfo {author} {\bibfnamefont {J.}~\bibnamefont {Gin\'e}},\ }\href {\doibase 10.1016/j.dark.2023.101256} {\bibfield  {journal} {\bibinfo  {journal} {Phys. Dark Univ.}\ }\textbf {\bibinfo {volume} {41}},\ \bibinfo {pages} {101256} (\bibinfo {year} {2023})},\ \Eprint {http://arxiv.org/abs/2210.09755} {arXiv:2210.09755 [gr-qc]} \BibitemShut {NoStop}%
\bibitem [{\citenamefont {Cvetic}\ and\ \citenamefont {Gubser}(1999{\natexlab{a}})}]{Cvetic:1999ne}%
  \BibitemOpen
  \bibfield  {author} {\bibinfo {author} {\bibfnamefont {M.}~\bibnamefont {Cvetic}}\ and\ \bibinfo {author} {\bibfnamefont {S.~S.}\ \bibnamefont {Gubser}},\ }\href {\doibase 10.1088/1126-6708/1999/04/024} {\bibfield  {journal} {\bibinfo  {journal} {JHEP}\ }\textbf {\bibinfo {volume} {04}},\ \bibinfo {pages} {024} (\bibinfo {year} {1999}{\natexlab{a}})},\ \Eprint {http://arxiv.org/abs/hep-th/9902195} {arXiv:hep-th/9902195} \BibitemShut {NoStop}%
\bibitem [{\citenamefont {Cvetic}\ and\ \citenamefont {Gubser}(1999{\natexlab{b}})}]{Cvetic:1999rb}%
  \BibitemOpen
  \bibfield  {author} {\bibinfo {author} {\bibfnamefont {M.}~\bibnamefont {Cvetic}}\ and\ \bibinfo {author} {\bibfnamefont {S.~S.}\ \bibnamefont {Gubser}},\ }\href {\doibase 10.1088/1126-6708/1999/07/010} {\bibfield  {journal} {\bibinfo  {journal} {JHEP}\ }\textbf {\bibinfo {volume} {07}},\ \bibinfo {pages} {010} (\bibinfo {year} {1999}{\natexlab{b}})},\ \Eprint {http://arxiv.org/abs/hep-th/9903132} {arXiv:hep-th/9903132} \BibitemShut {NoStop}%
\bibitem [{\citenamefont {Chamblin}\ \emph {et~al.}(1999{\natexlab{a}})\citenamefont {Chamblin}, \citenamefont {Emparan}, \citenamefont {Johnson},\ and\ \citenamefont {Myers}}]{Chamblin:1999hg}%
  \BibitemOpen
  \bibfield  {author} {\bibinfo {author} {\bibfnamefont {A.}~\bibnamefont {Chamblin}}, \bibinfo {author} {\bibfnamefont {R.}~\bibnamefont {Emparan}}, \bibinfo {author} {\bibfnamefont {C.~V.}\ \bibnamefont {Johnson}}, \ and\ \bibinfo {author} {\bibfnamefont {R.~C.}\ \bibnamefont {Myers}},\ }\href {\doibase 10.1103/PhysRevD.60.104026} {\bibfield  {journal} {\bibinfo  {journal} {Phys. Rev. D}\ }\textbf {\bibinfo {volume} {60}},\ \bibinfo {pages} {104026} (\bibinfo {year} {1999}{\natexlab{a}})},\ \Eprint {http://arxiv.org/abs/hep-th/9904197} {arXiv:hep-th/9904197} \BibitemShut {NoStop}%
\bibitem [{\citenamefont {Chamblin}\ \emph {et~al.}(1999{\natexlab{b}})\citenamefont {Chamblin}, \citenamefont {Emparan}, \citenamefont {Johnson},\ and\ \citenamefont {Myers}}]{Chamblin:1999tk}%
  \BibitemOpen
  \bibfield  {author} {\bibinfo {author} {\bibfnamefont {A.}~\bibnamefont {Chamblin}}, \bibinfo {author} {\bibfnamefont {R.}~\bibnamefont {Emparan}}, \bibinfo {author} {\bibfnamefont {C.~V.}\ \bibnamefont {Johnson}}, \ and\ \bibinfo {author} {\bibfnamefont {R.~C.}\ \bibnamefont {Myers}},\ }\href {\doibase 10.1103/PhysRevD.60.064018} {\bibfield  {journal} {\bibinfo  {journal} {Phys. Rev. D}\ }\textbf {\bibinfo {volume} {60}},\ \bibinfo {pages} {064018} (\bibinfo {year} {1999}{\natexlab{b}})},\ \Eprint {http://arxiv.org/abs/hep-th/9902170} {arXiv:hep-th/9902170} \BibitemShut {NoStop}%
\bibitem [{\citenamefont {Caldarelli}\ \emph {et~al.}(2000)\citenamefont {Caldarelli}, \citenamefont {Cognola},\ and\ \citenamefont {Klemm}}]{Caldarelli:1999xj}%
  \BibitemOpen
  \bibfield  {author} {\bibinfo {author} {\bibfnamefont {M.~M.}\ \bibnamefont {Caldarelli}}, \bibinfo {author} {\bibfnamefont {G.}~\bibnamefont {Cognola}}, \ and\ \bibinfo {author} {\bibfnamefont {D.}~\bibnamefont {Klemm}},\ }\href {\doibase 10.1088/0264-9381/17/2/310} {\bibfield  {journal} {\bibinfo  {journal} {Class. Quant. Grav.}\ }\textbf {\bibinfo {volume} {17}},\ \bibinfo {pages} {399} (\bibinfo {year} {2000})},\ \Eprint {http://arxiv.org/abs/hep-th/9908022} {arXiv:hep-th/9908022} \BibitemShut {NoStop}%
\bibitem [{\citenamefont {Kastor}\ \emph {et~al.}(2009)\citenamefont {Kastor}, \citenamefont {Ray},\ and\ \citenamefont {Traschen}}]{Kastor:2009wy}%
  \BibitemOpen
  \bibfield  {author} {\bibinfo {author} {\bibfnamefont {D.}~\bibnamefont {Kastor}}, \bibinfo {author} {\bibfnamefont {S.}~\bibnamefont {Ray}}, \ and\ \bibinfo {author} {\bibfnamefont {J.}~\bibnamefont {Traschen}},\ }\href {\doibase 10.1088/0264-9381/26/19/195011} {\bibfield  {journal} {\bibinfo  {journal} {Class. Quant. Grav.}\ }\textbf {\bibinfo {volume} {26}},\ \bibinfo {pages} {195011} (\bibinfo {year} {2009})},\ \Eprint {http://arxiv.org/abs/0904.2765} {arXiv:0904.2765 [hep-th]} \BibitemShut {NoStop}%
\bibitem [{\citenamefont {Dolan}(2011{\natexlab{a}})}]{Dolan:2010ha}%
  \BibitemOpen
  \bibfield  {author} {\bibinfo {author} {\bibfnamefont {B.~P.}\ \bibnamefont {Dolan}},\ }\href {\doibase 10.1088/0264-9381/28/12/125020} {\bibfield  {journal} {\bibinfo  {journal} {Class. Quant. Grav.}\ }\textbf {\bibinfo {volume} {28}},\ \bibinfo {pages} {125020} (\bibinfo {year} {2011}{\natexlab{a}})},\ \Eprint {http://arxiv.org/abs/1008.5023} {arXiv:1008.5023 [gr-qc]} \BibitemShut {NoStop}%
\bibitem [{\citenamefont {Dolan}(2011{\natexlab{b}})}]{Dolan:2011xt}%
  \BibitemOpen
  \bibfield  {author} {\bibinfo {author} {\bibfnamefont {B.~P.}\ \bibnamefont {Dolan}},\ }\href {\doibase 10.1088/0264-9381/28/23/235017} {\bibfield  {journal} {\bibinfo  {journal} {Class. Quant. Grav.}\ }\textbf {\bibinfo {volume} {28}},\ \bibinfo {pages} {235017} (\bibinfo {year} {2011}{\natexlab{b}})},\ \Eprint {http://arxiv.org/abs/1106.6260} {arXiv:1106.6260 [gr-qc]} \BibitemShut {NoStop}%
\bibitem [{\citenamefont {Dolan}(2011{\natexlab{c}})}]{Dolan:2011jm}%
  \BibitemOpen
  \bibfield  {author} {\bibinfo {author} {\bibfnamefont {B.~P.}\ \bibnamefont {Dolan}},\ }\href {\doibase 10.1103/PhysRevD.84.127503} {\bibfield  {journal} {\bibinfo  {journal} {Phys. Rev. D}\ }\textbf {\bibinfo {volume} {84}},\ \bibinfo {pages} {127503} (\bibinfo {year} {2011}{\natexlab{c}})},\ \Eprint {http://arxiv.org/abs/1109.0198} {arXiv:1109.0198 [gr-qc]} \BibitemShut {NoStop}%
\bibitem [{\citenamefont {Cvetic}\ \emph {et~al.}(2011)\citenamefont {Cvetic}, \citenamefont {Gibbons}, \citenamefont {Kubiznak},\ and\ \citenamefont {Pope}}]{Cvetic:2010jb}%
  \BibitemOpen
  \bibfield  {author} {\bibinfo {author} {\bibfnamefont {M.}~\bibnamefont {Cvetic}}, \bibinfo {author} {\bibfnamefont {G.~W.}\ \bibnamefont {Gibbons}}, \bibinfo {author} {\bibfnamefont {D.}~\bibnamefont {Kubiznak}}, \ and\ \bibinfo {author} {\bibfnamefont {C.~N.}\ \bibnamefont {Pope}},\ }\href {\doibase 10.1103/PhysRevD.84.024037} {\bibfield  {journal} {\bibinfo  {journal} {Phys. Rev. D}\ }\textbf {\bibinfo {volume} {84}},\ \bibinfo {pages} {024037} (\bibinfo {year} {2011})},\ \Eprint {http://arxiv.org/abs/1012.2888} {arXiv:1012.2888 [hep-th]} \BibitemShut {NoStop}%
\bibitem [{\citenamefont {Lu}\ \emph {et~al.}(2012)\citenamefont {Lu}, \citenamefont {Pang}, \citenamefont {Pope},\ and\ \citenamefont {Vazquez-Poritz}}]{Lu:2012xu}%
  \BibitemOpen
  \bibfield  {author} {\bibinfo {author} {\bibfnamefont {H.}~\bibnamefont {Lu}}, \bibinfo {author} {\bibfnamefont {Y.}~\bibnamefont {Pang}}, \bibinfo {author} {\bibfnamefont {C.~N.}\ \bibnamefont {Pope}}, \ and\ \bibinfo {author} {\bibfnamefont {J.~F.}\ \bibnamefont {Vazquez-Poritz}},\ }\href {\doibase 10.1103/PhysRevD.86.044011} {\bibfield  {journal} {\bibinfo  {journal} {Phys. Rev. D}\ }\textbf {\bibinfo {volume} {86}},\ \bibinfo {pages} {044011} (\bibinfo {year} {2012})},\ \Eprint {http://arxiv.org/abs/1204.1062} {arXiv:1204.1062 [hep-th]} \BibitemShut {NoStop}%
\bibitem [{\citenamefont {Kubiznak}\ and\ \citenamefont {Mann}(2012)}]{Kubiznak:2012wp}%
  \BibitemOpen
  \bibfield  {author} {\bibinfo {author} {\bibfnamefont {D.}~\bibnamefont {Kubiznak}}\ and\ \bibinfo {author} {\bibfnamefont {R.~B.}\ \bibnamefont {Mann}},\ }\href {\doibase 10.1007/JHEP07(2012)033} {\bibfield  {journal} {\bibinfo  {journal} {JHEP}\ }\textbf {\bibinfo {volume} {07}},\ \bibinfo {pages} {033} (\bibinfo {year} {2012})},\ \Eprint {http://arxiv.org/abs/1205.0559} {arXiv:1205.0559 [hep-th]} \BibitemShut {NoStop}%
\bibitem [{\citenamefont {Weinhold}(1975{\natexlab{a}})}]{Weinhold1}%
  \BibitemOpen
  \bibfield  {author} {\bibinfo {author} {\bibfnamefont {F.}~\bibnamefont {Weinhold}},\ }\href@noop {} {\bibfield  {journal} {\bibinfo  {journal} {J. Chem. Phys.}\ }\textbf {\bibinfo {volume} {63}},\ \bibinfo {pages} {2479} (\bibinfo {year} {1975}{\natexlab{a}})}\BibitemShut {NoStop}%
\bibitem [{\citenamefont {Weinhold}(1975{\natexlab{b}})}]{Weinhold2}%
  \BibitemOpen
  \bibfield  {author} {\bibinfo {author} {\bibfnamefont {F.}~\bibnamefont {Weinhold}},\ }\href@noop {} {\bibfield  {journal} {\bibinfo  {journal} {J. Chem. Phys.}\ }\textbf {\bibinfo {volume} {63}},\ \bibinfo {pages} {2484} (\bibinfo {year} {1975}{\natexlab{b}})}\BibitemShut {NoStop}%
\bibitem [{\citenamefont {Ruppeiner}(1979)}]{Rupp1}%
  \BibitemOpen
  \bibfield  {author} {\bibinfo {author} {\bibfnamefont {G.}~\bibnamefont {Ruppeiner}},\ }\href@noop {} {\bibfield  {journal} {\bibinfo  {journal} {Phys. Rev. A}\ }\textbf {\bibinfo {volume} {20}},\ \bibinfo {pages} {1608} (\bibinfo {year} {1979})}\BibitemShut {NoStop}%
\bibitem [{\citenamefont {Ruppeiner}(1995)}]{Rupp2}%
  \BibitemOpen
  \bibfield  {author} {\bibinfo {author} {\bibfnamefont {G.}~\bibnamefont {Ruppeiner}},\ }\href@noop {} {\bibfield  {journal} {\bibinfo  {journal} {Rev. Modern Phys.}\ }\textbf {\bibinfo {volume} {67}},\ \bibinfo {pages} {605} (\bibinfo {year} {1995})}\BibitemShut {NoStop}%
\bibitem [{\citenamefont {Salamon}\ and\ \citenamefont {Ihrig}(1984)}]{WR}%
  \BibitemOpen
  \bibfield  {author} {\bibinfo {author} {\bibfnamefont {J.}~\bibnamefont {Salamon}, \bibfnamefont {P.~Nulton}}\ and\ \bibinfo {author} {\bibfnamefont {E.}~\bibnamefont {Ihrig}},\ }\href@noop {} {\bibfield  {journal} {\bibinfo  {journal} {J. Chem. Phys.}\ }\textbf {\bibinfo {volume} {80}},\ \bibinfo {pages} {436} (\bibinfo {year} {1984})}\BibitemShut {NoStop}%
\bibitem [{\citenamefont {Quevedo}\ and\ \citenamefont {Tapias}(2014)}]{Leg}%
  \BibitemOpen
  \bibfield  {author} {\bibinfo {author} {\bibfnamefont {H.}~\bibnamefont {Quevedo}}\ and\ \bibinfo {author} {\bibfnamefont {D.}~\bibnamefont {Tapias}},\ }\href@noop {} {\bibfield  {journal} {\bibinfo  {journal} {J. Math. Chem.}\ }\textbf {\bibinfo {volume} {52}},\ \bibinfo {pages} {141} (\bibinfo {year} {2014})}\BibitemShut {NoStop}%
\bibitem [{\citenamefont {Quevedo}\ and\ \citenamefont {Quevedo}()}]{Quevedo:2011np}%
  \BibitemOpen
  \bibfield  {author} {\bibinfo {author} {\bibfnamefont {H.}~\bibnamefont {Quevedo}}\ and\ \bibinfo {author} {\bibfnamefont {M.~N.}\ \bibnamefont {Quevedo}},\ }\href@noop {} {\bibinfo  {journal} {arXiv:math-ph/1111.5056}\ }\BibitemShut {NoStop}%
\bibitem [{\citenamefont {Quevedo}(2007)}]{Que1}%
  \BibitemOpen
\bibfield  {journal} {  }\bibfield  {author} {\bibinfo {author} {\bibfnamefont {H.}~\bibnamefont {Quevedo}},\ }\href@noop {} {\bibfield  {journal} {\bibinfo  {journal} {J. Math. Phys.}\ }\textbf {\bibinfo {volume} {48}},\ \bibinfo {pages} {013506} (\bibinfo {year} {2007})}\BibitemShut {NoStop}%
\bibitem [{\citenamefont {Quevedo}(2008{\natexlab{b}})}]{Que2}%
  \BibitemOpen
  \bibfield  {author} {\bibinfo {author} {\bibfnamefont {A.}~\bibnamefont {Quevedo}, \bibfnamefont {H.~Sanchez}},\ }\href@noop {} {\bibfield  {journal} {\bibinfo  {journal} {JHEP.}\ }\textbf {\bibinfo {volume} {09}},\ \bibinfo {pages} {034} (\bibinfo {year} {2008}{\natexlab{b}})}\BibitemShut {NoStop}%
\bibitem [{\citenamefont {Hendi}\ \emph {et~al.}(2015{\natexlab{a}})\citenamefont {Hendi}, \citenamefont {Panahiyan}, \citenamefont {Eslam~Panah},\ and\ \citenamefont {Momennia}}]{Hendi:2015rja}%
  \BibitemOpen
  \bibfield  {author} {\bibinfo {author} {\bibfnamefont {S.~H.}\ \bibnamefont {Hendi}}, \bibinfo {author} {\bibfnamefont {S.}~\bibnamefont {Panahiyan}}, \bibinfo {author} {\bibfnamefont {B.}~\bibnamefont {Eslam~Panah}}, \ and\ \bibinfo {author} {\bibfnamefont {M.}~\bibnamefont {Momennia}},\ }\href {\doibase 10.1140/epjc/s10052-015-3701-5} {\bibfield  {journal} {\bibinfo  {journal} {Eur. Phys. J. C}\ }\textbf {\bibinfo {volume} {75}},\ \bibinfo {pages} {507} (\bibinfo {year} {2015}{\natexlab{a}})},\ \Eprint {http://arxiv.org/abs/1506.08092} {arXiv:1506.08092 [gr-qc]} \BibitemShut {NoStop}%
\bibitem [{\citenamefont {Hendi}\ \emph {et~al.}(2015{\natexlab{b}})\citenamefont {Hendi}, \citenamefont {Panahiyan},\ and\ \citenamefont {Eslam~Panah}}]{Hendi:2015fya}%
  \BibitemOpen
  \bibfield  {author} {\bibinfo {author} {\bibfnamefont {S.~H.}\ \bibnamefont {Hendi}}, \bibinfo {author} {\bibfnamefont {S.}~\bibnamefont {Panahiyan}}, \ and\ \bibinfo {author} {\bibfnamefont {B.}~\bibnamefont {Eslam~Panah}},\ }\href {\doibase 10.1155/2015/743086} {\bibfield  {journal} {\bibinfo  {journal} {Adv. High Energy Phys.}\ }\textbf {\bibinfo {volume} {2015}},\ \bibinfo {pages} {743086} (\bibinfo {year} {2015}{\natexlab{b}})},\ \Eprint {http://arxiv.org/abs/1509.07014} {arXiv:1509.07014 [gr-qc]} \BibitemShut {NoStop}%
\bibitem [{\citenamefont {Hendi}\ \emph {et~al.}(2015{\natexlab{c}})\citenamefont {Hendi}, \citenamefont {Sheykhi}, \citenamefont {Panahiyan},\ and\ \citenamefont {Eslam~Panah}}]{Hendi:2015xya}%
  \BibitemOpen
  \bibfield  {author} {\bibinfo {author} {\bibfnamefont {S.~H.}\ \bibnamefont {Hendi}}, \bibinfo {author} {\bibfnamefont {A.}~\bibnamefont {Sheykhi}}, \bibinfo {author} {\bibfnamefont {S.}~\bibnamefont {Panahiyan}}, \ and\ \bibinfo {author} {\bibfnamefont {B.}~\bibnamefont {Eslam~Panah}},\ }\href {\doibase 10.1103/PhysRevD.92.064028} {\bibfield  {journal} {\bibinfo  {journal} {Phys. Rev. D}\ }\textbf {\bibinfo {volume} {92}},\ \bibinfo {pages} {064028} (\bibinfo {year} {2015}{\natexlab{c}})},\ \Eprint {http://arxiv.org/abs/1509.08593} {arXiv:1509.08593 [hep-th]} \BibitemShut {NoStop}%
\bibitem [{\citenamefont {Hendi}\ \emph {et~al.}(2015{\natexlab{d}})\citenamefont {Hendi}, \citenamefont {Eslam~Panah},\ and\ \citenamefont {Panahiyan}}]{Hendi:2015hoa}%
  \BibitemOpen
  \bibfield  {author} {\bibinfo {author} {\bibfnamefont {S.~H.}\ \bibnamefont {Hendi}}, \bibinfo {author} {\bibfnamefont {B.}~\bibnamefont {Eslam~Panah}}, \ and\ \bibinfo {author} {\bibfnamefont {S.}~\bibnamefont {Panahiyan}},\ }\href {\doibase 10.1007/JHEP11(2015)157} {\bibfield  {journal} {\bibinfo  {journal} {JHEP}\ }\textbf {\bibinfo {volume} {11}},\ \bibinfo {pages} {157} (\bibinfo {year} {2015}{\natexlab{d}})},\ \Eprint {http://arxiv.org/abs/1508.01311} {arXiv:1508.01311 [hep-th]} \BibitemShut {NoStop}%
\bibitem [{\citenamefont {Hendi}\ \emph {et~al.}(2016)\citenamefont {Hendi}, \citenamefont {Panahiyan},\ and\ \citenamefont {Eslam~Panah}}]{Hendi:2015pda}%
  \BibitemOpen
  \bibfield  {author} {\bibinfo {author} {\bibfnamefont {S.~H.}\ \bibnamefont {Hendi}}, \bibinfo {author} {\bibfnamefont {S.}~\bibnamefont {Panahiyan}}, \ and\ \bibinfo {author} {\bibfnamefont {B.}~\bibnamefont {Eslam~Panah}},\ }\href {\doibase 10.1007/JHEP01(2016)129} {\bibfield  {journal} {\bibinfo  {journal} {JHEP}\ }\textbf {\bibinfo {volume} {01}},\ \bibinfo {pages} {129} (\bibinfo {year} {2016})},\ \Eprint {http://arxiv.org/abs/1507.06563} {arXiv:1507.06563 [hep-th]} \BibitemShut {NoStop}%
\bibitem [{\citenamefont {Mansoori}\ and\ \citenamefont {Mirza}(2014)}]{Mansoori:2013pna}%
  \BibitemOpen
  \bibfield  {author} {\bibinfo {author} {\bibfnamefont {S.~A.~H.}\ \bibnamefont {Mansoori}}\ and\ \bibinfo {author} {\bibfnamefont {B.}~\bibnamefont {Mirza}},\ }\href {\doibase 10.1140/epjc/s10052-013-2681-6} {\bibfield  {journal} {\bibinfo  {journal} {Eur. Phys. J. C}\ }\textbf {\bibinfo {volume} {74}},\ \bibinfo {pages} {2681} (\bibinfo {year} {2014})},\ \Eprint {http://arxiv.org/abs/1308.1543} {arXiv:1308.1543 [gr-qc]} \BibitemShut {NoStop}%
\bibitem [{\citenamefont {Mansoori}\ \emph {et~al.}(2015)\citenamefont {Mansoori}, \citenamefont {Mirza},\ and\ \citenamefont {Fazel}}]{Mansoori:2014oia}%
  \BibitemOpen
  \bibfield  {author} {\bibinfo {author} {\bibfnamefont {S.~A.~H.}\ \bibnamefont {Mansoori}}, \bibinfo {author} {\bibfnamefont {B.}~\bibnamefont {Mirza}}, \ and\ \bibinfo {author} {\bibfnamefont {M.}~\bibnamefont {Fazel}},\ }\href {\doibase 10.1007/JHEP04(2015)115} {\bibfield  {journal} {\bibinfo  {journal} {JHEP}\ }\textbf {\bibinfo {volume} {04}},\ \bibinfo {pages} {115} (\bibinfo {year} {2015})},\ \Eprint {http://arxiv.org/abs/1411.2582} {arXiv:1411.2582 [gr-qc]} \BibitemShut {NoStop}%
\bibitem [{\citenamefont {Sekhmani}\ \emph {et~al.}(2023)\citenamefont {Sekhmani}, \citenamefont {Gogoi}, \citenamefont {Baouahi},\ and\ \citenamefont {Dahiri}}]{Sekhmani:2023qqe}%
  \BibitemOpen
  \bibfield  {author} {\bibinfo {author} {\bibfnamefont {Y.}~\bibnamefont {Sekhmani}}, \bibinfo {author} {\bibfnamefont {D.~J.}\ \bibnamefont {Gogoi}}, \bibinfo {author} {\bibfnamefont {M.}~\bibnamefont {Baouahi}}, \ and\ \bibinfo {author} {\bibfnamefont {I.}~\bibnamefont {Dahiri}},\ }\href {\doibase 10.1088/1402-4896/acf7fb} {\bibfield  {journal} {\bibinfo  {journal} {Phys. Scripta}\ }\textbf {\bibinfo {volume} {98}},\ \bibinfo {pages} {105014} (\bibinfo {year} {2023})}\BibitemShut {NoStop}%
\bibitem [{\citenamefont {Aghanim}\ \emph {et~al.}(2020)\citenamefont {Aghanim} \emph {et~al.}}]{Planck:2018vyg}%
  \BibitemOpen
  \bibfield  {author} {\bibinfo {author} {\bibfnamefont {N.}~\bibnamefont {Aghanim}} \emph {et~al.} (\bibinfo {collaboration} {Planck}),\ }\href {\doibase 10.1051/0004-6361/201833910} {\bibfield  {journal} {\bibinfo  {journal} {Astron. Astrophys.}\ }\textbf {\bibinfo {volume} {641}},\ \bibinfo {pages} {A6} (\bibinfo {year} {2020})},\ \bibinfo {note} {[Erratum: Astron.Astrophys. 652, C4 (2021)]},\ \Eprint {http://arxiv.org/abs/1807.06209} {arXiv:1807.06209 [astro-ph.CO]} \BibitemShut {NoStop}%
\bibitem [{\citenamefont {Kiselev}(2003)}]{Kiselev:2002dx}%
  \BibitemOpen
  \bibfield  {author} {\bibinfo {author} {\bibfnamefont {V.~V.}\ \bibnamefont {Kiselev}},\ }\href {\doibase 10.1088/0264-9381/20/6/310} {\bibfield  {journal} {\bibinfo  {journal} {Class. Quant. Grav.}\ }\textbf {\bibinfo {volume} {20}},\ \bibinfo {pages} {1187} (\bibinfo {year} {2003})},\ \Eprint {http://arxiv.org/abs/gr-qc/0210040} {arXiv:gr-qc/0210040} \BibitemShut {NoStop}%
\bibitem [{\citenamefont {Fernando}(2012)}]{Fernando:2012ue}%
  \BibitemOpen
  \bibfield  {author} {\bibinfo {author} {\bibfnamefont {S.}~\bibnamefont {Fernando}},\ }\href {\doibase 10.1007/s10714-012-1368-x} {\bibfield  {journal} {\bibinfo  {journal} {Gen. Rel. Grav.}\ }\textbf {\bibinfo {volume} {44}},\ \bibinfo {pages} {1857} (\bibinfo {year} {2012})},\ \Eprint {http://arxiv.org/abs/1202.1502} {arXiv:1202.1502 [gr-qc]} \BibitemShut {NoStop}%
\bibitem [{\citenamefont {Malakolkalami}\ and\ \citenamefont {Ghaderi}(2015)}]{Malakolkalami:2015cza}%
  \BibitemOpen
  \bibfield  {author} {\bibinfo {author} {\bibfnamefont {B.}~\bibnamefont {Malakolkalami}}\ and\ \bibinfo {author} {\bibfnamefont {K.}~\bibnamefont {Ghaderi}},\ }\href {\doibase 10.1007/s10509-015-2340-5} {\bibfield  {journal} {\bibinfo  {journal} {Astrophys. Space Sci.}\ }\textbf {\bibinfo {volume} {357}},\ \bibinfo {pages} {112} (\bibinfo {year} {2015})}\BibitemShut {NoStop}%
\bibitem [{\citenamefont {Hussain}\ and\ \citenamefont {Ali}(2015)}]{Hussain:2014fca}%
  \BibitemOpen
  \bibfield  {author} {\bibinfo {author} {\bibfnamefont {I.}~\bibnamefont {Hussain}}\ and\ \bibinfo {author} {\bibfnamefont {S.}~\bibnamefont {Ali}},\ }\href {\doibase 10.1007/s10714-015-1883-7} {\bibfield  {journal} {\bibinfo  {journal} {Gen. Rel. Grav.}\ }\textbf {\bibinfo {volume} {47}},\ \bibinfo {pages} {34} (\bibinfo {year} {2015})},\ \Eprint {http://arxiv.org/abs/1408.3111} {arXiv:1408.3111 [gr-qc]} \BibitemShut {NoStop}%
\bibitem [{\citenamefont {Ghosh}\ \emph {et~al.}(2017)\citenamefont {Ghosh}, \citenamefont {Amir},\ and\ \citenamefont {Maharaj}}]{Ghosh:2016ddh}%
  \BibitemOpen
  \bibfield  {author} {\bibinfo {author} {\bibfnamefont {S.~G.}\ \bibnamefont {Ghosh}}, \bibinfo {author} {\bibfnamefont {M.}~\bibnamefont {Amir}}, \ and\ \bibinfo {author} {\bibfnamefont {S.~D.}\ \bibnamefont {Maharaj}},\ }\href {\doibase 10.1140/epjc/s10052-017-5099-8} {\bibfield  {journal} {\bibinfo  {journal} {Eur. Phys. J. C}\ }\textbf {\bibinfo {volume} {77}},\ \bibinfo {pages} {530} (\bibinfo {year} {2017})},\ \Eprint {http://arxiv.org/abs/1611.02936} {arXiv:1611.02936 [gr-qc]} \BibitemShut {NoStop}%
\bibitem [{\citenamefont {Ghosh}\ \emph {et~al.}(2018)\citenamefont {Ghosh}, \citenamefont {Maharaj}, \citenamefont {Baboolal},\ and\ \citenamefont {Lee}}]{Ghosh:2017cuq}%
  \BibitemOpen
  \bibfield  {author} {\bibinfo {author} {\bibfnamefont {S.~G.}\ \bibnamefont {Ghosh}}, \bibinfo {author} {\bibfnamefont {S.~D.}\ \bibnamefont {Maharaj}}, \bibinfo {author} {\bibfnamefont {D.}~\bibnamefont {Baboolal}}, \ and\ \bibinfo {author} {\bibfnamefont {T.-H.}\ \bibnamefont {Lee}},\ }\href {\doibase 10.1140/epjc/s10052-018-5570-1} {\bibfield  {journal} {\bibinfo  {journal} {Eur. Phys. J. C}\ }\textbf {\bibinfo {volume} {78}},\ \bibinfo {pages} {90} (\bibinfo {year} {2018})},\ \Eprint {http://arxiv.org/abs/1708.03884} {arXiv:1708.03884 [gr-qc]} \BibitemShut {NoStop}%
\bibitem [{\citenamefont {Hassaine}\ and\ \citenamefont {Martinez}(2007)}]{Hassaine:2007py}%
  \BibitemOpen
  \bibfield  {author} {\bibinfo {author} {\bibfnamefont {M.}~\bibnamefont {Hassaine}}\ and\ \bibinfo {author} {\bibfnamefont {C.}~\bibnamefont {Martinez}},\ }\href {\doibase 10.1103/PhysRevD.75.027502} {\bibfield  {journal} {\bibinfo  {journal} {Phys. Rev. D}\ }\textbf {\bibinfo {volume} {75}},\ \bibinfo {pages} {027502} (\bibinfo {year} {2007})},\ \Eprint {http://arxiv.org/abs/hep-th/0701058} {arXiv:hep-th/0701058} \BibitemShut {NoStop}%
\bibitem [{\citenamefont {Dey}(2004)}]{Dey:2004yt}%
  \BibitemOpen
  \bibfield  {author} {\bibinfo {author} {\bibfnamefont {T.~K.}\ \bibnamefont {Dey}},\ }\href {\doibase 10.1016/j.physletb.2004.06.047} {\bibfield  {journal} {\bibinfo  {journal} {Phys. Lett. B}\ }\textbf {\bibinfo {volume} {595}},\ \bibinfo {pages} {484} (\bibinfo {year} {2004})},\ \Eprint {http://arxiv.org/abs/hep-th/0406169} {arXiv:hep-th/0406169} \BibitemShut {NoStop}%
\bibitem [{\citenamefont {Cai}\ \emph {et~al.}(2004)\citenamefont {Cai}, \citenamefont {Pang},\ and\ \citenamefont {Wang}}]{Cai:2004eh}%
  \BibitemOpen
  \bibfield  {author} {\bibinfo {author} {\bibfnamefont {R.-G.}\ \bibnamefont {Cai}}, \bibinfo {author} {\bibfnamefont {D.-W.}\ \bibnamefont {Pang}}, \ and\ \bibinfo {author} {\bibfnamefont {A.}~\bibnamefont {Wang}},\ }\href {\doibase 10.1103/PhysRevD.70.124034} {\bibfield  {journal} {\bibinfo  {journal} {Phys. Rev. D}\ }\textbf {\bibinfo {volume} {70}},\ \bibinfo {pages} {124034} (\bibinfo {year} {2004})},\ \Eprint {http://arxiv.org/abs/hep-th/0410158} {arXiv:hep-th/0410158} \BibitemShut {NoStop}%
\bibitem [{\citenamefont {Bizon}(1990)}]{Bizon:1990sr}%
  \BibitemOpen
  \bibfield  {author} {\bibinfo {author} {\bibfnamefont {P.}~\bibnamefont {Bizon}},\ }\href {\doibase 10.1103/PhysRevLett.64.2844} {\bibfield  {journal} {\bibinfo  {journal} {Phys. Rev. Lett.}\ }\textbf {\bibinfo {volume} {64}},\ \bibinfo {pages} {2844} (\bibinfo {year} {1990})}\BibitemShut {NoStop}%
\bibitem [{\citenamefont {Li}\ \emph {et~al.}(2018)\citenamefont {Li}, \citenamefont {Yang},\ and\ \citenamefont {Wu}}]{Li:2018hhy}%
  \BibitemOpen
  \bibfield  {author} {\bibinfo {author} {\bibfnamefont {H.}~\bibnamefont {Li}}, \bibinfo {author} {\bibfnamefont {W.}~\bibnamefont {Yang}}, \ and\ \bibinfo {author} {\bibfnamefont {Y.}~\bibnamefont {Wu}},\ }\href {\doibase 10.1016/j.dark.2018.09.001} {\bibfield  {journal} {\bibinfo  {journal} {Phys. Dark Univ.}\ }\textbf {\bibinfo {volume} {22}},\ \bibinfo {pages} {60} (\bibinfo {year} {2018})}\BibitemShut {NoStop}%
\bibitem [{\citenamefont {Toledo}\ and\ \citenamefont {Bezerra}(2019)}]{Toledo:2019amt}%
  \BibitemOpen
  \bibfield  {author} {\bibinfo {author} {\bibfnamefont {J.~M.}\ \bibnamefont {Toledo}}\ and\ \bibinfo {author} {\bibfnamefont {V.~B.}\ \bibnamefont {Bezerra}},\ }\href {\doibase 10.1140/epjc/s10052-019-6616-8} {\bibfield  {journal} {\bibinfo  {journal} {Eur. Phys. J. C}\ }\textbf {\bibinfo {volume} {79}},\ \bibinfo {pages} {110} (\bibinfo {year} {2019})}\BibitemShut {NoStop}%
\bibitem [{\citenamefont {Gadbail}\ \emph {et~al.}(2022)\citenamefont {Gadbail}, \citenamefont {Arora},\ and\ \citenamefont {Sahoo}}]{Gadbail:2022qnw}%
  \BibitemOpen
  \bibfield  {author} {\bibinfo {author} {\bibfnamefont {G.~N.}\ \bibnamefont {Gadbail}}, \bibinfo {author} {\bibfnamefont {S.}~\bibnamefont {Arora}}, \ and\ \bibinfo {author} {\bibfnamefont {P.~K.}\ \bibnamefont {Sahoo}},\ }\href {\doibase 10.1016/j.dark.2022.101074} {\bibfield  {journal} {\bibinfo  {journal} {Phys. Dark Univ.}\ }\textbf {\bibinfo {volume} {37}},\ \bibinfo {pages} {101074} (\bibinfo {year} {2022})},\ \Eprint {http://arxiv.org/abs/2206.10336} {arXiv:2206.10336 [gr-qc]} \BibitemShut {NoStop}%
\bibitem [{\citenamefont {Bean}\ and\ \citenamefont {Dore}(2003)}]{Bean:2003ae}%
  \BibitemOpen
  \bibfield  {author} {\bibinfo {author} {\bibfnamefont {R.}~\bibnamefont {Bean}}\ and\ \bibinfo {author} {\bibfnamefont {O.}~\bibnamefont {Dore}},\ }\href {\doibase 10.1103/PhysRevD.68.023515} {\bibfield  {journal} {\bibinfo  {journal} {Phys. Rev. D}\ }\textbf {\bibinfo {volume} {68}},\ \bibinfo {pages} {023515} (\bibinfo {year} {2003})},\ \Eprint {http://arxiv.org/abs/astro-ph/0301308} {arXiv:astro-ph/0301308} \BibitemShut {NoStop}%
\bibitem [{\citenamefont {Chiba}(2002)}]{Chiba:2002mw}%
  \BibitemOpen
  \bibfield  {author} {\bibinfo {author} {\bibfnamefont {T.}~\bibnamefont {Chiba}},\ }\href {\doibase 10.1103/PhysRevD.66.063514} {\bibfield  {journal} {\bibinfo  {journal} {Phys. Rev. D}\ }\textbf {\bibinfo {volume} {66}},\ \bibinfo {pages} {063514} (\bibinfo {year} {2002})},\ \Eprint {http://arxiv.org/abs/astro-ph/0206298} {arXiv:astro-ph/0206298} \BibitemShut {NoStop}%
\bibitem [{\citenamefont {Gasperini}\ \emph {et~al.}(2002)\citenamefont {Gasperini}, \citenamefont {Piazza},\ and\ \citenamefont {Veneziano}}]{Gasperini:2001pc}%
  \BibitemOpen
  \bibfield  {author} {\bibinfo {author} {\bibfnamefont {M.}~\bibnamefont {Gasperini}}, \bibinfo {author} {\bibfnamefont {F.}~\bibnamefont {Piazza}}, \ and\ \bibinfo {author} {\bibfnamefont {G.}~\bibnamefont {Veneziano}},\ }\href {\doibase 10.1103/PhysRevD.65.023508} {\bibfield  {journal} {\bibinfo  {journal} {Phys. Rev. D}\ }\textbf {\bibinfo {volume} {65}},\ \bibinfo {pages} {023508} (\bibinfo {year} {2002})},\ \Eprint {http://arxiv.org/abs/gr-qc/0108016} {arXiv:gr-qc/0108016} \BibitemShut {NoStop}%
\bibitem [{\citenamefont {Horava}\ and\ \citenamefont {Minic}(2000)}]{Horava:2000tb}%
  \BibitemOpen
  \bibfield  {author} {\bibinfo {author} {\bibfnamefont {P.}~\bibnamefont {Horava}}\ and\ \bibinfo {author} {\bibfnamefont {D.}~\bibnamefont {Minic}},\ }\href {\doibase 10.1103/PhysRevLett.85.1610} {\bibfield  {journal} {\bibinfo  {journal} {Phys. Rev. Lett.}\ }\textbf {\bibinfo {volume} {85}},\ \bibinfo {pages} {1610} (\bibinfo {year} {2000})},\ \Eprint {http://arxiv.org/abs/hep-th/0001145} {arXiv:hep-th/0001145} \BibitemShut {NoStop}%
\bibitem [{\citenamefont {Kamenshchik}\ \emph {et~al.}(2001)\citenamefont {Kamenshchik}, \citenamefont {Moschella},\ and\ \citenamefont {Pasquier}}]{Kamenshchik:2001cp}%
  \BibitemOpen
  \bibfield  {author} {\bibinfo {author} {\bibfnamefont {A.~Y.}\ \bibnamefont {Kamenshchik}}, \bibinfo {author} {\bibfnamefont {U.}~\bibnamefont {Moschella}}, \ and\ \bibinfo {author} {\bibfnamefont {V.}~\bibnamefont {Pasquier}},\ }\href {\doibase 10.1016/S0370-2693(01)00571-8} {\bibfield  {journal} {\bibinfo  {journal} {Phys. Lett. B}\ }\textbf {\bibinfo {volume} {511}},\ \bibinfo {pages} {265} (\bibinfo {year} {2001})},\ \Eprint {http://arxiv.org/abs/gr-qc/0103004} {arXiv:gr-qc/0103004} \BibitemShut {NoStop}%
\bibitem [{\citenamefont {Chaubey}\ and\ \citenamefont {Shukla}(2015)}]{Chaubey:2014gja}%
  \BibitemOpen
  \bibfield  {author} {\bibinfo {author} {\bibfnamefont {R.}~\bibnamefont {Chaubey}}\ and\ \bibinfo {author} {\bibfnamefont {A.~K.}\ \bibnamefont {Shukla}},\ }\href {\doibase 10.1139/cjp-2014-0225} {\bibfield  {journal} {\bibinfo  {journal} {Can. J. Phys.}\ }\textbf {\bibinfo {volume} {93}},\ \bibinfo {pages} {68} (\bibinfo {year} {2015})}\BibitemShut {NoStop}%
\bibitem [{\citenamefont {Bilic}\ \emph {et~al.}(2002)\citenamefont {Bilic}, \citenamefont {Tupper},\ and\ \citenamefont {Viollier}}]{Bilic:2001cg}%
  \BibitemOpen
  \bibfield  {author} {\bibinfo {author} {\bibfnamefont {N.}~\bibnamefont {Bilic}}, \bibinfo {author} {\bibfnamefont {G.~B.}\ \bibnamefont {Tupper}}, \ and\ \bibinfo {author} {\bibfnamefont {R.~D.}\ \bibnamefont {Viollier}},\ }\href {\doibase 10.1016/S0370-2693(02)01716-1} {\bibfield  {journal} {\bibinfo  {journal} {Phys. Lett. B}\ }\textbf {\bibinfo {volume} {535}},\ \bibinfo {pages} {17} (\bibinfo {year} {2002})},\ \Eprint {http://arxiv.org/abs/astro-ph/0111325} {arXiv:astro-ph/0111325} \BibitemShut {NoStop}%
\bibitem [{\citenamefont {Bento}\ \emph {et~al.}(2002)\citenamefont {Bento}, \citenamefont {Bertolami},\ and\ \citenamefont {Sen}}]{Bento:2002ps}%
  \BibitemOpen
  \bibfield  {author} {\bibinfo {author} {\bibfnamefont {M.~C.}\ \bibnamefont {Bento}}, \bibinfo {author} {\bibfnamefont {O.}~\bibnamefont {Bertolami}}, \ and\ \bibinfo {author} {\bibfnamefont {A.~A.}\ \bibnamefont {Sen}},\ }\href {\doibase 10.1103/PhysRevD.66.043507} {\bibfield  {journal} {\bibinfo  {journal} {Phys. Rev. D}\ }\textbf {\bibinfo {volume} {66}},\ \bibinfo {pages} {043507} (\bibinfo {year} {2002})},\ \Eprint {http://arxiv.org/abs/gr-qc/0202064} {arXiv:gr-qc/0202064} \BibitemShut {NoStop}%
\bibitem [{\citenamefont {Carturan}\ and\ \citenamefont {Finelli}(2003)}]{Carturan:2002si}%
  \BibitemOpen
  \bibfield  {author} {\bibinfo {author} {\bibfnamefont {D.}~\bibnamefont {Carturan}}\ and\ \bibinfo {author} {\bibfnamefont {F.}~\bibnamefont {Finelli}},\ }\href {\doibase 10.1103/PhysRevD.68.103501} {\bibfield  {journal} {\bibinfo  {journal} {Phys. Rev. D}\ }\textbf {\bibinfo {volume} {68}},\ \bibinfo {pages} {103501} (\bibinfo {year} {2003})},\ \Eprint {http://arxiv.org/abs/astro-ph/0211626} {arXiv:astro-ph/0211626} \BibitemShut {NoStop}%
\bibitem [{\citenamefont {Amendola}\ \emph {et~al.}(2003)\citenamefont {Amendola}, \citenamefont {Finelli}, \citenamefont {Burigana},\ and\ \citenamefont {Carturan}}]{Amendola:2003bz}%
  \BibitemOpen
  \bibfield  {author} {\bibinfo {author} {\bibfnamefont {L.}~\bibnamefont {Amendola}}, \bibinfo {author} {\bibfnamefont {F.}~\bibnamefont {Finelli}}, \bibinfo {author} {\bibfnamefont {C.}~\bibnamefont {Burigana}}, \ and\ \bibinfo {author} {\bibfnamefont {D.}~\bibnamefont {Carturan}},\ }\href {\doibase 10.1088/1475-7516/2003/07/005} {\bibfield  {journal} {\bibinfo  {journal} {JCAP}\ }\textbf {\bibinfo {volume} {07}},\ \bibinfo {pages} {005} (\bibinfo {year} {2003})},\ \Eprint {http://arxiv.org/abs/astro-ph/0304325} {arXiv:astro-ph/0304325} \BibitemShut {NoStop}%
\bibitem [{\citenamefont {vom Marttens}\ \emph {et~al.}(2017)\citenamefont {vom Marttens}, \citenamefont {Casarini}, \citenamefont {Zimdahl}, \citenamefont {Hip\'olito-Ricaldi},\ and\ \citenamefont {Mota}}]{vomMarttens:2017cuz}%
  \BibitemOpen
  \bibfield  {author} {\bibinfo {author} {\bibfnamefont {R.~F.}\ \bibnamefont {vom Marttens}}, \bibinfo {author} {\bibfnamefont {L.}~\bibnamefont {Casarini}}, \bibinfo {author} {\bibfnamefont {W.}~\bibnamefont {Zimdahl}}, \bibinfo {author} {\bibfnamefont {W.~S.}\ \bibnamefont {Hip\'olito-Ricaldi}}, \ and\ \bibinfo {author} {\bibfnamefont {D.~F.}\ \bibnamefont {Mota}},\ }\href {\doibase 10.1016/j.dark.2017.02.001} {\bibfield  {journal} {\bibinfo  {journal} {Phys. Dark Univ.}\ }\textbf {\bibinfo {volume} {15}},\ \bibinfo {pages} {114} (\bibinfo {year} {2017})},\ \Eprint {http://arxiv.org/abs/1702.00651} {arXiv:1702.00651 [astro-ph.CO]} \BibitemShut {NoStop}%
\bibitem [{\citenamefont {Sengupta}\ \emph {et~al.}()\citenamefont {Sengupta}, \citenamefont {Paul}, \citenamefont {Paul},\ and\ \citenamefont {Kalam}}]{Sengupta:2023yxh}%
  \BibitemOpen
  \bibfield  {author} {\bibinfo {author} {\bibfnamefont {R.}~\bibnamefont {Sengupta}}, \bibinfo {author} {\bibfnamefont {P.}~\bibnamefont {Paul}}, \bibinfo {author} {\bibfnamefont {B.~C.}\ \bibnamefont {Paul}}, \ and\ \bibinfo {author} {\bibfnamefont {M.}~\bibnamefont {Kalam}},\ }\href@noop {} {\bibinfo  {journal} {2307.02602 [gr-qc]}\ }\BibitemShut {NoStop}%
\bibitem [{\citenamefont {Abdullah}\ \emph {et~al.}(2022)\citenamefont {Abdullah}, \citenamefont {El-Zant},\ and\ \citenamefont {Ellithi}}]{Abdullah:2021tee}%
  \BibitemOpen
\bibfield  {journal} {  }\bibfield  {author} {\bibinfo {author} {\bibfnamefont {A.}~\bibnamefont {Abdullah}}, \bibinfo {author} {\bibfnamefont {A.~A.}\ \bibnamefont {El-Zant}}, \ and\ \bibinfo {author} {\bibfnamefont {A.}~\bibnamefont {Ellithi}},\ }\href {\doibase 10.1103/PhysRevD.106.083524} {\bibfield  {journal} {\bibinfo  {journal} {Phys. Rev. D}\ }\textbf {\bibinfo {volume} {106}},\ \bibinfo {pages} {083524} (\bibinfo {year} {2022})},\ \Eprint {http://arxiv.org/abs/2108.03260} {arXiv:2108.03260 [astro-ph.CO]} \BibitemShut {NoStop}%
\bibitem [{\citenamefont {Li}\ \emph {et~al.}(2020)\citenamefont {Li}, \citenamefont {Chen},\ and\ \citenamefont {Xing}}]{Li:2019lhr}%
  \BibitemOpen
  \bibfield  {author} {\bibinfo {author} {\bibfnamefont {X.-Q.}\ \bibnamefont {Li}}, \bibinfo {author} {\bibfnamefont {B.}~\bibnamefont {Chen}}, \ and\ \bibinfo {author} {\bibfnamefont {L.-l.}\ \bibnamefont {Xing}},\ }\href {\doibase 10.1140/epjp/s13360-020-00231-z} {\bibfield  {journal} {\bibinfo  {journal} {Eur. Phys. J. Plus}\ }\textbf {\bibinfo {volume} {135}},\ \bibinfo {pages} {175} (\bibinfo {year} {2020})},\ \Eprint {http://arxiv.org/abs/1905.08156} {arXiv:1905.08156 [gr-qc]} \BibitemShut {NoStop}%
\bibitem [{\citenamefont {Li}\ \emph {et~al.}(2022{\natexlab{a}})\citenamefont {Li}, \citenamefont {Chen},\ and\ \citenamefont {Xing}}]{Li:2019ndh}%
  \BibitemOpen
  \bibfield  {author} {\bibinfo {author} {\bibfnamefont {X.-Q.}\ \bibnamefont {Li}}, \bibinfo {author} {\bibfnamefont {B.}~\bibnamefont {Chen}}, \ and\ \bibinfo {author} {\bibfnamefont {L.-l.}\ \bibnamefont {Xing}},\ }\href {\doibase 10.1140/epjp/s13360-022-03379-y} {\bibfield  {journal} {\bibinfo  {journal} {Eur. Phys. J. Plus}\ }\textbf {\bibinfo {volume} {137}},\ \bibinfo {pages} {1167} (\bibinfo {year} {2022}{\natexlab{a}})},\ \Eprint {http://arxiv.org/abs/1908.09827} {arXiv:1908.09827 [gr-qc]} \BibitemShut {NoStop}%
\bibitem [{\citenamefont {Li}\ \emph {et~al.}(2022{\natexlab{b}})\citenamefont {Li}, \citenamefont {Chen},\ and\ \citenamefont {Xing}}]{Li:2022csn}%
  \BibitemOpen
  \bibfield  {author} {\bibinfo {author} {\bibfnamefont {X.-Q.}\ \bibnamefont {Li}}, \bibinfo {author} {\bibfnamefont {B.}~\bibnamefont {Chen}}, \ and\ \bibinfo {author} {\bibfnamefont {L.-L.}\ \bibnamefont {Xing}},\ }\href {\doibase 10.1016/j.aop.2022.169125} {\bibfield  {journal} {\bibinfo  {journal} {Annals Phys.}\ }\textbf {\bibinfo {volume} {446}},\ \bibinfo {pages} {169125} (\bibinfo {year} {2022}{\natexlab{b}})}\BibitemShut {NoStop}%
\bibitem [{\citenamefont {Li}\ \emph {et~al.}(2023)\citenamefont {Li}, \citenamefont {Yan}, \citenamefont {Xing},\ and\ \citenamefont {Zhou}}]{Li:2023zfl}%
  \BibitemOpen
  \bibfield  {author} {\bibinfo {author} {\bibfnamefont {X.-Q.}\ \bibnamefont {Li}}, \bibinfo {author} {\bibfnamefont {H.-P.}\ \bibnamefont {Yan}}, \bibinfo {author} {\bibfnamefont {L.-L.}\ \bibnamefont {Xing}}, \ and\ \bibinfo {author} {\bibfnamefont {S.-W.}\ \bibnamefont {Zhou}},\ }\href {\doibase 10.1103/PhysRevD.107.104055} {\bibfield  {journal} {\bibinfo  {journal} {Phys. Rev. D}\ }\textbf {\bibinfo {volume} {107}},\ \bibinfo {pages} {104055} (\bibinfo {year} {2023})},\ \Eprint {http://arxiv.org/abs/2305.03028} {arXiv:2305.03028 [gr-qc]} \BibitemShut {NoStop}%
\bibitem [{\citenamefont {Aviles}\ \emph {et~al.}(2012)\citenamefont {Aviles}, \citenamefont {Bastarrachea-Almodovar}, \citenamefont {Campuzano},\ and\ \citenamefont {Quevedo}}]{Aviles:2012et}%
  \BibitemOpen
  \bibfield  {author} {\bibinfo {author} {\bibfnamefont {A.}~\bibnamefont {Aviles}}, \bibinfo {author} {\bibfnamefont {A.}~\bibnamefont {Bastarrachea-Almodovar}}, \bibinfo {author} {\bibfnamefont {L.}~\bibnamefont {Campuzano}}, \ and\ \bibinfo {author} {\bibfnamefont {H.}~\bibnamefont {Quevedo}},\ }\href {\doibase 10.1103/PhysRevD.86.063508} {\bibfield  {journal} {\bibinfo  {journal} {Phys. Rev. D}\ }\textbf {\bibinfo {volume} {86}},\ \bibinfo {pages} {063508} (\bibinfo {year} {2012})},\ \Eprint {http://arxiv.org/abs/1203.4637} {arXiv:1203.4637 [gr-qc]} \BibitemShut {NoStop}%
\bibitem [{\citenamefont {Debnath}\ \emph {et~al.}(2004)\citenamefont {Debnath}, \citenamefont {Banerjee},\ and\ \citenamefont {Chakraborty}}]{Debnath:2004cd}%
  \BibitemOpen
  \bibfield  {author} {\bibinfo {author} {\bibfnamefont {U.}~\bibnamefont {Debnath}}, \bibinfo {author} {\bibfnamefont {A.}~\bibnamefont {Banerjee}}, \ and\ \bibinfo {author} {\bibfnamefont {S.}~\bibnamefont {Chakraborty}},\ }\href {\doibase 10.1088/0264-9381/21/23/019} {\bibfield  {journal} {\bibinfo  {journal} {Class. Quant. Grav.}\ }\textbf {\bibinfo {volume} {21}},\ \bibinfo {pages} {5609} (\bibinfo {year} {2004})},\ \Eprint {http://arxiv.org/abs/gr-qc/0411015} {arXiv:gr-qc/0411015} \BibitemShut {NoStop}%
\bibitem [{\citenamefont {Benaoum}(2022)}]{Benaoum:2002zs}%
  \BibitemOpen
  \bibfield  {author} {\bibinfo {author} {\bibfnamefont {H.}~\bibnamefont {Benaoum}},\ }\href {\doibase 10.3390/universe8070340} {\bibfield  {journal} {\bibinfo  {journal} {Universe}\ }\textbf {\bibinfo {volume} {8}},\ \bibinfo {pages} {340} (\bibinfo {year} {2022})},\ \Eprint {http://arxiv.org/abs/hep-th/0205140} {arXiv:hep-th/0205140} \BibitemShut {NoStop}%
\bibitem [{\citenamefont {Chimento}(2004)}]{Chimento:2003ta}%
  \BibitemOpen
  \bibfield  {author} {\bibinfo {author} {\bibfnamefont {L.~P.}\ \bibnamefont {Chimento}},\ }\href {\doibase 10.1103/PhysRevD.69.123517} {\bibfield  {journal} {\bibinfo  {journal} {Phys. Rev. D}\ }\textbf {\bibinfo {volume} {69}},\ \bibinfo {pages} {123517} (\bibinfo {year} {2004})},\ \Eprint {http://arxiv.org/abs/astro-ph/0311613} {arXiv:astro-ph/0311613} \BibitemShut {NoStop}%
\bibitem [{\citenamefont {Sharif}\ \emph {et~al.}(2012)\citenamefont {Sharif}, \citenamefont {Yesmakhanova}, \citenamefont {Rani},\ and\ \citenamefont {Myrzakulov}}]{Sharif:2012cu}%
  \BibitemOpen
  \bibfield  {author} {\bibinfo {author} {\bibfnamefont {M.}~\bibnamefont {Sharif}}, \bibinfo {author} {\bibfnamefont {K.}~\bibnamefont {Yesmakhanova}}, \bibinfo {author} {\bibfnamefont {S.}~\bibnamefont {Rani}}, \ and\ \bibinfo {author} {\bibfnamefont {R.}~\bibnamefont {Myrzakulov}},\ }\href {\doibase 10.1140/epjc/s10052-012-2067-1} {\bibfield  {journal} {\bibinfo  {journal} {Eur. Phys. J. C}\ }\textbf {\bibinfo {volume} {72}},\ \bibinfo {pages} {2067} (\bibinfo {year} {2012})},\ \Eprint {http://arxiv.org/abs/1204.2181} {arXiv:1204.2181 [physics.gen-ph]} \BibitemShut {NoStop}%
\bibitem [{\citenamefont {Myrzakulov}\ \emph {et~al.}(2011)\citenamefont {Myrzakulov}, \citenamefont {Jamil},\ and\ \citenamefont {Razina}}]{Myrzakulov:2011cm}%
  \BibitemOpen
  \bibfield  {author} {\bibinfo {author} {\bibfnamefont {R.}~\bibnamefont {Myrzakulov}}, \bibinfo {author} {\bibfnamefont {M.}~\bibnamefont {Jamil}}, \ and\ \bibinfo {author} {\bibfnamefont {O.}~\bibnamefont {Razina}},\ }\href {\doibase 10.1007/s10509-011-0870-z} {\bibfield  {journal} {\bibinfo  {journal} {Astrophys. Space Sci.}\ }\textbf {\bibinfo {volume} {336}},\ \bibinfo {pages} {315} (\bibinfo {year} {2011})},\ \Eprint {http://arxiv.org/abs/1107.1008} {arXiv:1107.1008 [physics.gen-ph]} \BibitemShut {NoStop}%
\bibitem [{\citenamefont {Debnath}(2020)}]{Debnath:2019mzs}%
  \BibitemOpen
  \bibfield  {author} {\bibinfo {author} {\bibfnamefont {U.}~\bibnamefont {Debnath}},\ }\href {\doibase 10.1140/epjp/s13360-020-00416-6} {\bibfield  {journal} {\bibinfo  {journal} {Eur. Phys. J. Plus}\ }\textbf {\bibinfo {volume} {135}},\ \bibinfo {pages} {424} (\bibinfo {year} {2020})},\ \Eprint {http://arxiv.org/abs/1903.04379} {arXiv:1903.04379 [gr-qc]} \BibitemShut {NoStop}%
\bibitem [{\citenamefont {Debnath}\ \emph {et~al.}(2022)\citenamefont {Debnath}, \citenamefont {Pourhassan},\ and\ \citenamefont {Sakalli}}]{Debnath:2019yit}%
  \BibitemOpen
  \bibfield  {author} {\bibinfo {author} {\bibfnamefont {U.}~\bibnamefont {Debnath}}, \bibinfo {author} {\bibfnamefont {B.}~\bibnamefont {Pourhassan}}, \ and\ \bibinfo {author} {\bibfnamefont {I.}~\bibnamefont {Sakalli}},\ }\href {\doibase 10.1142/S0217732322500857} {\bibfield  {journal} {\bibinfo  {journal} {Mod. Phys. Lett. A}\ }\textbf {\bibinfo {volume} {37}},\ \bibinfo {pages} {2250085} (\bibinfo {year} {2022})},\ \Eprint {http://arxiv.org/abs/1910.00466} {arXiv:1910.00466 [gr-qc]} \BibitemShut {NoStop}%
\bibitem [{\citenamefont {Abramowitz}(1965)}]{Abra}%
  \BibitemOpen
  \bibfield  {author} {\bibinfo {author} {\bibfnamefont {I.}~\bibnamefont {Abramowitz}, \bibfnamefont {M.~Stegun}},\ }\href@noop {} {\emph {\bibinfo {title} {{Handbook of Mathematical Functions}}}}\ (\bibinfo {year} {1965})\BibitemShut {NoStop}%
\bibitem [{\citenamefont {Balart}\ and\ \citenamefont {Vagenas}(2014)}]{Balart:2014cga}%
  \BibitemOpen
  \bibfield  {author} {\bibinfo {author} {\bibfnamefont {L.}~\bibnamefont {Balart}}\ and\ \bibinfo {author} {\bibfnamefont {E.~C.}\ \bibnamefont {Vagenas}},\ }\href {\doibase 10.1103/PhysRevD.90.124045} {\bibfield  {journal} {\bibinfo  {journal} {Phys. Rev. D}\ }\textbf {\bibinfo {volume} {90}},\ \bibinfo {pages} {124045} (\bibinfo {year} {2014})},\ \Eprint {http://arxiv.org/abs/1408.0306} {arXiv:1408.0306 [gr-qc]} \BibitemShut {NoStop}%
\bibitem [{\citenamefont {Hawking}\ and\ \citenamefont {Ellis}(2023)}]{Hawking:1973uf}%
  \BibitemOpen
  \bibfield  {author} {\bibinfo {author} {\bibfnamefont {S.~W.}\ \bibnamefont {Hawking}}\ and\ \bibinfo {author} {\bibfnamefont {G.~F.~R.}\ \bibnamefont {Ellis}},\ }\href {\doibase 10.1017/9781009253161} {\emph {\bibinfo {title} {{The Large Scale Structure of Space-Time}}}},\ Cambridge Monographs on Mathematical Physics\ (\bibinfo  {publisher} {Cambridge University Press},\ \bibinfo {year} {2023})\BibitemShut {NoStop}%
\bibitem [{\citenamefont {Ghosh}\ and\ \citenamefont {Kothawala}(2008)}]{Ghosh:2008zza}%
  \BibitemOpen
  \bibfield  {author} {\bibinfo {author} {\bibfnamefont {S.~G.}\ \bibnamefont {Ghosh}}\ and\ \bibinfo {author} {\bibfnamefont {D.}~\bibnamefont {Kothawala}},\ }\href {\doibase 10.1007/s10714-007-0511-6} {\bibfield  {journal} {\bibinfo  {journal} {Gen. Rel. Grav.}\ }\textbf {\bibinfo {volume} {40}},\ \bibinfo {pages} {9} (\bibinfo {year} {2008})},\ \Eprint {http://arxiv.org/abs/0801.4342} {arXiv:0801.4342 [gr-qc]} \BibitemShut {NoStop}%
\bibitem [{\citenamefont {Kothawala}\ and\ \citenamefont {Ghosh}(2004)}]{Kothawala:2004fy}%
  \BibitemOpen
  \bibfield  {author} {\bibinfo {author} {\bibfnamefont {D.}~\bibnamefont {Kothawala}}\ and\ \bibinfo {author} {\bibfnamefont {S.~G.}\ \bibnamefont {Ghosh}},\ }\href {\doibase 10.1103/PhysRevD.70.104010} {\bibfield  {journal} {\bibinfo  {journal} {Phys. Rev. D}\ }\textbf {\bibinfo {volume} {70}},\ \bibinfo {pages} {104010} (\bibinfo {year} {2004})},\ \Eprint {http://arxiv.org/abs/1007.2500} {arXiv:1007.2500 [gr-qc]} \BibitemShut {NoStop}%
\bibitem [{\citenamefont {Ghosh}\ and\ \citenamefont {Kumar}(2020)}]{Ghosh:2020syx}%
  \BibitemOpen
  \bibfield  {author} {\bibinfo {author} {\bibfnamefont {S.~G.}\ \bibnamefont {Ghosh}}\ and\ \bibinfo {author} {\bibfnamefont {R.}~\bibnamefont {Kumar}},\ }\href {\doibase 10.1088/1361-6382/abc134} {\bibfield  {journal} {\bibinfo  {journal} {Class. Quant. Grav.}\ }\textbf {\bibinfo {volume} {37}},\ \bibinfo {pages} {245008} (\bibinfo {year} {2020})},\ \Eprint {http://arxiv.org/abs/2003.12291} {arXiv:2003.12291 [gr-qc]} \BibitemShut {NoStop}%
\bibitem [{\citenamefont {Kubiznak}\ \emph {et~al.}(2017)\citenamefont {Kubiznak}, \citenamefont {Mann},\ and\ \citenamefont {Teo}}]{Kubiznak:2016qmn}%
  \BibitemOpen
  \bibfield  {author} {\bibinfo {author} {\bibfnamefont {D.}~\bibnamefont {Kubiznak}}, \bibinfo {author} {\bibfnamefont {R.~B.}\ \bibnamefont {Mann}}, \ and\ \bibinfo {author} {\bibfnamefont {M.}~\bibnamefont {Teo}},\ }\href {\doibase 10.1088/1361-6382/aa5c69} {\bibfield  {journal} {\bibinfo  {journal} {Class. Quant. Grav.}\ }\textbf {\bibinfo {volume} {34}},\ \bibinfo {pages} {063001} (\bibinfo {year} {2017})},\ \Eprint {http://arxiv.org/abs/1608.06147} {arXiv:1608.06147 [hep-th]} \BibitemShut {NoStop}%
\bibitem [{\citenamefont {Cai}\ and\ \citenamefont {Soh}(1999)}]{Cai:1998vy}%
  \BibitemOpen
  \bibfield  {author} {\bibinfo {author} {\bibfnamefont {R.-G.}\ \bibnamefont {Cai}}\ and\ \bibinfo {author} {\bibfnamefont {K.-S.}\ \bibnamefont {Soh}},\ }\href {\doibase 10.1103/PhysRevD.59.044013} {\bibfield  {journal} {\bibinfo  {journal} {Phys. Rev. D}\ }\textbf {\bibinfo {volume} {59}},\ \bibinfo {pages} {044013} (\bibinfo {year} {1999})},\ \Eprint {http://arxiv.org/abs/gr-qc/9808067} {arXiv:gr-qc/9808067} \BibitemShut {NoStop}%
\bibitem [{\citenamefont {Bekenstein}(1972)}]{Bekenstein:1972tm}%
  \BibitemOpen
  \bibfield  {author} {\bibinfo {author} {\bibfnamefont {J.~D.}\ \bibnamefont {Bekenstein}},\ }\href {\doibase 10.1007/BF02757029} {\bibfield  {journal} {\bibinfo  {journal} {Lett. Nuovo Cim.}\ }\textbf {\bibinfo {volume} {4}},\ \bibinfo {pages} {737} (\bibinfo {year} {1972})}\BibitemShut {NoStop}%
\bibitem [{\citenamefont {Bekenstein}(1974)}]{Bekenstein:1974ax}%
  \BibitemOpen
  \bibfield  {author} {\bibinfo {author} {\bibfnamefont {J.~D.}\ \bibnamefont {Bekenstein}},\ }\href {\doibase 10.1103/PhysRevD.9.3292} {\bibfield  {journal} {\bibinfo  {journal} {Phys. Rev. D}\ }\textbf {\bibinfo {volume} {9}},\ \bibinfo {pages} {3292} (\bibinfo {year} {1974})}\BibitemShut {NoStop}%
\bibitem [{\citenamefont {Gunasekaran}\ \emph {et~al.}(2012)\citenamefont {Gunasekaran}, \citenamefont {Mann},\ and\ \citenamefont {Kubiznak}}]{Gunasekaran:2012dq}%
  \BibitemOpen
  \bibfield  {author} {\bibinfo {author} {\bibfnamefont {S.}~\bibnamefont {Gunasekaran}}, \bibinfo {author} {\bibfnamefont {R.~B.}\ \bibnamefont {Mann}}, \ and\ \bibinfo {author} {\bibfnamefont {D.}~\bibnamefont {Kubiznak}},\ }\href {\doibase 10.1007/JHEP11(2012)110} {\bibfield  {journal} {\bibinfo  {journal} {JHEP}\ }\textbf {\bibinfo {volume} {11}},\ \bibinfo {pages} {110} (\bibinfo {year} {2012})},\ \Eprint {http://arxiv.org/abs/1208.6251} {arXiv:1208.6251 [hep-th]} \BibitemShut {NoStop}%
\bibitem [{\citenamefont {Brown}\ and\ \citenamefont {Teitelboim}(1987)}]{Brown:1987dd}%
  \BibitemOpen
  \bibfield  {author} {\bibinfo {author} {\bibfnamefont {J.~D.}\ \bibnamefont {Brown}}\ and\ \bibinfo {author} {\bibfnamefont {C.}~\bibnamefont {Teitelboim}},\ }\href {\doibase 10.1016/0370-2693(87)91190-7} {\bibfield  {journal} {\bibinfo  {journal} {Phys. Lett. B}\ }\textbf {\bibinfo {volume} {195}},\ \bibinfo {pages} {177} (\bibinfo {year} {1987})}\BibitemShut {NoStop}%
\bibitem [{\citenamefont {Altamirano}\ \emph {et~al.}(2014)\citenamefont {Altamirano}, \citenamefont {Kubiznak}, \citenamefont {Mann},\ and\ \citenamefont {Sherkatghanad}}]{Altamirano:2014tva}%
  \BibitemOpen
  \bibfield  {author} {\bibinfo {author} {\bibfnamefont {N.}~\bibnamefont {Altamirano}}, \bibinfo {author} {\bibfnamefont {D.}~\bibnamefont {Kubiznak}}, \bibinfo {author} {\bibfnamefont {R.~B.}\ \bibnamefont {Mann}}, \ and\ \bibinfo {author} {\bibfnamefont {Z.}~\bibnamefont {Sherkatghanad}},\ }\href {\doibase 10.3390/galaxies2010089} {\bibfield  {journal} {\bibinfo  {journal} {Galaxies}\ }\textbf {\bibinfo {volume} {2}},\ \bibinfo {pages} {89} (\bibinfo {year} {2014})},\ \Eprint {http://arxiv.org/abs/1401.2586} {arXiv:1401.2586 [hep-th]} \BibitemShut {NoStop}%
\bibitem [{\citenamefont {Kumar}\ \emph {et~al.}(2020)\citenamefont {Kumar}, \citenamefont {Ghosh},\ and\ \citenamefont {Maharaj}}]{Kumar:2020cve}%
  \BibitemOpen
  \bibfield  {author} {\bibinfo {author} {\bibfnamefont {A.}~\bibnamefont {Kumar}}, \bibinfo {author} {\bibfnamefont {S.~G.}\ \bibnamefont {Ghosh}}, \ and\ \bibinfo {author} {\bibfnamefont {S.~D.}\ \bibnamefont {Maharaj}},\ }\href {\doibase 10.1016/j.dark.2020.100634} {\bibfield  {journal} {\bibinfo  {journal} {Phys. Dark Univ.}\ }\textbf {\bibinfo {volume} {30}},\ \bibinfo {pages} {100634} (\bibinfo {year} {2020})},\ \Eprint {http://arxiv.org/abs/2106.15925} {arXiv:2106.15925 [gr-qc]} \BibitemShut {NoStop}%
\bibitem [{\citenamefont {Sahabandu}\ \emph {et~al.}(2006)\citenamefont {Sahabandu}, \citenamefont {Suranyi}, \citenamefont {Vaz},\ and\ \citenamefont {Wijewardhana}}]{Sahabandu:2005ma}%
  \BibitemOpen
  \bibfield  {author} {\bibinfo {author} {\bibfnamefont {C.}~\bibnamefont {Sahabandu}}, \bibinfo {author} {\bibfnamefont {P.}~\bibnamefont {Suranyi}}, \bibinfo {author} {\bibfnamefont {C.}~\bibnamefont {Vaz}}, \ and\ \bibinfo {author} {\bibfnamefont {L.~C.~R.}\ \bibnamefont {Wijewardhana}},\ }\href {\doibase 10.1103/PhysRevD.73.044009} {\bibfield  {journal} {\bibinfo  {journal} {Phys. Rev. D}\ }\textbf {\bibinfo {volume} {73}},\ \bibinfo {pages} {044009} (\bibinfo {year} {2006})},\ \Eprint {http://arxiv.org/abs/gr-qc/0509102} {arXiv:gr-qc/0509102} \BibitemShut {NoStop}%
\bibitem [{\citenamefont {Cai}(2004)}]{Cai:2003kt}%
  \BibitemOpen
  \bibfield  {author} {\bibinfo {author} {\bibfnamefont {R.-G.}\ \bibnamefont {Cai}},\ }\href {\doibase 10.1016/j.physletb.2004.01.015} {\bibfield  {journal} {\bibinfo  {journal} {Phys. Lett. B}\ }\textbf {\bibinfo {volume} {582}},\ \bibinfo {pages} {237} (\bibinfo {year} {2004})},\ \Eprint {http://arxiv.org/abs/hep-th/0311240} {arXiv:hep-th/0311240} \BibitemShut {NoStop}%
\bibitem [{\citenamefont {Tzikas}(2019)}]{Tzikas:2018cvs}%
  \BibitemOpen
  \bibfield  {author} {\bibinfo {author} {\bibfnamefont {A.~G.}\ \bibnamefont {Tzikas}},\ }\href {\doibase 10.1016/j.physletb.2018.11.036} {\bibfield  {journal} {\bibinfo  {journal} {Phys. Lett. B}\ }\textbf {\bibinfo {volume} {788}},\ \bibinfo {pages} {219} (\bibinfo {year} {2019})},\ \Eprint {http://arxiv.org/abs/1811.01104} {arXiv:1811.01104 [gr-qc]} \BibitemShut {NoStop}%
\bibitem [{\citenamefont {Soroushfar}\ \emph {et~al.}(2016)\citenamefont {Soroushfar}, \citenamefont {Saffari},\ and\ \citenamefont {Kamvar}}]{Soroushfar:2016nbu}%
  \BibitemOpen
  \bibfield  {author} {\bibinfo {author} {\bibfnamefont {S.}~\bibnamefont {Soroushfar}}, \bibinfo {author} {\bibfnamefont {R.}~\bibnamefont {Saffari}}, \ and\ \bibinfo {author} {\bibfnamefont {N.}~\bibnamefont {Kamvar}},\ }\href {\doibase 10.1140/epjc/s10052-016-4311-6} {\bibfield  {journal} {\bibinfo  {journal} {Eur. Phys. J. C}\ }\textbf {\bibinfo {volume} {76}},\ \bibinfo {pages} {476} (\bibinfo {year} {2016})},\ \Eprint {http://arxiv.org/abs/1605.00767} {arXiv:1605.00767 [gr-qc]} \BibitemShut {NoStop}%
\bibitem [{\citenamefont {Mrugala}(1984)}]{Mrugala1984OnEO}%
  \BibitemOpen
  \bibfield  {author} {\bibinfo {author} {\bibfnamefont {R.}~\bibnamefont {Mrugala}},\ }\href {https://api.semanticscholar.org/CorpusID:122879706} {\bibfield  {journal} {\bibinfo  {journal} {Physica A-statistical Mechanics and Its Applications}\ }\textbf {\bibinfo {volume} {125}},\ \bibinfo {pages} {631} (\bibinfo {year} {1984})}\BibitemShut {NoStop}%
\end{thebibliography}%

\end{document}